\PassOptionsToPackage{pdfpagelabels=false}{hyperref} 
\documentclass[fleqn]{mnras}

\usepackage{graphicx}	
\usepackage{amsmath}	
\usepackage{amssymb}	
\usepackage{color}
\usepackage{float}
\usepackage{ulem}
\usepackage{rotating}
\usepackage{booktabs}
\usepackage{tabularx}
\usepackage{longtable}
\usepackage{natbib}
\usepackage{xcolor}
\def\eq{\begin{equation}}
\def\en{\end{equation}}

\def\lesssim{\raisebox{-0.3ex}{\mbox{$\stackrel{<}{_\sim} \,$}}}

\def\etal{{\it et al}\thinspace}

\def\apj{{\it Ap.J.}\thinspace}
\def\apjl{{\it Ap.J. Lett.}\thinspace}
\def\apjs{{\it Ap.J. Suppl.}\thinspace}
\def\aj{{\it A.J.}\thinspace}

\def\aap{{\it A\&A}\thinspace}
\def\aaps{{\it A\&A Suppl.}\thinspace}
\def\mnras{{\it MNRAS}\thinspace}
\def\P3hat{{\mathaccent 94 P}_3}

\def\nat{{\it Nature}\thinspace}
\def\deg{^{\circ}}
\def\etal{{\it et al.}\thinspace}

\def\eg{{\it e.g.,}\thinspace}

\title[Emission Beam Geometry at Low Frequency]{Radio Pulsar Emission-Beam Geometry at Low Frequency:  LOFAR High Band Survey Sources Studied using Arecibo at 1.4 GHz and 327 MHz\thanks{This paper is dedicated to our colleagues at the Institute for
Astronomy, Kharkiv, Ukraine}}

\author[Wahl, Rankin, Venkataraman, \&  Olszanski]
{Haley Wahl$^{1, 2, 3}$\thanks{E-mail: hmw0023@mix.wvu.edu}, Joanna Rankin$^{3,4}$, Arun Venkataraman{$^5$}, Timothy Olszanski$^{1, 2, 3}$\\
$^{1}$Department of Physics and Astronomy, West Virginia University, P.O. Box 6315, Morgantown, WV 26505\\
$^{2}$Center for Gravitational Waves and Cosmology, West Virginia University, Morgantown, WV 26505\\
$^{3}$Physics Department, University of Vermont, Burlington, VT 05405, USA \\
$^{4}$Anton Pannekoek Institute for Astronomy, University of Amsterdam, Science Park 904, 1098 XH Amsterdam \\
$^{5}$Arecibo Observatory, bo. La Esperanza, P.P. Box 53995, Arecibo, Puerto Rico, 0612}

\date{Accepted XXX. Received XXX; in original form XXX}

\pubyear{XX}

\begin{document}
\label{firstpage}
\pagerange{\pageref{firstpage}--\pageref{lastpage}}
\maketitle

\begin{abstract}
This paper continues our study of radio pulsar emission-beam configurations with the primary intent of extending study to the lowest possible frequencies. Here we focus on a group of 133 more recently discovered pulsars, most of which were included in the (100-200 MHz) LOFAR High Band Survey, observed with Arecibo at 1.4 GHz and 327 MHz, and some observed at decameter wavelengths.  Our analysis framework is the core/double-cone beam model, and we took opportunity to apply it as widely as possible, both conceptually and quantitatively, while highlighting situations where modeling is difficult, or where its premises may be violated.  In the great majority of pulsars, beam forms consistent with the core/double-cone model were identified.  Moreover, we found that each pulsar's beam structure remained largely constant over the frequency range available; where profile variations were observed, they were attributable to different component spectra and in some instances to varying conal beam sizes.  As an Arecibo population, many or most of the objects tend to fall in the Galactic anticenter region and/or at high Galactic latitudes, so overall it includes a number of nearer, older pulsars.  We found a number of interesting or unusual characteristics in some of the pulsars that would benefit from additional study.  The scattering levels encountered for this group are low to moderate, apart from a few pulsars lying in directions more toward the inner Galaxy.
\end{abstract}

\begin{keywords}
stars: pulsars: general; physical data and processes: polarizationradiation mechanisms: non-thermalISM: structure
\end{keywords}



\section{Introduction}
The narrow emission beams of radio pulsars provide particular challenges because each shines toward the Earth only owing to a pulsar's accidental alignment.  Therefore, there is typically no direct means of discerning just what portion of the entire beam our sightline encounters.  Multiple efforts to understand pulsar beamforms began shortly after the discovery of pulsars, and this history is reviewed in recent publications by both \citet{paperI} and \citet{paperII}. Nonetheless, a comprehensive understanding of pulsar beamforms has remained elusive, but crucial to comprehending the physical mechanisms of pulsar dynamics and emission.

In what follows we provide analyses intended to assess the efficacy of the core/double-cone beam model at frequencies down to the 100-MHz band, largely completing our efforts to study the population of Arecibo pulsars (declinations between about --1.\degr\ and 38\degr) surveyed using the LOFAR telescope's High Band by \citet{bilous2016}. Groups of Arecibo ``B'' pulsars were analysed in \citet[Paper I]{paperI} and \citet[Paper II]{paperII} using 1.4-GHz and 327-MHz observations, and this work similarly treats a remaining population of more recently discovered ``J'' objects.  Our overarching goal in these works is to identify the physical implications of pulsar beamform variations with radio frequency.    

The Pushchino Radio Astronomy Observatory (PRAO) has long pioneered 103/111-MHz studies of pulsar emission using their Large Phased Array (LPA).  Surveys by \citet[hereafter KL99]{kuzmin1999} and \citet[MM10]{malov2010} have provided a foundation for our overall work.  More recently, the Low Frequency Array (LOFAR) in the Netherlands has produced an abundance of high-quality profiles with their High Band Survey \citep[hereafter BKK+, PHS+]{bilous2016, pilia2016} in the 100-200 MHz band as well as others in their Low Band \citep{Kondratiev16,bilous2019,Bondonneau}.  In addition decametric observations are available from the Kharkiv Array \citep{ZVKU13,KZUS22} as well as from the Long Wavelenth Array \citep{KTSD23}.

A radio pulsar emission-beam model with a central ``core'' pencil beam and two concentric conal beams has proven useful and largely successful both qualitatively and quantitatively in efforts to model the beam geometry at frequencies around 1 GHz \citep{rankin93} and its appendix \citep{rankin93b}, (together hereafter ET VI); see \S\ref{sec:ccbeams} below.  Few attempts, however, have been made to explore the systematics of pulsar beam geometry over the entire radio spectrum\footnote{\citet{Olszanski2019} studied the beam geometries of a group of Arecibo pulsars at 4.5 GHz}.  Here, we present analyses aimed at elucidating the multiband beam geometry of a brighter group of ``J'' pulsars within the Arecibo sky, most of which with observations down to the 100-MHz band.


In this work, \S\ref{sec:obs} describes the Arecibo observations, \S\ref{sec:ccbeams} reviews the geometry and theory of core and conal beams, \S\ref{sec:models} describes how our beaming models are computed and displayed,  \S\ref{sec:scattering} discusses scattering and its effects at low frequencies, \S\ref{sec:discussion} the overall analysis and conclusions, and \S\ref{sec:summary} gives a short summary. The main text of the paper reviews our analysis while the tables, model plots, and notes are given in the Appendix.  In this Appendix, we discuss the interpretation and beam geometry of each pulsar, and Figures~\ref{figA102}--\ref{figA120} show the results of analyses clarifying the beam configurations.  Figs.~\ref{figA1}--\ref{figA40} then give the beam-model plots (Fig. 1 provides an example) and (mostly) Arecibo profiles on which they are based.  The supplementary material provides the Appendix and ascii versions of the three tables.

\section{Observations}\label{sec:obs}
We present observations carried out using the upgraded Arecibo Telescope in Puerto Rico with its Gregorian feed system, 327-MHz (``P-band") or 1100-1700-MHz (``L-band") receivers, and either the Wideband Arecibo Pulsar Processors (WAPPs) or Mock spectrometer backends.  At P-band four 12.5-MHz bands were used across the 50 MHz available.  Four nominally 100-MHz bands centered at 1170, 1420, 1520  and 1620 MHz were used at L-band, and the lower three were usually free enough of radio frequency interference (RFI) such that they could be added together to give about 300-MHz bandwidth nominally at 1400 MHz.  The four Stokes parameters were calibrated from the auto- and cross-voltage correlations computed by the spectrometers, corrected for interstellar Faraday rotation, various instrumental polarization effects, and dispersion.  The resolutions of the observations are usually about a milliperiod as indicated by the sample numbers in Table~A1.  For the lower frequency observations, please consult the paper of origin as referenced in this table.  

Where our Arecibo observations were poor or missing we made use of other published compendia \citep{barr,foster,han2009,JK18,mcewen,sgg+95,Weltevrede2007} and other studies \citep{brinkman_freire_rankin_stovall,Janssen09,lorimer2006}.

The observations and geometrical models of the pulsars are presented in the tables and figures of the Appendix.  Table~A1 describes each pulsar's period, dispersion measure (DM) and rotation measure (RM), and then gives the sources for the observations and measurements at each frequency (see the sample below in Table~\ref{tab1}).  Table~A2 gives the physical parameters of each pulsar, its period and spindown rate, energy loss rate, spindown age, surface magnetic field, the acceleration parameter {$B_{12}/P^2$} and the reciprocal of \citet{pulsar_magnetosphere_book}'s $Q$ parameter (1/Q=$0.5\ \dot P_{-15}^{0.4} P^{-1.1}$) as in Papers I and II. The Gaussian fits use Michael Kramer's bfit code \citep{kramer+94,kramer94}.  The geometrical models are given in Table~A3 as will be described below.  Plots then follow showing the behavior of the geometrical model over the frequency interval for which observations are available, as well as Arecibo polarized average profiles where available.

\section{Core and Conal Beams}
\label{sec:ccbeams}
A full recent discussion of the core/double-cone beam model and its use in computing geometric beam models is given in \citet{paperIV}.

Canonical pulsar average profiles are observed to have up to five components \citep{rankin1983a}, leading to the conception of the core/double-cone beam model \citep{backer}. Pulsar profiles then divide into two families depending on whether core or conal emission is dominant at about 1 GHz.  Core single {\textbf S$_{t}$} profiles consist of an isolated core component, often flanked by a pair of outriding conal components at high frequency, triple {\textbf T} profiles show a core and conal component pair over a wide band, or five-component {\textbf M} profiles have a central core component flanked by both an inner and outer pair of conal components. 

By contrast, conal profiles can be single {\textbf S$_{d}$} or double {\textbf D} when a single cone is involved, or triple c{\textbf T} or quadruple c{\textbf Q} when the sightline encounters both conal beams. Outer cones tend to have an increasing radius with wavelength, while inner cones tend to show little spectral variation.  Periodic modulation often associated with subpulse ``drift'' is a usual property of conal emission and assists in defining a pulsar's beam configuration \citep[\eg][]{et3}. 

Profile classes tend to evolve with frequency in  characteristic ways:  {\textbf S$_{t}$} profiles often show conal outriders at high frequency, whereas {\textbf S$_{d}$} profiles often broaden and bifurcate at low frequency.  {\textbf T} profiles tend to show their three components over a broad band, but their relative intensities can change greatly.  {\textbf M} profiles usually show their five components most clearly at meter wavelengths, while at high frequency they become conflated into a ``boxy'' form, and at low frequency they become triple because the inner cone often weakens relative to the outer one.

Application of spherical geometry to the measured profile dimensions provides a means of computing the angular beam dimensions---resulting in a quantitative emission-beam model for a given pulsar.  Two key angles describing the geometry are the magnetic colatitude (angle between the rotation and magnetic axes) $\alpha$ and the sightline-circle radius (the angle between the rotation axis and the observer’s sightline) $\zeta$, where the sightline impact angle $\beta$ = $\zeta-\alpha$.\footnote{$\alpha$ values are defined between 0\degr\ and 90\degr, as $\alpha$ cannot be distinguished from 90\degr\ -- $\alpha$ in the parlance of \citet{everett}.}  The three beams are found to have particular angular dimensions at 1 GHz in terms of a pulsar's polar cap angular diameter, {$\Delta_{PC}$} = $2.45\degr P^{-1/2}$ \citep[ET IV]{rankin1990}.  The outside half-power radii of the inner and outer cones, {$\rho_{i}$} and {$\rho_{o}$} are $4.33\degr P^{-1/2}$ and $5.75\degr P^{-1/2}$ (ET VIb).  

$\alpha$ can be estimated from the core-component width when present, as its half-power width at 1 GHz, $W_{\rm core}$ has been shown to scale as {$\Delta_{\rm PC}/\sin\alpha$} (ET IV).  The sightline impact angle $\beta$ can then be estimated from the polarization position angle (PPA) sweep rate $R$ [=$|d\chi/d\varphi|$ = $\sin\alpha/\sin\beta$]. Conal beam radii can similarly be estimated from the outside half-power width of a conal component or conal component pair at 1 GHz $W_{\rm cone}$ together with $\alpha$ and $\beta$ using eq.(4) in ET VIa:  
\begin{equation} \label{eq1}
    \rho_{i,o} = \text{cos} ^{-1} [ \text{cos }\beta  - 2\text{sin }\alpha \text{ sin }\zeta \text{ sin }^{2} (W_{i,o}/4)]
\end{equation}
where $W_{i,o}$ is the total half-power width of the inner or outer conal component or pair measured in degrees longitude.  The characteristic height of the emission can then be computed assuming dipolarity using eq.(6). 

The outflowing plasma responsible for a pulsar's emission is partly or fully generated by a polar ``gap" \citep{ruderman}, just above the stellar surface.  \citet{Timokhin} find that this plasma is generated in one of two pair-formation-front (PFF) configurations:  for the younger, energetic part of the pulsar population, pairs are created at some 100 m above the polar cap in a central, uniform (1-D) gap potential---thus a 2-D PFF, but for older pulsars the PFF has a lower, annular shape extending up along the polar fluxtube, thus having a 3-D cup shape.  

An approximate boundary between the two PFF geometries is plotted on the $P$-$\dot P$ diagram of Fig~\ref{fig1}, so that the more energetic pulsars are to the top left and those less so at the bottom right.  Its dependence is $\dot P=$3.95$\times$$10^{-15}P^{11/4}$.  Pulsars with dominant core emission tend to lie to the upper left of the boundary, while the conal population falls to the lower right.  In the parlance of ET VI, the division corresponds to an acceleration potential parameter $B_{12}/P^2$ of about 2.5, which in turn represents an energy loss $\dot E$ of 10$^{32.5}$ ergs/s.  This delineation also squares well with \citet{Weltevrede2008}'s observation that high energy pulsars have distinct properties and \citet{basu2016}'s demonstration that conal drifting occurs only for pulsars with $\dot E$ less than about $10^{32}$ ergs/s.  
As previously mentioned, a sample of physical parameters is given in Table~\ref{tab2} and in full in Table~A2. 


\begin{figure}
\begin{center}
\includegraphics[width=75mm]{J1720+2150_Bmodel.ps}
\caption{Sample core/double-cone beam model display for pulsar J1720+2150 (see Fig.~\ref{figA15}).  Curves for the scaled outer and inner conal radii and core width are shown---the former by $\sqrt{P}$ and the latter by $\sqrt{P}\sin\alpha$.  Conal error bars reflect the rms of 10\% uncertainties in both the profile widths and PPA rate (see text), and the core errors also reflect a 10\% error.  Further, J1720+2150 is a case where no scattering time has been measured and published, so an average value \citep{kuzmin2001} is used---and the 10x average orange (dot-dashed) line is just visible at the lower right. The plots are logarithmic on both axes and labels are shown only for orders of 1, 2 and 5.}
\label{fig20}
\end{center}
\end{figure}

\section{Computation and Presentation of Geometric Models} 
\label{sec:models}
Two observational quantities underlie the computation of conal radii at each frequency and thus the beam model overall:  the conal component width(s) and the polarization position angle (PPA) sweep rate $R$; the former gives the angular size of the conal beam(s) while the latter gives the impact angle $\beta$ [=$\sin\alpha/R$] showing how the sightline crosses the beam(s).  Figures~\ref{figA1}--\ref{figA40} show our Arecibo (or other) profiles and Table~A1 describes them as well as referencing any 100-MHz band or below published profiles.  Following the analysis procedures of ET VI, we have measured outside conal half-power (3 db or FWHM) widths and half-power core widths wherever possible. The measurements are given in Table~A3 for the 1.4-GHz and 327-MHz bands and for the decametric regime.  

These provide the bases for computing geometrical beaming models for each pulsar, which are also shown in the above figures and Table~A3.  However, we do not plot these directly.  Rather we use the widths to model the core and conal beam geometry as above, but here emphasizing as low a frequency range as possible. The model results are given in Table~A3 for the 1-GHz and  decameter band regimes.  $W_{c}$,  $\alpha$, $R$ and $\beta$ are the 1-GHz core width, the magnetic colatitude, the PPA sweep rate and the sightline impact angle;  $W_i$, $W_o$ and $\rho_i$, $\rho_o$ are the respective inner and outer conal component widths and the respective beam radii, at 1 GHz, 327 MHz, and the lowest frequency values in the 100-MHz or below bands.  

Core radiation is found empirically to have a bivariate Gaussian (von Mises) beamform such that its 1-GHz (and often invariant) width measures $\alpha$ but provides no sightline impact angle information. If a pulsar has a core component, we attempt to use its width at around 1-GHz to estimate the magnetic colatitude $\alpha$, and when this is possible the $\alpha$ value is bolded in Appendix Tables~A3 (see the sample below in Table~\ref{tab3}).  $\beta$ is then estimated from $\alpha$ and $R$ as above. The outside half-power (3 db) widths of conal components or pairs are measured, and the spherical geometry above then used to estimate the outside half-power conal beam radii.  Where $\alpha$ can be measured, the value is used, when not an $\alpha$ value is estimated by using the established conal radius or characteristic emission height for an inner or outer cone.  These conal radii and core widths are then computed for different frequencies wherever possible.  

We present our results in terms of the core and conal beam dimensions as a function of frequency.  The results of the model for each pulsar are then plotted in Figures~\ref{figA1} to \ref{figA40} (Fig 1 provides an example).  The plots are logarithmic on both axes, and labels are given only for values in orders of 1, 2 and 5. For each pulsar the plotted values represent the {\textit scaled} inner and outer conal beam radii and the core angular width, respectively.  The scaling plots each pulsar's beam dimensions as if it were an orthogonal rotator with a 1-sec rotation period---thus the conal beam radii are scaled by a factor of $\sqrt{P}$ and the core width (diameter) by $\sqrt{P}\sin{\alpha}$. This scaling then gives each pulsar the same expected model beam dimensions, so that similarities and differences can more readily be identified.  The scaled outer and inner conal radii are plotted with blue and cyan lines and the core diameter in red.  The nominal values of the three beam dimensions at 1 GHz are shown in each plot by a small triangle.  Please see the text and figures of \citet{paperI} or \citet{paperII} for a fuller explanation. 

Estimating and propagating the observational errors in the width values is very difficult.  Instead of quoting the individual measurement errors, we provide error bars reflecting the beam radii errors for a 10\% uncertainties in the conal width values, the PPA sweep rate, and the error in the scaled core width.  The conal error bars shown reflect the {\it rms} of the first two sources with the former indicated in the lower bar and the latter in the upper one.

\begin{table*}
\caption{Sample: Observation Information; see Appendix Table A1} 
\begin{tabular}{lccc|ccc|ccc|ccc}
\hline
 Pulsar & P & DM & RM  & MJD & $N_{pulses}$ & Bins & MJD & $N_{pulses}$ & Bins & References  \\
            & (s) & pc/$cm^{3}$ & $rad\ m^{2}$ &   & & & & & & & &\\
    \hline
    & & & & & \mbox{\textbf{(L-band)}} & & & \mbox{\textbf{(P-band)}} & &  \mbox{\textbf{($\lesssim$100 MHz)}} \\
    \hline\hline
J0006+1834  &   0.69   &    11.4   &   --20.9  &   58887   &   3009    &   512 &   57844   &   1036    &   1024    &  BKK+ \\
J0030+0451  &   .0049  &     4.3    &  +1.2     & \multicolumn{3}{c|}{\textbf{\citet{SBM+22}}} & ---   & ---    & ---    & KTSD, 35 \\
J0051+0423  &   0.35   &    13.9    &   --6.3   &   58887   &  16743    & 1108  &   58902   &   14844   &   1182    &  PHS+; ZVKU \\
J0122+1416  &  1.39    &    17.7    &   --15    &   58972   &   1665 &   256    &   58950   &  2586  &  1036  &       \\
J0137+1654  &   0.41   &    26.6    &   --15.7  &   58886   &  16743    & 1024  &   58902   &   2886    &   1020    &  BKK+ \\
\\[-0pt]
J0139+3336  &  1.25    &    21.2    &   --42    &   58965   &   1185 &   1031   &   58951   &  1436  &  1040  &           \\   
J0152+0948  &   2.75   &    21.9    &   --0.2   &   58299   &   1025    & 1056  &   57844   &   325     &   1024    &  BKK+ \\
J0245+1433  &   2.13   &    29.5    &    ---    &   ---     &   ---     & ---   &   57844   &   1028    &   1013    &  PRAO \\
J0329+1654  &   0.89   &    42.1    &   +5.8    &   56932   &   1028    & 1116  &   57123   &   1116    &   1049    &  BKK+ \\
J0337+1715  & 0.0027   &    21.3    &   +29.0   &   56760   &  365928   & 256   &   56584   &   365914 & 256  &  \\
\hline
\end{tabular}
\label{tab1}
\end{table*}

\begin{table}
\setlength{\tabcolsep}{2pt}
\caption{Sample: Pulsar Parameters; see Appendix Table A2}
\begin{center}
\begin{tabular}{lccccccc}
\hline
 Pulsar &  P & $\dot{P}$ & $\dot{E}$ & $\tau$ & $B_{surf}$ & $B_{12}/P^2$ & 1/Q  \\
 (B1950) & (s) & ($10^{-15}$ & ($10^{32}$  & (Myr) & ($10^{12}$ &   &   \\
 & & s/s) & ergs/s) & & G) &   &    \\
\hline
\hline
J0006+1834 & 0.694 & 2.10 & 2.479 & 5.2 & 1.22 & 2.5 & 1.0 \\
J0030+0451 & .0049 & 1e-5 & 35. & 7580. & 2.3e-4 & 9.5 & 1.8 \\
J0051+0423 & 0.355 & 2e-3 & .062 & 803.0 & 0.05 & 0.4 & 0.1 \\
J0122+1416 & 1.389 & 3.80 & .560 & 5.8 & 2.33 & 1.2 & 0.6 \\
J0137+1654 & 0.415 & 0.01 & .068 & 537.3 & 0.07 & 0.4 & 0.2 \\
\\
J0139+3336 & 1.248 & 2.06 & .420 & 9.6 & 1.62 & 1.0 & 0.5 \\
J0152+0948 & 2.747 & 1.70 & .032 & 25.6 & 2.19 & 0.3 & 0.2 \\
J0245+1433 & 2.128 & 2.36 & .097 & 14.3 & 2.27 & 0.5 & 0.3	\\
J0329+1654 & 0.893 & 0.22 & .119 & 65.8 & 0.44 & 0.6 & 0.3 \\
J0337+1715 & 0.003 & 0.00 & 340. & 2450.0 & 0.00 & 29.7 & 4.1 \\
\hline
\end{tabular}
\end{center}
\label{tab2}
Notes: Values from the ATNF Pulsar Catalog \citep{ATNF}, \textbf{Version 1.67}.
\end{table}

\setlength{\tabcolsep}{3pt}
\begin{table*}
\caption{Sample: Profile Geometry Information; see Appendix Table A3}  
\begin{tabular}{lc|ccc|ccccc|ccccc|cccc}
    \toprule
    Pulsar &  Class & $\alpha$ & $R$ & $\beta$ & $W_c $ & $W_i$ & $\rho_i$ & $W_o$  & $\rho_o$ & $W_c $ & $W_i$ & $\rho_i$ & $W_o$  & $\rho_o$ & $W_i$ & $\rho_i$ & $W_o$  & $\rho_o$  \\
          &   & (\degr) & (\degr/\degr) & (\degr) & (\degr) & (\degr) & (\degr) & (\degr) & (\degr) & (\degr) & (\degr) & (\degr) & (\degr) & (\degr) & (\degr) & (\degr) & (\degr) & (\degr) \\
    \midrule
    & & \multicolumn{3}{c|}{(1-GHz  Geometry)} & \multicolumn{5}{c|}{(1.4-GHz Beam Sizes)} & \multicolumn{5}{c|}{(327-MHz Beam Sizes)} & \multicolumn{4}{c}{($\lesssim$100-MHz Beam Sizes)} \\
    \midrule
    \midrule
J0006+1834 & D? & 7 & -1 & -7.0 &  --- &  --- &  --- & $\sim$45 & 7.0 &  --- &  --- &  --- & $\sim$180 & 7.0 &  --- &  --- & $\sim$180 & 7.0 \\
J0051+0423 & cT & 23 & +3.2 & +7.1 &  --- &  --- & 9.7 & 29.1 & 9.7 &  --- & 4.8 & 7.2 & 33.2 & 10.3 & 6.2 & 7.3 & 36.5 & 10.9 \\
J0122+1416 & Sd? & 35 & -10 & +3.3 &  --- & $\sim$6 & 3.7 &  --- &  --- &  --- & 6.0 & 3.7 &  --- &  --- &  --- &  --- &  --- &  --- \\
J0137+1654 & cQ?? & 18 & +3.6 & +4.8 &  --- &  --- &  --- & 45 & 9.0 &  --- &  --- &  --- & 60.4 & 11.2 & 15.6 &  --- &  --- &  --- \\
J0139+3336 & D? & 23 & -6 & +3.7 &  --- & 4.8 & 3.9 &  --- &  --- &  --- & $\sim$5 & 3.9 & $\sim$13 &  --- &  --- &  --- &  --- &  --- \\
\\[-8pt] 
J0152+0948 & D & 42 & $\infty$ & 0.0 &  --- & 10.4 & 0.0 & 10.4 & 3.5 &  --- & 13.4 &  --- & 13.4 & 4.5 &  --- &  --- & 17.3 & 5.8 \\
J0329+1654 & Sd?? & 37 & +6.4 & +5.4 &  --- &  --- & 6.0 & 8.3 & 6.0 &  --- &  --- &  --- & 10.1 & 6.3 &  --- &  --- & 13.2 & 6.8 \\
J0337+1715 & M & {\bf 44} & $\infty$ & 0 & $\sim$68 & 166 & 54.3 & 240 & 73.3 & $\sim$68 & 163 & 53.5 & 252.0 & 75.8 &  --- &  --- &  --- &  --- \\
J0348+0432 & T & {\bf 56} & -8.45 & +5.6 & $\sim$15 &  --- &  --- & 66 & 28.6 &  --- &  --- &  --- & 86 & 36.8 &  --- &  --- &  --- &  --- \\
J0417+35 & D?? & 79 & -9 & +6.3 &  --- &  --- & 7.2 & 7.0 & 7.2 &  --- &  --- &  --- & 9.2 & 7.7 &  --- &  --- & 11.1 & 8.3 \\
\bottomrule
\end{tabular}
\label{tab3}
\end{table*}

\section{Low Frequency Scattering Effects}\label{sec:scattering}
Interpretation of pulsar profiles at lower frequencies requires us to assess how severely they might be distorted by scattering in the interstellar medium---and this is best done on the basis of scattering measurements \citep[\eg][]{kuzmin_LL2007}, however these are available for only a handful of the pulsars under study here.  These are shown on the model plots as double-hatched orange regions where the boundary reflects the scattering timescale at that frequency in rotational degrees. For pulsars having no scattering study, we use the mean scattering level \citep{kuzmin2001} determined from the dispersion measure (DM), though some pulsars have scattering levels up to about 10 times greater or smaller than the average level.  Our model plots show the average scattering level (where applicable) as yellow single hatching and with an orange line indicating ten times this value as a rough upper limit.

\section{Analysis and Discussion} 
\label{sec:discussion}
\noindent\textit{\textbf{Augmented LOFAR High Band Survey Population}}:  \citet{bilous2016} surveyed 194 pulsars at declinations greater than 8\degr\ and ecliptic latitudes $>$3\degr\ and provided good quality decametric profiles for many.  Here, as in Papers I and II we include only pulsars within the Arecibo sky, but Appendix A of \citet{paperIV} treats objects elsewhere.  Of the BKK+ population, about 30 lack useful parallel observations at higher frequencies, and unsurprisingly, some 38 were not detected in the 100-MHz band.  However, the remaining 164 provide total power profiles to compare with observations at higher frequencies.
These substantially augment the PRAO 111/103-MHz profiles published first by \citet{kuzmin1999} (96) and as improved by \citet{malov2010} (106) with a number of duplications---and \citet{pilia2016} have provided LOFAR High Band profiles on another 80 pulsars.  
    
In total then, we have some 100-MHZ-band information on 225 pulsars down to 8\degr\ declination and on 54 more down to the Arecibo southern limit of --1.5\degr.  In addition, we have information on another 133 down to the --35\degr\ declination limit of the Jodrell Bank telescope as well as about a dozen that have been observed down to 200 MHz in the far south \citep[see][]{paperIV}.  This is the overall population we study and interpret in this and the three previous publications.  

A $P$-$\dot P$ diagram showing the characteristics of the LOFAR High Band population (augmented by 54 Arecibo objects studied in Papers I and II as well as this paper) is given in Fig.~\ref{fig1}. Drawing on the above information, this population includes pulsars missed by LOFAR but accessible using Arecibo down to its southern declination limit.  

The above diagram is striking by its lack of energetic core-dominated pulsars (shown by red triangles) and relatively rich in conal pulsars of all varieties.  One can compare a similar diagram showing the full sky population of ``B'' pulsars \citep[fig. 4]{paperIV} where roughly equal numbers of core and conal dominated pulsars are seen.  
\vskip 0.1in

\begin{figure}
\begin{center}
\includegraphics[width=75mm,angle=0.]{HB_Galactic_Distribution.ps}
\caption{Plot showing the distribution of the LOFAR High Band Survey pulsar population (164; circles) on the sky in Galactic coordinates, and others (112; rectangles) studied here down to the --1.5\degr\ Arecibo southern declination limit.  Pulsars with {$\dot E$} greater or less than 10$^{32.5}$ ergs/s are shown with red or green symbols, respectively.  Clearly the core-emission dominated energetic objects lie both closer to the Galactic plane and Galactic Center than their less energetic conal cousins.}  
\label{fig2}
\end{center}
\end{figure}

\noindent\textit{\textbf{1-GHz Core/double-cone Modeling Results}}:  The 133 pulsars considered here show beam configurations across all of the core/double-cone model classes.  However, only about a dozen of this group have $\dot E$ values $\ge$ $10^{32.5}$ ergs/s and either core-cone triple \textbf{T} or core-single {\bf S$_t$} profiles.  The remainder tend to have profiles dominated by conal emission---that is, conal single \textbf{S$_d$}, double \textbf{D}, triple c\textbf{T}, or quadruple c\textbf{Q} geometries. Indeed, stars with c\textbf{T/Q} geometries are surprisingly abundant within this group, whether one or the other often very difficult to discern.  We were able to construct quantitative beam geometry models for almost all of the pulsars, though some are better established than others on the basis of the available information.  Lack of reliable PPA rate estimates was a limiting factor in a number of cases, either due to low fractional linear polarization or difficulty interpreting it.  Usually it was possible to trace a fixed number of profile components across the three bands, sometimes despite very different spectral behavior.  However, in a few cases the profiles were possibly dominated by different profile modes, and this may well account for our difficulties in identifying their beam configurations.  Conal profiles tend to show periodic modulation, so fluctuation spectral features can often be useful in identifying conal emission.
\vskip 0.1in

\begin{figure}
\begin{center}
\includegraphics[width=75mm,angle=0.]{HB_p_pdot_fancy.ps}
\caption{P-$\dot P$ Diagram showing the distribution of the LOFAR High Band Survey pulsar population (194)---augmented by Arecibo pulsars studied here and in Papers I and II (110)---in relation to the PFF boundary. Core-emission-dominated pulsars tend to lie to the upper left of the \citet{Timokhin} boundary line, whereas those with mainly conal emission fall to the lower right (see text). Pulsars with core-cone triple {\textbf T} are distributed across both regions.  This population has a smaller proportion of energetic core-dominated pulsars (32\%) than seen in the sky overall.  Of particular interest are the two energetic pulsars that seem to be conal emitters, J0631+1036 and B1951+32.}
\label{fig1}
\end{center}
\end{figure}

\noindent\textit{\textbf{Emission Beam Evolution at Low Frequency}}:  A large fraction of the Augmented LOFAR High Band Survey pulsars 60\% (169) have been observed in the 100-200-MHz band, and scattering curtails detection of many in this regime.  Of these, a smaller number (64) have been observed in the decametric band below 100 MHz, and most of these lie in the Galactic anticenter direction. This population exhibits patterns of core and conal beam spectral evolution similar to those found earlier in other populations---
\begin{itemize}
\item Core beams are found to have angular sizes similar to those of a pulsar's polar cap.  These sizes tend to vary little with frequency, but the similarity is most marked around 1 GHz.  Scattering can broaden a core component's width at lower frequencies, whereas above 1 GHz conal power often seems to appear on the wings of core components.
\item Here again in this population we find that a few core beams narrow from their 1-GHz values in a mid-frequency range and sometime recover at lower frequencies.  The narrowing may represent the weakening or absence of half the component as if the differently polarized leading and trailing parts of the core have different spectra.  We saw earlier instances \citep[\eg]{paperIV} of these parts being displaced and partially resolved (\eg B1409--62), and in several cases (\eg B1046--58) the leading and trailing parts of the core have different amplitudes.   
\item Conal beams are found to assume their characteristic radii around 1 GHz when $\alpha$ is determined by a core component width.  Inner cones usually have fairly constant radii and outer ones often escalate with wavelength, but examples of inner conal growth with wavelength and outer conal constancy are also seen.  And we reemphasize that inner/outer cones can usually be distinguished in triple and five-component profiles where a core beam is present.  However, in a few cases outer cone models would require  $\alpha > 90\deg$, and thus an inner cone is indicated. 
\item Good examples of c\textbf{T} or c\textbf{Q} profiles then provide interesting opportunities to study the inner and outer cones in relation to each other.  In several cases neither cone shows much width increase with wavelength. 
\end{itemize}

\noindent\textit{\textbf{Galactic Distribution of LOFAR High Band Pulsars}}:  In an earlier paper \citep{paperIV} we considered the entire population of 487 ``B'' pulsars that represent surveys covering the entire sky, and its Fig.~2 shows the distribution centered on the Galactic Center direction with most pulsars lying close to the Galactic plane as expected.  

A similar distribution plot for the (augmented) LOFAR High Band Survey population is shown in Fig.~\ref{fig2}, and clearly a great deal of the sky was inaccessible to both the LOFAR and Arecibo instruments.  LOFAR was able to observe pulsars north of declination 8\degr\ (circles); whereas Arecibo was restricted to declinations between 37.5 and --1.5\degr.  Neither then had access to the rich southern Galactic plane or regions adjacent to the Galactic center.

Nonetheless, the \textit{rms} Galactic latitudes of this northern instrument population are not very different than those for core- and conal dominated pulsars over entire sky.  What is very different, however, is the proportion of core-dominated pulsars at only 32\% relative to about 50\% overall.    
\vskip 0.1in

\noindent\textit{\textbf{New Decametric Population of Pulsars and RRATs}}:  In this work we have attempted to use meter- and decimeter- wavelength observations to better understand pulsar emission properties in the decameter band.  A possibility for taking the opposite approach seems to be emerging.  The second UTR-2 pulsar census recently published by \citet{KZUS22} provides a largely new population of pulsars detected at extremely low frequencies, a number of which have not been adequately studied at higher frequencies.  Moreover, the PRAO cite \citep{prao} lists several dozen new pulsars and RRATs, some of which have been timed well enough to have good positions.  Further, the LWA is coming into use and bridges these two regimes.  We have much to learn about objects that are strong enough to be discovered in the decameter band, some perhaps with steep spectra that preclude observation at higher frequencies. 
\vskip 0.1in

\noindent\textit{\textbf{Pulsars With Interesting Characteristics}}

\noindent{\textbf{J0245+1433}}:  The pulsar \citep{dsm+13} has a barely resolved double profile at 327 MHz that also exhibits strong antisymmetric circular polarization (Fig.~\ref{figA102}).  It is thus an unusual hybrid of core and conal properties, and observations over a broader band are needed to better understand it. 
\vskip 0.1in
\noindent\textit{\textbf{J0329+1654}} shows consistently poor profiles, probably due to its being a RRAT \citep{BT21}. 
\vskip 0.1in
\noindent\textit{\textbf{J0337+1715}}, the famous 2.7-ms MSP in a triple system \citep{Ransom_2014} seems to have a core/double-cone profile   \citep{Rankin_2017}. 
\vskip 0.1in
\noindent{\textbf{J0348+0432}} \citep{antoniadis,lynch2013} appears to be another MSP with a core-cone profile as many or most seem not to \citep[see][]{Rankin_2017}.  Further study in single pulses is needed to fully explore its different properties. 
\vskip 0.1in
\noindent{\textbf{J0538+2817}}:  We confirm the two modes of this pulsar at 1.4 GHz \citep{acj+96,vonHoensbroech1997}.  Moding is unusual in such an energetic pulsar, so further study is needed to try to understand what are the causes.  
\vskip 0.1in
\noindent{\textbf{J0609+2130}} seems to be another MSP with a core-cone profile.  Detailed and sensitive single pulse study is needed to fully explore its characteristics.
\vskip 0.1in
\noindent{\textbf{J0628+0909}} is a RRAT \citep[see][]{McLaughlin06} that appears to have a core/cone profile.  
\vskip 0.1in
\noindent{\textbf{J0815+0939}} is the well known four-component profile that seems to exhibit some bi-drifting \citep{teixeira}.  
\vskip 0.1in
\noindent\textit{\textbf{J0843+0719}} seems to have a steep spectrum, so our several 1.4-GHz observations were all non-detections.  While our 327-MHz profile is similar to that of \citet{Burgay+06} no observations seem to exist at lower frequencies.  
\vskip 0.1in
\noindent{\textbf{J1022+1001}} is a 16-ms MSP that again seems have a core/cone {\bf T} beam geometry \citep{xilouris98,kkwj98,stairs99,Dai2015} and can be detected down into the 100-MHz band.  
\vskip 0.1in
\noindent{\textbf{J1313+0931}} shows a probable conal quadruple profile with a strong 11.6-$P$ modulation; see Fig.~\ref{figA109}.  Such pulsars deserve further detailed study because their properties reflect the relationship of the inner and outer cones.  
\vskip 0.1in
\noindent{\textbf{J1404+1159}} exhibits at strong modulation of drifting subpulses with a 4.7-$P$ period; see  Fig.~\ref{figA110}.  
\vskip 0.1in
\noindent{\textbf{J1649+2533}} seems to show both frequent nulls of a few pulses and a 2.5-$P$ modulation when emitting (Fig.~\ref{figA112}).  Can this be understood in terms of a carousel structure?  
\vskip 0.1in
\noindent{\textbf{J1720+2150}} has a clear five-component core/double-cone profile much like that of B1237+25---and shares some other properties as well.  Further detailed single-pulse study would be useful and rewarding. 
\vskip 0.1in
\noindent{\textbf{J1819+1305}} is another excellent example of a conal quadruple profile, and it shows two modes \citep{RankinWright} in a manner similar to several other similar pulsars; see Fig.~\ref{figA113}.
\vskip 0.1in
\noindent{\textbf{J1842+0257}} shows a complex mixture of organized 3-$P$ drifting and long nulls; see Fig.~\ref{figA114}.
\vskip 0.1in
\noindent{\textbf{J1843+2024}}: The harmonic resolved fluctuation spectrum of this pulsar \citep{Champion2005} shows a bright narrow feature that is the harmonic of a weaker one (see Fig.~\ref{figA115}), reflecting a highly stable even-odd modulation as in pulsar B0943+10 \citep{deshpande}.  Could further study identify a carousel structure? 
\vskip 0.1in
\noindent{\textbf{J1908+0734}}: Strangely, no polarimetry has been published for this pulsar that is detectable from higher frequencies \citep{Bhat04,camilo_nice} down to 20/25 MHz \citep{ZVKU13}.
\vskip 0.1in
\noindent{\textbf{J2111+2106}}: This very slow pulsar shows surprising evidence of core emission in its otherwise conal profile.   
\vskip 0.1in
\noindent{\textbf{J2139+2242}} provides another fine example of a bright drifter with a conal quadruple profile (see Fig.~\ref{figA120}).  Detailed single pulse study will surely discover new effects and show the relationship between its two emission cones.
\vskip 0.1in

\noindent\textit{\textbf{Scattering Levels of LOFAR High Band Population}}: While rather few of the pulsars considered in this paper have measured scattering times, many more are available for the objects in Papers I. II and \citet{paperIV}.  From the perspective of northern instruments such as Arecibo, LOFAR or PRAO, a meaningful average scattering level can be computed as was done in \citet{kuzmin2001}.  However, for pulsars close to the Galactic Center direction, we saw in \citet[fig. 5]{paperIV} that scattering levels range over many orders of magnitude and no meaningful average can be obtained.  
\vskip 0.1in

\section{Summary}
\label{sec:summary}
We have analyzed the Arecibo 1.4-GHz and 327-MHz polarized profiles together with the available observations at lower (and occasionally higher) frequencies of 133 pulsars, most of which were included in the LOFAR High Band Survey \citep{bilous2016}.  This group of more recently discovered pulsars complements similar analyses of both the most studied pulsars within the Arecibo sky in Paper I, and a further group of less studied objects in Paper II, most of which appear in the LOFAR survey and thus have been observed down into the 100-MHz band or below by a variety of investigators.  A further large group of objects lying outside the Arecibo sky was similarly studied \citep{paperIV}, and a number of these are also included in the LOFAR High Band Survey.  These efforts then attempt to fully parallel the LOFAR High Band Survey with complementary observations at longer wavelengths wherever possible in order to study the frequency evolution of pulsar emission beams comprehensively.  



HMW gratefully acknowledges a Sikora Summer Research Fellowship and thanks Anna Bilous for useful discussions regarding the LOFAR work. We especially thank our colleagues who maintain the ATNF Pulsar Catalog and the European Pulsar Network Database as this work drew heavily on them both.  Much of the work was made possible by support from the US National Science Foundation grants AST 99-87654 and 18-14397. The Arecibo Observatory is operated by the University of Central Florida under a cooperative agreement with the US National Science Foundation, and in alliance with Yang Enterprises and the Ana G. M\'endez-Universidad Metropolitana. This work made use of the NASA ADS astronomical data system.

The profiles will be available on the European Pulsar Network download site, and the pulse sequences can be obtained by corresponding with the lead author.




%
%

\bibliography{biblio.bib}

\bsp	

\appendix
\setcounter{figure}{0}
\renewcommand{\thefigure}{A\arabic{figure}}
\renewcommand{\thetable}{A\arabic{table}}
\setcounter{table}{0}
\renewcommand{\thefootnote}{A\arabic{footnote}}
\setcounter{footnote}{0}
\twocolumn

\section{Group B Pulsar Tables, Models and Notes, Corrected}

\noindent\textit{\textbf{J0006+1834}}: This pulsar has a wide, apparently double or conal quadruple profile with a strong leading pair of components and weak trailing power as seen at 327 MHz.  The longitude scales are different here to show this weak trailing power, and it cannot be detected either at 1.4 GHz or 149 MHz; however, the bright component is seen at 111 MHz \citep{prao}. The dimensions of the entire profile cannot then be measured accurately.  More difficult is that the PPA rate is close to --1\degr/\degr\ at 327 MHz, and the geometry depends critically on the sign of $\beta$.  If positive, $\rho$ increases dramatically with wavelength, but flattens at low frequency, so not in the usual manner.  If negative, $\rho = \alpha$ independent of the profile width. In short, little can be said of this pulsar's geometry.  The sinusoidal LOFAR 149-MHz profile provides little to go on, and scattering is expected to have negligible effect.  
\vskip 0.1in

\noindent\textit{\textbf{J0030+0451}}: This 4.9 msec MSP was observed very nicely at 1.3 GHz by \citet{SBM+22}, and \citet{KTSD23} have used the LWA to obtain profiles down to 35 MHz.  The high frequency profile shows two broad features spaced by roughly half a rotation period with a PPA traverse suggestive of a single-pole interpulsar.  The central part of the stronger roughly triple ``main pulse'' is in turn suggestive of core emission but as usual is too narrow to reflect a dipolar polar cap.  We provide an aspirational triple model in order to show the evolution of the putative conal profile over the large band available. 
\vskip 0.1in

\noindent\textit{\textbf{J0051+0423}}: PSR J0051+0423 has a three-component profile with some evidence of structure in the trailing component.  Fluctuation spectra often show a feature with a 2.4-$P$ period that modulates all three features.  We have thus explored whether a conal triple c{\bf T} beam geometry is compatible with the profile dimensions and find that it is so.  The pulsar is detected both at 111 MHz \citep{prao} and 20/25 MHz \citep{ZVKU13} but no measurements are possible.  However, the four LWA decameter observations \citep{KTSD23} provide high quality profiles down to 35 MHz.
\vskip 0.1in    

\noindent\textit{\textbf{J0122+1416}} has a well defined PPA rate and apparently a single profile in both bands with no evidence of width escalation.  We thus model it using an inner conal single geometry.  The pulsar is detected both at 111 MHz \citep{prao} where the width may be compatible with those at higher frequencies and at 20/25 MHz \citep{KZUS22} where the larger width may reflect scattering.
\vskip 0.1in 

\noindent\textit{\textbf{J0137+1654}}: A broad double profile is seen in this pulsar at 1.4 GHz, with seemingly three components at 327 MHz, and a single component with a leading ``bump'' is seen at LOFAR frequencies.  The weak leading 327-MHz component is seen in several observations, so is definitely present. It clearly declines markedly with wavelength, thus only the trailing features are seen in the LOFAR profiles.  We thus model the profile as having a conal quadruple c{\bf Q} structure; but the inner and outer features are conflated at 1.4 GHz, no leading inner conal feature is discernible at 327 MHz, and the leading component is so weak in the LOFAR profiles that the outer conal geometry cannot be traced into the decameter region.  Single pulses show that the emission comes in bursts of a few 10s of pulses interspersed by long nulls.  The pulsar is detected at 111 MHz \citep{prao} but no measurements are possible.  
\vskip 0.1in

\noindent\textit{\textbf{J0139+3336}}: Neither profile permits detailed interpretation; however, both suggest two conflated brighter features with possibly a further pair on the ``wings'' at 327 MHz---while  the main pair maintains about the same width.  The PPA rate appears to be about --6\degr/\degr.  We model the main features with an inner conal double configuration.  
\vskip 0.1in

\noindent\textit{\textbf{J0152+0948}}: Two components are seen at all three frequencies with the leading feature having a steeper relative spectrum in this pulsar.  The linear polarization suggests a highly central PPA traverse.  We model this as a {\bf D} profile.
\vskip 0.1in

\begin{figure}
\begin{center}
\includegraphics[width=65mm,angle=0.]{pJ0245+1433.57844p4a.ps}
\caption{Arecibo polarization profile of pulsar J0245+1433 showing its perplexingly large antisymmetric circular polarization in a profile that might well be conal.}
\label{figA102}
\end{center}
\end{figure}
\noindent\textit{\textbf{J0245+1433 (formerly J0244+14; \cite{dsm+13})}}: The 327-MHz Arecibo profile in Fig.~\ref{figA102} seems to be the only one available, and it shows a double structure with a chaotic PPA behavior.  Striking is the substantial, symmetrical antisymmetric $V$ that is often characteristic of core emission and rare in conal profiles.  However, this pulsar falls well into the conal emission dominated region on the basis of its $\dot E$ of only about $10^{31}$ ergs/s.  More information is clearly needed to work out its beam structure.  This pulsar has been weakly detected at 111 MHz (Valerij Malofeev, private communication), so it is possible that more study will prove to be revealing.
\vskip 0.1in

\noindent\textit{\textbf{J0329+1654}}: PSR J0329+1645 has poor LOFAR and 1.4-GHz profiles, probably due to its being a RRAT \citep{BT21} where it is also seen at 111 MHz. It seems to have a single symmetrical component across the three bands, so we model it as having a {\bf S$_d$} structure.  Scattering seems minimal at 149 MHz.
\vskip 0.1in

\noindent\textit{\textbf{J0337+1715}} is the famous 2.7-ms MSP in a triple system \citep{Ransom_2014} that also seems to have a core/double-cone profile.  We report the analysis in \cite{Rankin_2017} where a central traverse is assumed given that no PPA slope can be determined.  The beam dimensions and model emission heights are smaller than for other pulsars in part because the velocity-or-light cylinder radius is only 130 km for this MSP.
\vskip 0.1in

\begin{figure}
\begin{center}
\includegraphics[width=65mm,angle=0.]{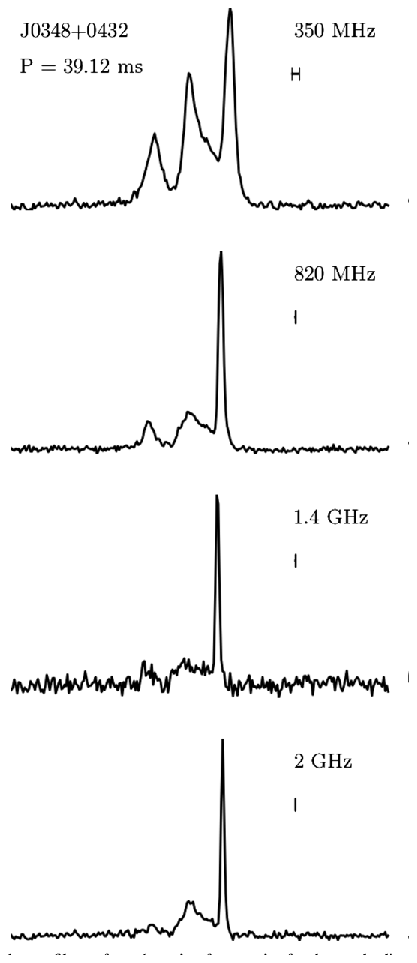}
\caption{Total power, full period profiles of pulsar J0348+0432 from \cite{lynch2013}.  }
\label{figA103}
\end{center}
\end{figure}
\noindent\textit{\textbf{J0348+0432}}: Our profile is not suitable for exploring the geometry, so we have replicated the four-frequency, full period, total power profiles from \cite{lynch2013}.  Also, the polarization is very nicely measured in the \cite{antoniadis} Fig. 5, where the PPA rate is --8.45\degr/\degr.  The profile evolution suggests an outer cone, and this requires an $\alpha$ value near 56\degr.  This in turn requires a core width near 14\degr, which is plausible, but impossible to determine accurately from the profiles in Fig.~\ref{figA103}, but the \citet{mcewen} 350-MHz profile provides the best opportunity.  Apparently, this is another MSP with a cone-cone triple {\bf T} beam geometry. 
\vskip 0.1in

\noindent\textit{\textbf{J0417+35}}: At LOFAR frequencies, the profile has two components, the leading much stronger than the trailing. At P-band, the trailing component decreases in height to become almost a "bump" on the side of the profile. We use a {\bf D} profile model.  The PPA rate is poorly determined, and scattering seems minimal at 149 MHz.
The pulsar is detected at 111 MHz \citep{prao} but no measurements are possible.

\begin{figure}
\begin{center}
\includegraphics[width=65mm,angle=-90.]{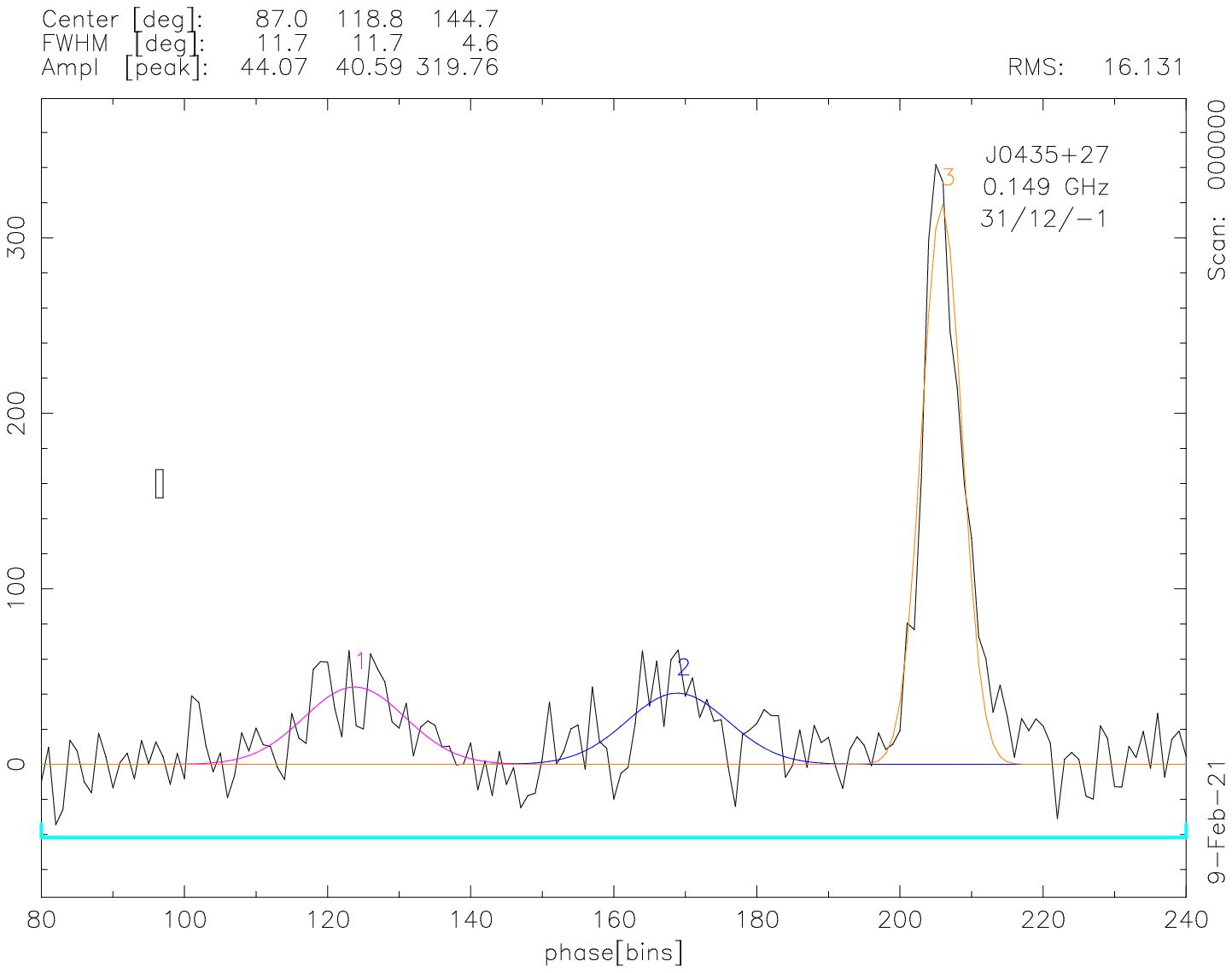}
\caption{J0435+2749: Fitted Gaussian-component model for BKK+'s 149-MHz observation.}
\label{figA104}
\end{center}
\end{figure}
\vskip 0.1in
\noindent\textit{\textbf{J0435+2749}}: Following on \citet{brinkman_freire_rankin_stovall}'s {\bf T} classification, we now see that the LOFAR profile shows a narrow third component, while the central one seems to be broader.  Close inspection of several 327-MHz profiles show dual central features that somewhat broaden the trailing component.  These can be seen in the weak $L$ curve.  Therefore, having explored the quantitative geometry, we now believe a conal quadruple c{\textbf{Q}} beam model is most appropriate for this pulsar.  The PPA rate is difficult to discern accurately, but a shallow negative traverse is most likely.  The LOFAR inner cone width cannot be measured accurately, but a value around 25\degr\ is compatible with those at the higher frequencies---though the fit in Fig.~\ref{figA104} gives a substantially narrower value.  
\vskip 0.1in

\noindent\textit{\textbf{J0457+2333}}: The two profiles are only of moderate quality, but they suggest a ``boxy'' form at 1.4 GHz that may show a core/double cone at 327 MHz---similar to the evolution of many {\bf M} structures.  We  can only guess the PPA rate, but a central traverse is suggested at the lower frequency.  To support this model a core width of about 7\degr\ is needed, and this cannot be verified in the profile structure; nor can we clearly discern the inner cone width.  Higher quality profiles are needed to improve the beam model.
\vskip 0.1in

\noindent\textit{\textbf{J0533+0402}}: This pulsar seems to be a conal single {\bf S$_d$} pulsar, but its geometry remains inaccurate.  The 1.4-GHz PPA traverse suggests a very central sightline traverse, which is incompatible with this classification; the 327-MHz one suggests an oblique traverse but gives no useful slope.  Fortunately, the \citet{JK18} profile shows a traverse that may be about  +11\degr/\degr, making $\alpha$ upwards of 90\degr\ for an outer cone.  This is then compatible with the Pushchino profile that shows clear bifurcation and a larger width suggesting an outer conal geometry.
\vskip 0.1in

\noindent\textit{\textbf{J0538+2817}} is a fast pulsar with broad, nearly fully linear polarized profiles.  Moreover, the two Lband profiles are very very different, reflecting the two quasi-stable profile forms identified by \citet{acj+96}. Similar moding seems to occur at 4.9 GHz \citep{vonHoensbroech1997}.  Moding is unusual in such an energetic pulsar, so this is a potentially important example.  Both of the profiles are comprised by a number of features, but we were barely able to distinguish the core and conal features or to trace their evolution.  Moreover, the core width can be accurately measured but appears incomplete, especially at the lower frequency, so we use a large 14\degr\ value.  Overall, the model is therefore aspirational.  This is an interesting pulsar that deserves detailed study.
\vskip 0.1in

\noindent\textit{\textbf{J0540+3207}}: Our several profiles suggest three features at 327 MHz and four at 1.4 GHz.  The PPA rate is poorly determined, but the needed +10\degr/\degr\ is plausible.  We thus model the beam geometry with a {c\bf T}/{c\bf Q} geometry. The pulsar is detected at 111 MHz \citep{prao} but no measurements are possible.
\vskip 0.1in

\noindent\textit{\textbf{J0546+2441}}: All our profiles for this slow pulsars are narrow, and the best show clear structure with hints of core emission in $V$.  The features are too conflated to measure with any confidence and the PPA slopes are inconsistent and the rate thus uncertain.  We model the profiles in an exemplary manner as a core single while regarding its emission as a conflated mixture of core and conal radiation.  The pulsar is detected at 111 MHz \citep{prao} but no measurements are possible.
\vskip 0.1in

\noindent\textit{\textbf{J0609+2130}}: The 1.4-GHz profile of this MSP shows a squarish triple form with little indication of the PPA rate; whereas, the 327-MHz profile is single with a width of 20\degr\ or less.  The central feature seems to have a width just larger than the polar cap diameter---and the width at 327 MHz seems only a little greater.  If we interpret the middle component as a core with a width of about 14\degr---and assume a central sightline traverse with $\beta$ $\sim$ 0, the dimensions conform to a core/inner conal single {\bf S$_t$} beam geometry.  
\vskip 0.1in

\noindent\textit{\textbf{J0611+30}}: Evidence of one component across all three bands with little change in width is seen in this pulsar.  We model it as a conal single {\bf S$_d$} configuration.  The pulsar is well detected at 111 MHz \citep{prao} but only crude width measurements are possible; the \citet{bilous2019} profiles give better values.  Scattering sets in at the very lowest frequencies.  
\vskip 0.1in

\noindent\textit{\textbf{J0613+3731}}: With profiles only at 327-MHz and below, we conclude that this is an inner cone triple configuration.  No higher frequency profile seems to have been published.  The conal emission is conflated at all frequencies down to and including 25 MHz \citep{KZUS22}.  The \citet{prao} profile is not well resolved; however, a much better one was kindly supplied (Valerij Malofeev, private communication) the four decametric LWA profiles \citep{KTSD23} trace the core-component below 50 MHz where scattering sets in and becomes dominant.
\vskip 0.1in

\noindent\textit{\textbf{J0621+0336}} seems to have a barely resolved inner conal double {\bf D} profile.  The early edge linear polarization seems to be a different OPM than that prominent on the trailing part of the profile---thus the moderate positive PPA slope.  A 350-MHz profile is available in the \citet{mcewen} compendium which seems to show a compatible barely double form, but its scale cannot be measured more accurately than the nominal half width they tabulate.  
\vskip 0.1in

\noindent\textit{\textbf{J0621+1002}}: the MSP has a very broad core-cone triple {\bf T} profile.  The PPA slope is negative, very shallow and difficult to determine accurately, but the earliest portion under the core is steepest and may have a slope near --0.7\degr/\degr.  A 410-MHz profile \citep{stairs99} shows a nearly identical width pointing to an inner cone geometry.  The about 25\degr\ L-band profile asymmetry suggest an aberration/retardation height ($c\Delta t/2$) of some 300 km.
\vskip 0.1in

\onecolumn
\begin{longtable}{lccc|ccc|ccc|ccc}
\caption{Observation Information.\\
Pulsar name, period, dispersion and rotation measure; then MJD, length and bins in the L- and Pband observations; finally references to observations at 100 MHz and below.}  \label{tabA1} \\

\hline
 Pulsar & P & DM & RM  & MJD & $N_{pulses}$ & Bins & MJD & $N_{pulses}$ & Bins & References  \\
            & (s) & pc/$cm^{3}$ & $rad\ m^{2}$ &   & & & & & & & &\\
    \hline
    & & & & & \mbox{\textbf{(L-band)}} & & & \mbox{\textbf{(P-band)}} & &  \mbox{\textbf{($\lesssim$100 MHz)}} \\
    \hline\hline
\endfirsthead
\hline
 Pulsar & P & DM & RM  & MJD & $N_{pulses}$ & Bins & MJD & $N_{pulses}$ & Bins & References  \\
            & (s) & pc/$cm^{3}$ & $rad\ m^{2}$ &   & & & & & & & &\\
    \hline
    & & & & & \mbox{\textbf{(L-band)}} & & & \mbox{\textbf{(P-band)}} & &  \mbox{\textbf{($\lesssim$100 MHz)}} \\
    \hline\hline
\endhead
J0006+1834  &   0.69   &    11.4   &   --20.9  &   58887   &   3009    &   512 &   57844   &   1036    &   1024    &  BKK+ \\
J0030+0451  &   .0049  &     4.3    &  +1.2     & \multicolumn{3}{c|}{\textbf{\citet{SBM+22}}} & ---   & ---    & ---    & KTSD, 35 \\
J0051+0423  &   0.35   &    13.9    &   --6.3   &   58887   &  16743    & 1108  &   58902   &   14844   &   1182    &  PHS+; ZVKU \\
J0122+1416  &  1.39    &    17.7    &   --15    &   58972   &   1665 &   256    &   58950   &  2586  &  1036  &  KZUS \\
J0137+1654  &   0.41   &    26.6    &   --15.7  &   58886   &  16743    & 1024  &   58902   &   2886    &   1020    &  BKK+ \\
\\[-0pt]
J0139+3336  &  1.25    &    21.2    &   --42    &   58965   &   1185 &   1031   &   58951   &  1436  &  1040  &           \\   
J0152+0948  &   2.75   &    21.9    &   --0.2   &   58299   &   1025    & 1056  &   57844   &   325     &   1024    &  BKK+ \\
J0245+1433  &   2.13   &    29.5    &    ---    &   ---     &   ---     & ---   &   57844   &   1028    &   1013    &  PRAO; KZUS \\
J0329+1654  &   0.89   &    42.1    &   +5.8    &   56932   &   1028    & 1116  &   57123   &   1116    &   1049    &  BKK+ \\
J0337+1715  & 0.0027   &    21.3    &   +29.0   &   56760   &  365928   & 256   &   56584   &   365914 & 256  &  \\
\\[-0pt]
J0348+0432  & 0.039    &    40.5    &   +49.5   &   57131   &   15326   & 927  & \multicolumn{3}{c|}{\textbf{\citet{lynch2013}}} &\\
J0417+35    &   0.65   &    51      &   +42.4   &   56932   &   1038    &  1091 &   57123   &   1068    &   1044    &  BKK+ \\
J0435+2749  &   1.20   &    53      &   +2.6    &   54541   &   3065    & 1274  &   53861   &   3677    &   637     &  BKK+ \\
J0457+2333  &  0.51    &    59      &   +23     &   56932   &   1183 &    256   &   57844   &   1647    &   256     &       \\  
J0533+0402  &   0.96   &    83.7    &    --59   &   54540   &   1028    &  1024 &   52861   &   1038    &   940     &   MM10 \\   
\\[-0pt]
J0538+2817 &  0.14   &      39.6    &   +40.2   &   57004/113   &  4189   &  1118   &   58390   & 16025  &   739  &        \\ 
J0540+3207 &  0.52   &      62.0    &   +12.8   &   57004   &  3428   & 1023    & 57123     &   1061 &   1030   &   \\ 
J0546+2441 &  2.84   &      73.8    &   +21.4   &   57004   &     251 &   1024  &   54016   &   1428  &  1024  &\\
J0609+2130 &  0.056  &      38.7    &    +2     &   57130   &   10764 &  128    &   55104   &   5291 &  185   &   \\
J0611+30   &  1.41   &     45.3    &   +16     &   58678   &    986  &  128  &   58436   &   2263 &  1024  &  BKK+ \\
\\[-0pt]
J0613+3731 & 0.62    &     19.0    &   +16.    &   ---     &   ---   &   --- & 57123  &  1937 &   1037 & PRAO; KTSD; KZUS, 20 \\
J0621+0336 & 0.270   &     72.6    &   +49.1   &   57004   &    2217 & 1054 &   \multicolumn{3}{c|}{\textbf{\citet{mcewen}}} & \\
J0621+1002 & 0.029   &     36.5    &	+53.2   &   56572   &   40138  &  958 &  \multicolumn{3}{c|}{\textbf{\citet{stairs99}}}   &  \\ 
J0623+0340 & 0.614   &     54.     &   ---     &   57005   &    7815  & 1198  &   --- &   ---   &   --- &     \\ 
J0625+1015 &  0.50   &  	78.     &   +47.0   &   54540   &   1204 &   512  &   53861   &   256  &  1080  &   \\
\\[-0pt]
J0627+0649 &  0.35   &      86.6    &   +178.5  &   56932   &   1726 &  1323  &   57123   &  1725  &` 1106  &        \\
J0628+0909 &  1.24   &      88.3    &   +123    &   57006   &   2450 &  1212  &    --- &   ---   &   --- &    \\
J0630--0046&  0.68   &      97.3    &   +39     &   56932   &   1023 &  1134 & \multicolumn{3}{c|}{\textbf{\citet{mcewen}}} &  \\
J0646+0905 &  0.90   &      149.    &   ---     &   57004   &   711  &  1063 &   ---     &   ---   &   --- &   \\
J0647+0913 &  1.24   &      154.7   &   ---     &   57005   &    966 &  1205 &   ---     &   ---   &   --- &   \\
\\[-0pt]
J0658+0022 &  0.56   &      122.    &   ---     &   57257   &    882 &   512  &   ---     &   ---   &   --- &  \\
J0711+0931 &  1.21   &      45.     &   +67     &   54541   &   1030  & 1185  &   53490   &   514     &   2048    &  BKK+ \\
J0806+08   &  2.06   &      46.7    &   ---     &   57200   &   1022  & 1031   &  ---     &   ---   &   --- &   \\ 
J0815+0939 &   0.65   &    52.7     &   +55.5   &   57516   &   12730   &   512 &   53375   &   3070    &   586     &  BKK+ \\
J0843+0719  &   1.37   &    36.6    & +44.6 & \multicolumn{3}{c|}{\textbf{\citet{Burgay+06}}} & 56964 & 826 &    1332 & \\
\\[-0pt]
J0848+1640 &  0.45   &      38.5    &   +41     &   57200   &   3973 &   512    &   57537   &   1126    &   2647     &    \\
J0928+0614 &  2.06   &      50.5    &   --14    &   57200   &   1023 &   256    &   57539   &   1045    &   1083    &       \\  
J0943+2253  &   0.53   &    25.1    &   +8      &   57377   &   13682   & 1024  &   57379   &   7104    &   1024    &  BKK+ \\
J0947+2740  &   0.85   &    29.     &   +24     &   57347   &   3172    & 1024  &   57379   &   3683    &   1024    &  BKK+ \\
J1022+1001  &  0.016   &    10.3    &   +2.4    &   56577   &  109423   &  780  &   56576   &  109380   &    141 &  PHS+; MM10 \\  
\\[-0pt]
J1046+0304 &  0.33   &      25.0    &   --1.1   &   57229   &    1833   &  256  &   57539   &   3677    &    256 &     \\
J1147+0829 &  0.33   &      25.0    &   +33     &   57524   &     455   &  256  &   57537   &   2230    &    1083    &       \\ 
J1238+2152  &  1.12    &    17.5    &   +6      &   57592   &   3672    & 1092  &   56576   &   1026    &   1090    &  BKK+; ZVKU \\
J1246+2253  &  0.47    &    17.9    &  --0.7    &   57307   &   1885    & 925   &   56576   &   1265    &   925     &  BKK+ \\
J1313+0931  &  0.85    &    12.     &   +1.7    &   53779   &   1601    &  998  &   53778   &   2829    &   828     &  BKK+ \\
\\[-0pt]
J1404+1159  &  2.65   &     18.5    &   +4.     &   57564   &   3192    & 1024  &   57565   &   3435    &   1024    &  PRAO, 111 \\
J1503+2111  &  3.31   &     11.8    &   --10    &   53779   &   155     & 3236  &   58185   &   1039    &   1004    &  BKK+ \\
J1538+2345  &  3.45   &     14.9    &    ---    &   57312   &   464     & 1077  &   ---     &   ---     &   ---     &       \\ 
J1549+2113  &  1.26   &     24.8    &   +7      &   55107   &   1030    & 1232  &   53994   &   1030    &   1024    &  BKK+ \\
J1612+2008  &  0.43   &     19.5    &   +22     &   58552   &   8152    &  833  &  57313    &   5928    &    415    &  BKK+ \\
\\[-0pt]
J1627+1419  &  0.49   &     33.8    &   +27     &   56406   &   1221    & 1024  &  57955    &   2898  &        1024   &    BKK+ \\
J1645+1012  &  0.41   &     36.3    &   +36.2   &   54842   &   1457    & 1604  &  53377    &   1461  & 547   &   PHS+; BKK+\\
J1649+2533  &  1.02   &     35.5    &   +23.4   &   55637   &   2382    & 1269  &  54631    &   1988  & 725   &   BKK+\\
J1652+2651  &  0.91   &     40.8    &   +31.8   &   55107   &   1026    & 1017  &  53778    &   722  &  2233    &   BKK+\\
J1713+0747  & 0.005   &     15.9    &	+8.7    &   56768   &  131291   & 146       \\ 
\\[-0pt]
J1720+2150  &  1.62   &     41.1    &   +48     &   55975   &   2103    & 1009  &  55973    &   1903 &  1575    &   BKK+\\
J1739+0612  &  0.23   &     101.5   &   ---     &   56605   &   2561    & 1024      \\ 
J1741+2758  &  1.36   &     29.3    &   +65     &   56419   &   1062    & 885   &   52931   &   882   & 1024    &  BKK+; ZVKU \\
J1746+2245  &  3.47   &     52.     &   +77     &   54540   &   519     & 1024  &   53777   &   519   & 3383    &  BKK+\\
J1746+2540  &  1.06   &     51.5    &   +82     &   52738   &   617     & 1033  &   52931   &   1132  & 1024   &   BKK+\\
\\[-0pt]
J1752+2359  &  0.41   &     36.     &   +88     &   55107   &   2054    & 1597  &   53994   &   1311 &  997  &  BKK+\\
J1758+3030  &  0.95   &     34.9    &   +69     &   54842   &   1035    & 925   &   54631   &   2047 &  1184   &  BKK+\\
J1807+0756 &  0.46   &      89.3	&   +37     &   56605   &   640     & 512   &   55433   &   232    & 906    &       \\
J1811+0702 &  0.46   &     57.8     &   +139.4  &   56769   &   1294 & 1024  &  \multicolumn{3}{c|}{\textbf{Foster \etal\ (1995)}}& MM10  \\
J1813+1822 &  0.34   &     60.8     &   +120    &   56769?  &   1130    & 1024  &   58779   &   3554  & 1086    & BKK+  \\
\\[-0pt]
J1814+1130  &  0.75   &     65.     &   +77     &   56605   &    793    & 1024  &   55283   &   4172  & 1024    &  BKK+ \\
J1819+1305  &  1.06   &     64.9    &   +114    &   57688   &   1026    & 1024  &   53778   &   3394 &  706 &  BKK+ \\
J1821+1715  &  1.37   &     60.5    &   +107.5  &   58552   &   445    & 1138  &   53967   &   1023 &  1024    &  BKK+\\
J1822+0705  &  0.76   &     135.9   &   +159    &   56769   &    656 & 1024 &\multicolumn{3}{c|}{\textbf{Foster \etal\ (1995)}}& MM10  \\
J1828+1359  &  0.74   &      56.    &   +116.6  &   56769   &   1086    &  512  &   53966   &   1024 &  988  &  BKK+ \\
\\[-0pt]
J1829+0000  & 0.199   &     114     &   --53    &   56768   &   3007    & 512 & \multicolumn{3}{c|}{\textbf{\citet{mcewen}}} & \\
J1837+1221  &  1.96   &     100.6   &  +173     &   58779   &   1013    & 1033  &   53378   &    997 &  853  &  BKK+ \\
J1837--0045 &  0.62   &      86.98	&  +130     &   56415   &   1032    & 1205  &   53778   &   1446 & 1205  &   \\
J1838+1650  &  1.90   &     33.9    &  +71.2    &   57688   &   625  &   1024   &  57690   &   1028 &  1024     &  BKK+\\
J1842+0257  & 3.088   &    +148.1   &  +65.3    &   57113 & 513 & 1029 & \multicolumn{3}{c|}{\textbf{\citet{mcewen}}} & \\
\\[-0pt]
J1843+2024  &  3.41   &     85.3    &  +151     &   57533   &   1068 & 1024 &\multicolumn{3}{c|}{\textbf{Champion \etal\ (2005)}}& BKK+ \\
J1848+0826  &  0.57   &     90.8    &  +249     &   52738   &   1825    & 1024    &   58274   &   1546 &  1061    &  BKK+\\
J1849+2423  &  0.28   &     62.2    &  +18      &   56419   &   2175    & 1024    &   57690   &   2172 &  1017     &  BKK+\\
J1850+0026  &  1.082  &    201.4    &  --17.0   &   56415 & 1030 & 1056 & \multicolumn{3}{c|}{\textbf{\citet{mcewen}}} & \\
J1851--0053 &  1.409  &     24 &  --11.2   &   54540 & 1029 & 1375 & \multicolumn{3}{c|}{\textbf{\citet{mcewen}}} & ZVKU \\
\\[-0pt]
J1859+1526  &  0.93   &     97.5    &   +333    &   57115   &   829     & 1037    &  55433  &    1063 &  912    &   BKK+\\
J1900+30    &  0.60   &     65.     &   +122    &   58432   &   5311    & 1024   &  57878   &   2055  & 1197   &  BKK+\\
J1901+1306  &  1.83   &     75.0    &   +276    &   52738   &   1038    & 512     &       52947   &   982  &      1024   &  BKK+\\
J1903+2225  &  0.65   &     109.2   &   +52.3   &   58294  &  3312   &   1033   &  55432  &     1075    &   1122 &  BKK+  \\
J1906+1854  &  1.02   &     156.7   &   +387    &   57886   &   998 &   1131    &   57690  &     1022   &     1024    &  BKK+\\
\\[-0pt]
J1908+0734  &  0.21   &     10.1    &   --19    &   ---     &   ---  &  ---     &    ---    &   ---     & ---   &  ZVKU  \\
J1908+2351  &  0.38   &     101.5   &   +56     &   57886   &   1031 &  1080    &    57690   &  1583   &     1024    &  BKK+\\
J1909+1859  &  0.54   &     64.5    &   +217.9  &   57120   &   1104 &  1059    &   57882   &   1099  &   1078      &  BKK+\\
J1910+0714  &  2.71   &     124.1   &   +147    &   56445   &   1076 &  1004    &   53777   &   553 &   1024    &           \\
J1911+1758  &  0.46   &     49.0    &   +155    &   57131   &   2887 &  1046    &   57882   &   1973 &    1081     &  BKK+\\
\\[-0pt]
J1912+2525  &  0.62   &     37.8    &   +29     &   57886   &   1023 &  1214    &   57882   &   1041  &   1236      & BKK+\\
J1913+3732  &  0.85   &     69.     &   +0.2    &   58300   &   2177  & 1025    &   58294   &   1052  &   1022   &  BKK+\\
J1913+0446 &  1.62   &      109.1   &  --99.6   &   55637   &    706  & 1024    &   58640   &   582 &   1070    &   \\
J1914+0219 &  0.458  &      233.8   &   +275.1  &   57131   &   1721 & 1040 & \multicolumn{3}{c|}{\textbf{\citet{mcewen}}} & \\
J1915+0227 &  0.317  &      192.6   &   +162    &   57131   &   2482 & 813 & \multicolumn{3}{c|}{\textbf{\citet{mcewen}}} & \\
\\[-0pt]
J1915+0738 &  1.54   & 	    39      &   --2.8   &   54842   &   1037  & 1506    &   53966   &   705 &   1024    &   \\
J1919+0134 &  1.60   &      191.9   &   +40     &   54539   &   1028  & 1572    &   52949   &  1028 &    783    &       \\
J1937+2950 &  1.66   &      113.8   &   ---     &   \multicolumn{3}{c|}{\textbf{\citet{Janssen09}}} & \multicolumn{3}{c|}{\textbf{\citet{Janssen09}}} & \\
J1938+0650 &  1.12   &      70.8    &   +95.6   &   55982   &   1025  & 1095    &   53777   &   1034    &   1095   &      \\
J1941+1026  &  0.91   &     138.9   &   --117   &   55633   &   1038 &  1024    &   58775   &    1016  &    1017  &    BKK+\\
\\[-0pt]
J1951+1123  &  5.09   &     31.3    &   --53    &   58779   &   546  &  1018    &   55433   &   912 &    1024     &  BKK+\\
J1953+1149  &  0.85   &     140.    &   --11    &   57886   &   1495  & 1063    &   53777   &   1250 &   1048     &  BKK+\\
J1956+0838  &  0.31   &     68.2    &   --115   &   58278   &   3943 &  1048    &   58274   &   3945 &   1121     &  BKK+\\
J1957+2831  & 0.308   &     138.99  &   --41    &   57981   &   1059 & 1024 & \multicolumn{3}{c|}{\textbf{\citet{mcewen}}} & \\ 
J1959+3620  &  0.41   &     273.    &   +61.4   &   57981   &   1017 &  1024    & \multicolumn{3}{c|}{\textbf{Barr \etal\ (2013)}}&  BKK+ \\
\\[-0pt]
J2002+1637  &  0.28   &     94.4    &   --38.1  &   56604   &   2602 &  1024    &   55283   &   1929 &    1001    &  BKK+\\
J2007+0809  &  0.33   &     53.9    &  --133.7  &   58573   &   9863 &  256     &   58274   &   1718 &    256    &  BKK+\\
J2007+0910  &  0.46   &     48.6    &  --79.3   &   57639   &   2178 &  1024    &   53778   &   1304 &  895    &  BKK+\\
J2008+2513  &  0.59   &     60.6    & --119.2   &   57109   &   1034 &  1150    &   53778   &   1271 &  841     &   BKK+\\
J2010+2845  & 0.565   &    112.5    &  --233.5  &   57115   &   2117 &  1028 & \multicolumn{3}{c|}{\textbf{\citet{mcewen}}} &\\
\\[-0pt]
J2015+2524  &  2.30   &     13.     &  --31.4   &   55633   &   1042 &  1024     &  55432   &   1034 &  1024     &  BKK+\\
J2016+1948  & 0.065   &     33.8    &  --123    &   57115   &   9233 & 1002 & \multicolumn{3}{c|}{\textbf{\citet{mcewen}}} & \\
J2017+2043  &  0.54   &     61.5    &  --163    &   56604   &   1115 &  1024    &   55283   &   874 &    1049   &  BKK+\\
J2033+0042  & 5.01    &     37.8    &  --71.4   & \multicolumn{3}{c|}{\textbf{\citet{lynch2013}}} & 57567 & 847 & 1024 & \\
J2036+2835  &  1.36   &     99.     &  --152    &   57688   &    688 &   1024    &   57690   &   830 &     1024  &   BKK+\\
\\[-0pt]
J2040+1657  &  0.87   &     50.7    &   --90    &   54541   &   1029 &  1024    &   54632   &   1028 &  845   &  BKK+\\
J2043+2740  &  0.096 &      21.     &   --96.4  &   57109   &   6240 &  1063     &  57690   & 6240 &     621   &   BKK+\\
J2045+0912  &  0.40   &     31.8    &   --86.5  &   57639   &    2526 & 1024    &   53778   &    1515 &     988   &   BKK+\\
J2048+2255  &  0.28   &     68.8    &   --189   &   56585   &   3520 &  1109    &   53967   &    1306 &     709    &  BKK+\\
J2111+2106  &  3.95   &     59.7    &   --75.3  &   57317   &   263 &   1930    &   57591   &  1679 &    1024   &   BKK+\\
\\[-0pt]
J2139+2242  &  1.08   &     44.1    &   --89.9  &   58288   &   1057 &  1041    &   56353   &    2898 &    1056   &   BKK+\\
J2151+2315  &  0.59   &     23.6    &   --25.9  &   56768   &   1044 &  1159    &   54632   &   1044  & 1024 &   BKK+\\
J2155+2813  &  1.61   &     77.4    &  --130.5  &   54540   &   1025 &  1024    &   54632   &   1024 &  946  &   BKK+\\ 
J2156+2618  &  0.50   &     48.8    &   --75    &   57109   &   1203 &  1038    &   53377   &   1202 &  973  &   BKK+\\
J2205+1444  &  0.94   &     36.7    &   --22.9  &   56604   &   1064 &  1024    &   54632   &   1034 &  1831 &   BKK+\\      
\\[-0pt]
J2215+1538  &  0.37   &     29.3    &   --20.4  &   56605   &   1602 &  1024    &   54632   &   1604 &  1100 &   BKK+\\
J2222+2923  &  0.28   &     49.4    &   --96.8  &   56768   &   2130 &  1024    &   56361   &   2054 &  1039 &  BKK+\\
J2234+2114  &  1.36   &     35.1    &   --96.9  &   54540   &   1030 &  1326    &   56361   &   1536 &  1024 &   BKK+\\
J2243+1518  &  0.60   &     39.8    &   --37    &   54839   &   1040 &  1326    &   54632   &   1035 &  1024 &   BKK+\\
J2248--0101 &  0.59   &     23.6    &   --7     &   54839   &   2746 &  1020    &   55639   &    741 &  786  & PHS+ \\
\\[-0pt]
J2253+1516  &  0.79   &     29.2    &   --27    &   54839   &   1035 &  1056    &   55639   &   1071 &  1024 &   BKK+\\
J2307+2225  &  0.54   &     7.8     &   --6     &   54540   &   1086 &  1046    &   52948   &   888  & 1050  & BKK+; ZVKU \\
J2355+2246 &  1.84   &      23.1    &   --56    &   58965   &    972  &  256    &   58951   &    972    &   1024    &       \\  
\hline
\end{longtable}
Notes: BKK+: \cite{bilous2016}; JK18: \cite{JK18}; KKWJ: \cite{kkwj98}; KTSD: \citet{KTSD23}; KZUS, \citet{KZUS22}; MM10: \cite{malov2010}; PHS+: \cite{pilia2016}; PRAO: Valerij Maloveev (private communication); SGG+95: \cite{sgg+95}; ZVKU: \cite{ZVKU13}. Values from the ATNF Pulsar Catalog \citep{ATNF}.

Reference parameters for PSRs J0245+1433, J0848+1640, J0928+0614, J1147+0829 and J1844+0036, which do not appear in the ATNF Pulsar Catalogue, were obtained from the discoverers (Kevin Stovall, private communication) as refinements to the measurements of \citet{dsm+13}.

\twocolumn

\begin{table}
\setlength{\tabcolsep}{2pt}
\caption{Pulsar Parameters.
Name, period, period derivative, energy loss rate, spindown age, surface magnetic field, acceleration parameter, and the \citet{pulsar_magnetosphere_book} $1/Q$ parameter.}
\begin{center}
\begin{tabular}{lccccccc}
\hline
 Pulsar &  P & $\dot{P}$ & $\dot{E}$ & $\tau$ & $B_{surf}$ & $B_{12}/P^2$ & 1/Q  \\
  & (s) & ($10^{-15}$ & ($10^{32}$  & (Myr) & ($10^{12}$ &   &   \\
 & & s/s) & ergs/s) & & G) &   &    \\
\hline
\hline

J0006+1834 & 0.694 & 2.10 & 2.479 & 5.2 & 1.22 & 2.5 & 1.0 \\
J0030+0451 & .0049 & 1e-5 & 35. & 7580. & 2.3e-4 & 9.5 & 1.8 \\
J0051+0423 & 0.355 & 2e-3 & 0.062 & 803.0 & 0.05 & 0.4 & 0.1 \\
J0122+1416 & 1.389 & 3.80 & 0.560 & 5.8 & 2.33 & 1.2 & 0.6 \\
J0137+1654 & 0.415 & 0.01 & 0.068 & 537.3 & 0.07 & 0.4 & 0.2 \\
\\
J0139+3336 & 1.248 & 2.06 & 0.420 & 9.6 & 1.62 & 1.0 & 0.5 \\
J0152+0948 & 2.747 & 1.70 & 0.032 & 25.6 & 2.19 & 0.3 & 0.2 \\
J0245+1433 & 2.128 & 2.36 & 0.097 & 14.3 & 2.27 & 0.5 & 0.3	\\
J0329+1654 & 0.893 & 0.22 & 0.119 & 65.8 & 0.44 & 0.6 & 0.3 \\
J0337+1715 & 0.003 & 0.00 & 340. & 2450 & 0.00 & 29.7 & 4.1 \\
\\
J0348+0432 & 0.039 & 0.00 & 1.60 & 2570 & 0.00 & 2.0 & 0.6 \\
J0417+35 & 0.654 &  --- &  --- &  --- &  --- &  --- &  --- \\
J0435+2749 & 0.326 & 0.01 & 0.093 & 632.8 & 0.05 & 0.5 & 0.3 \\
J0457+2333 & 0.505 & 0.01 & 0.046 & 538.3 & 0.09 & 0.3 & 0.2 \\
J0533+0402 & 0.963 & 0.16 & 0.071 & 95.4 & 0.40 & 0.4 & 0.3 \\
\\
J0538+2817 & 0.143 & 3.67 & 490. & 0.6 & 0.73 & 35.8 & 7.1 \\
J0540+3207 & 0.524 & 0.45 & 1.20 & 18.5 & 0.49 & 1.8 & 0.7 \\
J0546+2441 & 2.844 & 7.65 & 0.130 & 5.9 & 4.72 & 0.6 & 0.4 \\
J0609+2130 & 0.056 & 0.00 & 0.540 & 3755 & 0.00 & 1.2 & 0.4 \\
J0611+30 & 1.412 &  --- &  --- &  --- &  --- &  --- &  --- \\
\\
J0613+3731 & 0.619 & 3.24 & 5.40 & 3.0 & 1.43 & 3.7 & 1.4 \\
J0621+0336 & 0.270 & 0.01 & 0.150 & 591.0 & 0.04 & 0.6 & 0.3 \\
J0621+1002 & 0.029 & 0.00 & 0.780 & 9670 & 0.00 & 1.4 & 0.5 \\
J0623+0340 & 0.614 & 0.09 & 0.150 & 110.0 & 0.24 & 0.6 & 0.3 \\
J0625+1015 & 0.498 & 5.77 & 18.42 & 1.4 & 1.72 & 6.9 & 2.2 \\
\\
J0627+0649 & 0.347 & 1.70 & 16.0 & 3.2 & 0.78 & 6.5 & 2.0 \\
J0628+0909 & 1.241 & 0.55 & 0.110 & 35.9 & 0.84 & 0.5 & 0.3 \\
J0630--0046 & 0.681 & 3.75 & 4.70 & 2.9 & 1.62 & 3.5 & 1.3 \\
J0646+0905 & 0.904 & 0.74 & 0.390 & 19.5 & 0.83 & 1.0 & 0.5 \\
J0647+0913 & 1.235 & 6.42 & 1.30 & 3.1 & 2.85 & 1.9 & 0.8 \\
\\
J0658+0022 & 0.563 & 9.15 & 20.0 & 1.0 & 2.30 & 7.2 & 2.3 \\
J0711+0931 & 1.214 & 0.40 & 0.088 & 48.1 & 0.71 & 0.5 & 0.3 \\
J0806+08 & 2.063 & 0.30 & 0.013 & 110.5 & 0.79 & 0.2 & 0.1 \\
J0815+0939 & 0.645 & 0.14 & 0.204 & 73.5 & 0.30 & 0.7 & 0.4 \\
J0843+0719 & 1.366 &  --- &  --- &  --- &  --- &  --- &  --- \\
\\
J0848+1640 & 0.452 & 0.11 & 0.490 & 62.4 & 0.23 & 1.1 & 0.5 \\
J0928+0614 & 2.060 & 1.79 & 0.081 & 18.2 & 1.94 & 0.5 & 0.3 \\
J0943+2253 & 0.533 & 0.09 & 0.234 & 94.0 & 0.22 & 0.8 & 0.4 \\
J0947+2740 & 0.851 & 0.43 & 0.275 & 31.4 & 0.61 & 0.8 & 0.4 \\
J1022+1001 & 0.016 & 4e-5 & 3.80 & 6020 & 8e-4 & 3.2 & 0.8 \\
\\
J1046+0304 & 0.326 & 0.12 & 1.40 & 41.6 & 0.20 & 1.9 & 0.7 \\
J1147+0829 & 1.625 & 2.16 & 0.199 & 11.9 & 1.90 & 0.7 & 0.4 \\
J1238+2152 & 1.119 & 1.45 & 0.408 & 12.3 & 1.29 & 1.0 & 0.5 \\
J1246+2253 & 0.474 & 0.09 & 0.330 & 84.5 & 0.21 & 0.9 & 0.4 \\
J1313+0931 & 0.849 & 0.80 & 0.516 & 16.8 & 0.83 & 1.2 & 0.5 \\
\\
J1404+1159 & 2.650 & 1.38 & 0.029 & 30.5 & 1.93 & 0.3 & 0.2 \\
J1503+2111 & 3.314 & 0.14 & 0.002 & 375.1 & 0.69 & 0.1 & 0.1 \\
J1538+2345 & 3.449 & 6.89 & 0.066 & 7.9 & 4.93 & 0.4 & 0.3 \\
J1549+2113 & 1.262 & 0.85 & 0.168 & 23.4 & 1.05 & 0.7 & 0.4 \\
J1612+2008 & 0.427 & 0.04 & 0.189 & 181.4 & 0.13 & 0.7 & 0.3 \\
\end{tabular}
\end{center}
\label{tabA2}
\end{table}
\begin{table}
\setlength{\tabcolsep}{2pt}
{Table A2. Pulsar Parameters (cont'd)}
\begin{center}
\begin{tabular}{lccccccc}
\hline
 Pulsar &  P & $\dot{P}$ & $\dot{E}$ & $\tau$ & $B_{surf}$ & $B_{12}/P^2$ & 1/Q  \\
      & (s) & ($10^{-15}$ & ($10^{32}$  & (Myr) & ($10^{12}$ &   &   \\
 & & s/s) & ergs/s) & & G) &   &    \\
\hline
\hline
J1627+1419 & 0.491 & 0.39 & 1.312 & 19.8 & 0.44 & 1.8 & 0.8 \\
J1645+1012 & 0.411 & 0.08 & 0.463 & 80.1 & 0.18 & 1.1 & 0.5 \\
J1649+2533 & 1.015 & 0.56 & 0.211 & 28.8 & 0.76 & 0.7 & 0.4 \\
J1652+2651 & 0.916 & 0.65 & 0.336 & 22.2 & 0.78 & 0.9 & 0.5 \\
J1713+0747 & 0.005 & 8e-6 & 35.0 & 8490 & 2e-4 & 9.6 & 1.8 \\
\\
J1720+2150 & 1.616 & 0.74 & 0.069 & 34.6 & 1.11 & 0.4 & 0.3 \\
J1739+0612 & 0.234 & 0.16 & 4.80 & 23.7 & 0.19 & 3.5 & 1.2 \\
J1741+2758 & 1.361 & 1.84 & 0.288 & 11.7 & 1.60 & 0.9 & 0.5 \\
J1746+2245 & 3.465 & 4.92 & 0.047 & 11.2 & 4.18 & 0.3 & 0.2 \\
J1746+2540 & 1.058 & 1.05 & 0.349 & 16.0 & 1.07 & 1.0 & 0.5 \\
\\
J1752+2359 & 0.409 & 0.64 & 3.707 & 10.1 & 0.52 & 3.1 & 1.1 \\
J1758+3030 & 0.947 & 0.72 & 0.334 & 20.8 & 0.84 & 0.9 & 0.5 \\
J1807+0756 & 0.464 & 0.13 & 0.510 & 56.9 & 0.25 & 1.2 & 0.5 \\
J1811+0702 & 0.462 & 2.69 & 11.0 & 2.7 & 1.13 & 5.3 & 1.7 \\
J1813+1822 & 0.336 & 0.02 & 0.218 & 253.8 & 0.09 & 0.8 & 0.4 \\
\\
J1814+1130 & 0.751 & 1.66 & 1.546 & 7.2 & 1.13 & 2.0 & 0.8 \\
J1819+1305 & 1.060 & 0.36 & 0.119 & 46.8 & 0.62 & 0.6 & 0.3 \\
J1821+1715 & 1.367 & 0.87 & 0.135 & 24.8 & 1.10 & 0.6 & 0.3 \\
J1822+0705 & 0.760 & 1.13 & 1.00 & 10.6 & 0.94 & 1.6 & 0.7 \\
J1828+1359 & 0.742 & 0.73 & 0.705 & 16.1 & 0.74 & 1.4 & 0.6 \\
\\
J1829+0000 & 0.199 & 0.52 & 26.0 & 6.0 & 0.33 & 8.2 & 2.3 \\
J1837+1221 & 1.964 & 6.20 & 0.323 & 5.0 & 3.53 & 0.9 & 0.5 \\
J1837--0045 & 0.617 & 1.68 & 2.80 & 5.8 & 1.03 & 2.7 & 1.0 \\
J1838+1650 & 1.902 & 2.68 & 0.154 & 11.3 & 2.28 & 0.6 & 0.4 \\
J1842+0257 & 3.088 & 29.59 & 0.40 & 1.7 & 9.67 & 1.0 & 0.6 \\
\\
J1843+2024 & 3.407 & 1.04 & 0.010 & 51.9 & 1.90 & 0.2 & 0.1 \\
J1848+0826 & 0.329 & 0.27 & 3.026 & 19.1 & 0.30 & 2.8 & 1.0 \\
J1849+2423 & 0.276 & 0.15 & 2.876 & 28.6 & 0.21 & 2.7 & 1.0 \\
J1850+0026 & 1.082 & 0.36 & 0.110 & 47.7 & 0.63 & 0.5 & 0.3 \\
J1851--0053 & 1.409 & 0.87 & 0.120 & 25.6 & 1.12 & 0.6 & 0.3 \\
\\
J1859+1526 & 0.934 & 3.93 & 1.902 & 3.8 & 1.94 & 2.2 & 0.9 \\
J1900+30 & 0.602 &  --- &  --- &  --- &  --- &  --- &  --- \\
J1901+1306 & 1.831 & 0.13 & 0.008 & 221.4 & 0.50 & 0.1 & 0.1 \\
J1903+2225 & 0.651 & 0.45 & 0.636 & 23.2 & 0.54 & 1.3 & 0.6 \\
J1906+1854 & 1.019 & 0.21 & 0.077 & 78.6 & 0.46 & 0.4 & 0.3 \\
\\
J1908+0734 & 0.212 & 0.83 & 34.0 & 4.1  &  0.42 & 9.4 & 2.6 \\
J1908+2351 & 0.378 & 0.02 & 0.125 & 351.9 & 0.08 & 0.6 & 0.3 \\
J1909+1859 & 0.542 & 0.10 & 0.240 & 88.6 & 0.23 & 0.8 & 0.4 \\
J1910+0714 & 2.712 & 6.12 & 0.120 & 7.0 & 4.12 & 0.6 & 0.3 \\
J1911+1758 & 0.460 & 0.01 & 0.053 & 556.8 & 0.08 & 0.4 & 0.2 \\
\\
J1912+2525 & 0.622 & 0.23 & 0.369 & 43.8 & 0.38 & 1.0 & 0.5 \\
J1913+3732 & 0.851 & 1.38 & 0.883 & 9.8 & 1.10 & 1.5 & 0.7 \\
J1913+0446 & 1.616 & 279. & 26.0 & 0.1 & 21.50 & 8.2 & 2.8 \\
J1914+0219 & 0.458 & 1.02 & 4.20 & 7.1 & 0.69 & 3.3 & 1.2 \\
J1915+0227 & 0.317 & 0.30 & 3.70 & 16.8 & 0.31 & 3.1 & 1.1 \\
\\
J1915+0738 & 1.543 & 3.31 & 0.360 & 7.4 & 2.29 & 1.0 & 0.5 \\
J1919+0134 & 1.604 & 0.59 & 0.056 & 43.1 & 0.98 & 0.4 & 0.2 \\
J1937+2950 & 1.657 & 3.48 & 0.302 & 7.5 & 2.43 & 0.9 & 0.5 \\
J1938+0650 & 1.122 & 5.68 & 1.60 & 3.1 & 2.55 & 2.0 & 0.9 \\
J1941+1026 & 0.905 & 1.00 & 0.530 & 14.4 & 0.96 & 1.2 & 0.6 \\
\end{tabular}
\end{center}
\end{table}
\begin{table}
\setlength{\tabcolsep}{2pt}
Table A2. {Pulsar Parameters (cont'd)}
\begin{center}
\begin{tabular}{lccccccc}
\hline
 Pulsar &  P & $\dot{P}$ & $\dot{E}$ & $\tau$ & $B_{surf}$ & $B_{12}/P^2$ & 1/Q  \\
     & (s) & ($10^{-15}$ & ($10^{32}$  & (Myr) & ($10^{12}$ &   &   \\
 & & s/s) & ergs/s) & & G) &   &    \\
\hline
\hline
J1951+1123 & 5.094 & 3.03 & 0.009 & 26.6 & 3.98 & 0.2 & 0.1 \\
J1953+1149 & 0.852 & 2.87 & 1.83 & 4.7 & 1.58 & 2.2 & 0.9 \\
J1956+0838 & 0.304 & 0.22 & 3.093 & 21.9 & 0.26 & 2.8 & 1.0 \\
J1957+2831 & 0.308 & 3.11 & 42.0 & 1.6 & 0.99 & 10.5 & 2.9 \\
J1959+3620 & 0.406 & 0.04 & 0.212 & 178.7 & 0.12 & 0.7 & 0.4 \\
\\
J2002+1637 & 0.276 & 0.23 & 4.202 & 19.5 & 0.25 & 3.3 & 1.1 \\
J2007+0809 & 0.326 & 0.14 & 1.565 & 37.7 & 0.21 & 2.0 & 0.8 \\
J2007+0910 & 0.459 & 0.33 & 1.360 & 21.9 & 0.40 & 1.9 & 0.8 \\
J2008+2513 & 0.589 & 5.40 & 10.41 & 1.7 & 1.80 & 5.2 & 1.8 \\
J2010+2845 & 0.565 & 0.09 & 0.200 & 98.7 & 0.23 & 0.7 & 0.4 \\
\\
J2015+2524 & 2.303 & 0.51 & 0.016 & 71.6 & 1.10 & 0.2 & 0.2 \\
J2016+1948 & 0.065 & 0.00 & 0.580 & 2570 & 0.01 & 1.2 & 0.4 \\
J2017+2043 & 0.537 & 1.00 & 2.53 & 8.5 & 0.74 & 2.6 & 1.0 \\
J2033+0042 & 5.013 & 9.69 & 0.030 & 8.2 & 7.05 & 0.3 & 0.2 \\
J2036+2835 & 1.359 & 2.09 & 0.329 & 10.3 & 1.71 & 0.9 & 0.5 \\
\\
J2040+1657 & 0.866 & 0.59 & 0.362 & 23.1 & 0.73 & 1.0 & 0.5 \\
J2043+2740 & 0.096 & 1.27 & 564.4 & 1.2 & 0.35 & 38.3 & 7.2 \\
J2045+0912 & 0.396 & 0.20 & 1.245 & 32.1 & 0.28 & 1.8 & 0.7 \\
J2048+2255 & 0.284 & 0.02 & 0.262 & 296.7 & 0.07 & 0.8 & 0.4 \\
J2111+2106 & 3.954 & 3.24 & 0.021 & 19.3 & 3.62 & 0.2 & 0.2 \\
\\
J2139+2242 & 1.084 & 1.42 & 0.441 & 12.1 & 1.26 & 1.1 & 0.5 \\
J2151+2315 & 0.594 & 0.71 & 1.337 & 13.3 & 0.66 & 1.9 & 0.8 \\
J2155+2813 & 1.609 & 0.92 & 0.087 & 27.8 & 1.23 & 0.5 & 0.3 \\
J2156+2618 & 0.498 & 0.01 & 0.045 & 555.8 & 0.09 & 0.3 & 0.2 \\
J2205+1444 & 0.938 & 0.90 & 0.431 & 16.5 & 0.93 & 1.1 & 0.5 \\
\\
J2215+1538 & 0.374 & 2.37 & 17.84 & 2.5 & 0.95 & 6.8 & 2.1 \\
J2222+2923 & 0.281 & 0.01 & 0.109 & 722.6 & 0.04 & 0.5 & 0.3 \\
J2234+2114 & 1.359 & 0.22 & 0.035 & 97.0 & 0.56 & 0.3 & 0.2 \\
J2243+1518 & 0.597 & 0.11 & 0.209 & 84.1 & 0.26 & 0.7 & 0.4 \\
J2248--0101 & 0.477 & 0.66 & 2.400 & 11.5 & 0.57 & 2.5 & 1.0 \\
\\
J2253+1516 & 0.792 & 0.07 & 0.053 & 188.8 & 0.23 & 0.4 & 0.2 \\
J2307+2225 & 0.536 & 0.01 & 0.022 & 975.8 & 0.07 & 0.2 & 0.1 \\
J2355+2246 & 1.841 & 3.78 & 0.240 & 7.7 & 2.67 & 0.8 & 0.4 \\
\hline
\end{tabular}
\end{center}
\end{table}

\noindent\textit{\textbf{J0623+0340}} appears to have a double cone {c\bf T} profile, despite the PPA rate being difficult to estimate.  Interestingly, the emission comes in widely separated bursts of 50-100 pulses as shown in Fig.~\ref{figA105}.  The pulsar is detected at 111 MHz \citep{prao} but no measurements are possible.
\begin{figure}
\begin{center}
\includegraphics[height=85mm,angle=-90.]{J0623+0340l_prof_30p_avrgs.ps}
\caption{A 7815 pulse sequence of pulsar J0623+0340 showing 30-pulse averages.  
Note that the emission comes in widely separated bursts of 50-100 pulses.}
\hspace{-0.55 cm}
\label{figA105}
\end{center}
\end{figure}
\vskip 0.1in

\noindent\textit{\textbf{J0625+1015}}: Very probably dominant core emission, with possible weak conal emission on the profile edges.  No indication of the PPA rate is discernible.  We model the beam with a core-single geometry.  
\vskip 0.1in

\noindent\textit{\textbf{J0627+0649}}: Our 1.4 GHz profiles are suggestive of a core/double-cone structure; however, we see only some inflections in our lone 327-MHz profile.  Further, the fractional linear $L/I$ is unusually large, and the PPA rate is both well defined and relatively shallow---resulting in an oblique rather than central sightline traverse.  We model only an outer conal geometry, and this would require a core width of about 11\degr, a possible but uncomfortably large value given the scale of the 1.4-GHz profile.  The higher frequency profile shows a potential core/double-cone structure, but additionally no reasonable inner cone solution can be achieve with such a shallow PPA rate. 
\vskip 0.1in

\noindent\textit{\textbf{J0628+0909}}: Our three narrow 1.4-GHz profiles of this RRAT \citep{McLaughlin06} show slightly different variations of a similar three-component structure, where the central feature surprisingly shows strong hints of antisymmetric $V$, and putative conal outriders show as inflections in the total power profiles or structure in the $L/I$ one. Its PPA rate seems to be fairly steep and +ve.  If the core width is just larger that that of the polar cap, a core/inner conal model can be computed. One can however see that the profiles are formed of a small fraction of strong pulses, so its profiles of even several thousand pulses may not be very stable.  
\vskip 0.1in

\noindent\textit{\textbf{J0630--0046}} We have only the 1.4-GHz polarimetry to go on.  It seems dominated by the trailing conal component, and the leading component together with the PPA traverse suggest that they are conal.  The \citet{mcewen} profile suggests a larger width, but its poor quality provides no clarity.  We thus model it as an inner conal double, but weak core emission may occur in the saddle region.
\vskip 0.1in

\noindent\textit{\textbf{J0646+0905}}: we model the pulsar roughly with an inner conal single {\bf S$_d$} configuration, but it could also be outer conal with a larger $\alpha$ value.  No clear drift feature is seen in fluctuation spectra.  The $L$ and PPA traverse 
are asymmetric as in other B0809+74-like profiles, possibly suggesting some ``missing'' early emission.
\vskip 0.1in

\noindent\textit{\textbf{J0647+0913}}: One of our 3 observations shows a deep 3.9-$P$ 
modulation, whereas the other 2 do not, suggesting that the pulsar is conal and has 
several modes.  Single pulses show that the leading-edge ``kink'' on the profile is 
produced by two populations of subpulses at different longitudes.  The trailing part 
of the profile is too weak to see clearly, either in the pulse stack or the profile, but 
this strong leading/weak trailing behavior is seen in many conal quadruple pulsars.
Another issue is the unclear PPA traverse, but we guestimate the rate to be around 
+15\degr/\degr\ and model the geometry with a double cone {c\bf Q} configuration.
\vskip 0.1in

\noindent\textit{\textbf{J0658+0022}}: one can only guess that the pulsar has a core single 
{\bf S$_t$} configuration.  
\vskip 0.1in

\noindent\textit{\textbf{J0711+0931}}: The pulsar shows a usual outer conal {\bf S$_d$} evolution and configuration.  Scattering may broaden the LOFAR profile slightly.  
\vskip 0.1in

\noindent\textit{\textbf{J0806+08}} has a double-cone structure and perhaps a weak core component, thus we model it. The PPA traverse seems to have the usual shallow edges and steep center, so we assume a highly central sightline traverse.
\vskip 0.1in

\noindent\textit{\textbf{J0815+0939}}: This pulsar became well known as the first four-component pulsar.  The double-cone model seems fairly successful here, and at 149 MHz we see that as usual the outer conal components are stronger than the inner conal ones. The conal quadruple c{\bf Q} classification seems inescapable here, however not without some problems:  the inner cone dimensions seem to vary substantially and other issues are discussed in \citet{teixeira}. 
\begin{figure}
\begin{center}
\includegraphics[width=65mm,angle=-90.]{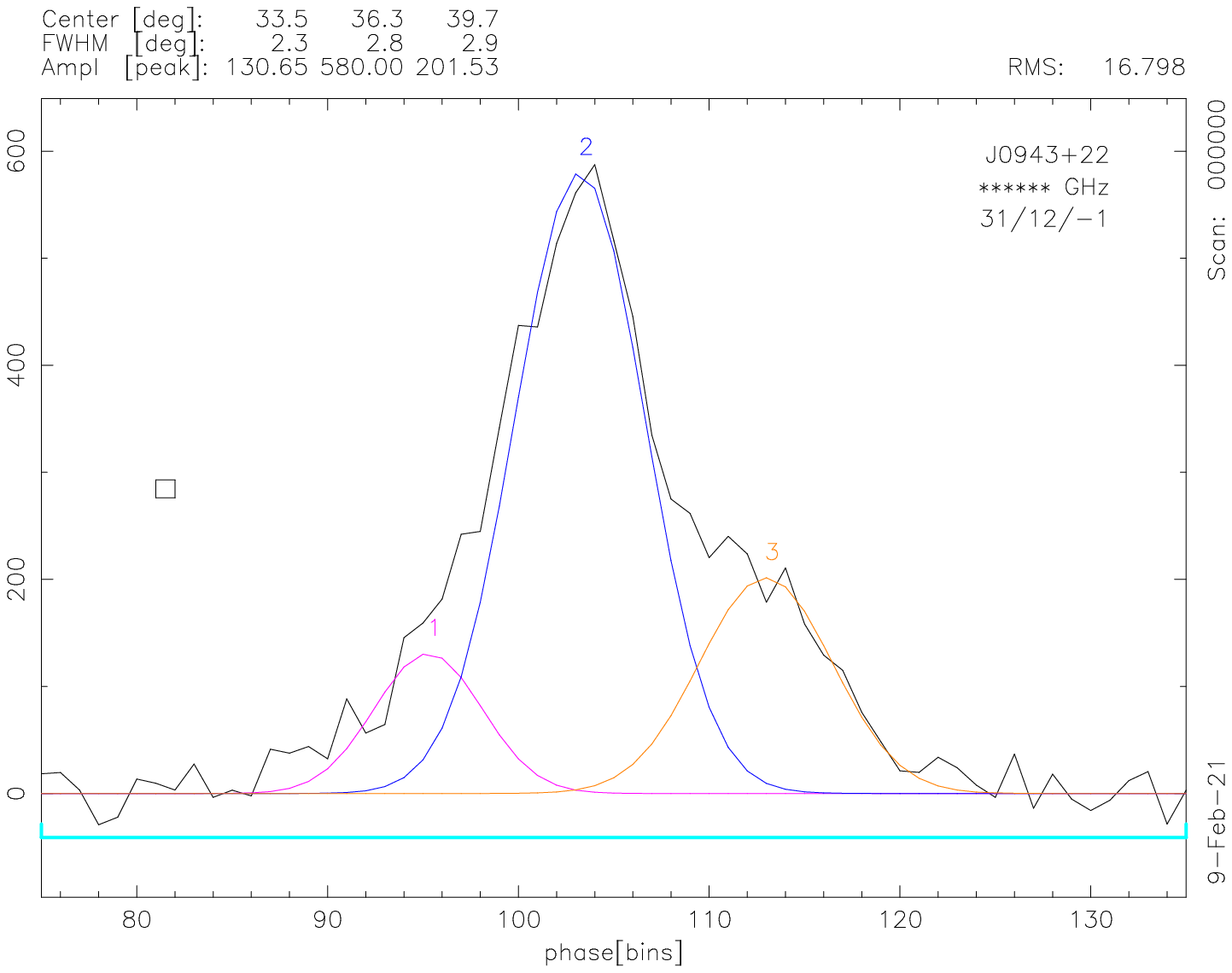}
\caption{J0943+2253: Fitted Gaussian-component model for BKK+'s 149-MHz observation.}
\label{figA106}
\end{center}
\end{figure}
\vskip 0.1in

\noindent\textit{\textbf{J0843+0719}}: The pulsar seems to have a steep spectrum, so our several 1.4-GHz observations were all non-detections, but fortunately \citet{Burgay+06} provide a total power profile of about the same width as ours at 327 MHz.  The three features are clearly seen at 149 MHz as shown in Fig.~\ref{figA106}.  The PPA rate is very well defined, but no $\dot P$ has been measured. We thus model it with an inner conal single geometry.  
\vskip 0.1in

\noindent\textit{\textbf{J0848+1640}}: The pair of well resolved components seem to refect a standard outer conal double configuration, and we so model as such.  
\vskip 0.1in

\noindent\textit{\textbf{J0928+0614}}: Another pulsar with a steep spectrum, so there is little information in the 1.4-GHz profile.  Our two 328-MHz profiles show a complex structure with a central feature that seems to be labeled by $V$ in the core manner as well as features what could represent two cones (although no clear trailing inner cone feature is seen, but it is often conflated with the outer one).  The PPA rate is also uncertain but an estimate is possible.  We thus model the lower frequency profile asperationally as potentially having an {\bf M} configuration and assume that the high frequency geometry is similar.  The inner cone dimension is only a guess. 
\vskip 0.1in

\noindent\textit{\textbf{J0943+2253}}: Noting \citet{brinkman_freire_rankin_stovall}'s {\bf D/T?} classification, this pulsar seems to have several conflated features at all three frequencies; see Fig.~\ref{figA106}.  We now see that its emission consists of bursts of several hundred pulses interspursed by nulls that can persist for similar intervals.  Fluctuation spectra within the bursts reveal a deep nearly odd-even (1.97 $P$) stationary modulation.  We thus model it as a conal triple c{\bf T}, and this structure is easiest to see at 327 MHz.  The pulsar is detected at 111 MHz \citep{prao} but no measurements are possible.
\vskip 0.1in

\begin{figure}
\begin{center}
\includegraphics[width=65mm,angle=-90.]{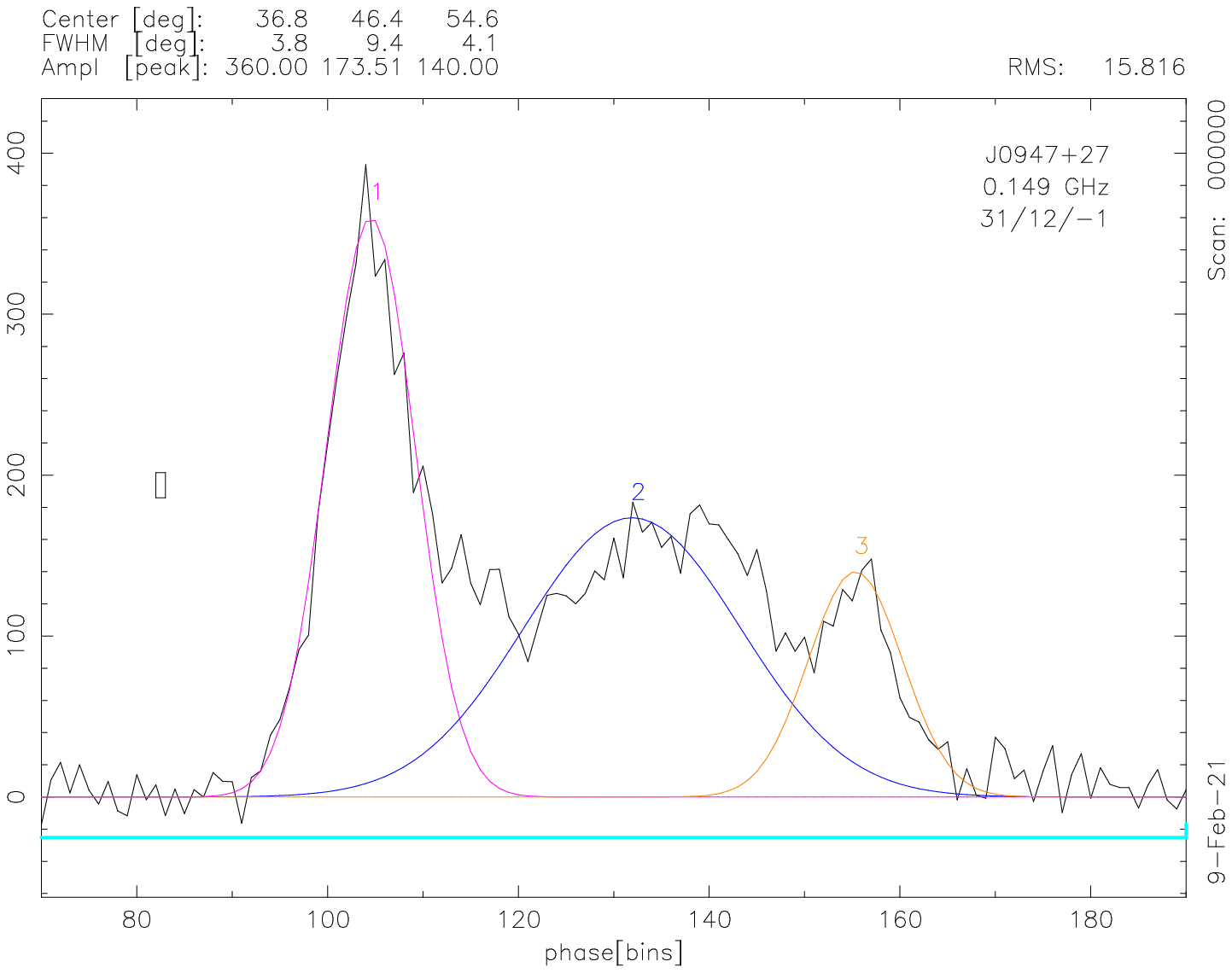}
\caption{J0947+2740: Fitted Gaussian-component model for BKK+'s 149-MHz observation.}
\label{figA107}
\end{center}
\end{figure}
\vskip 0.1in
\noindent\textit{\textbf{J0947+2740}}: Noting \citet{brinkman_freire_rankin_stovall}'s {\bf T/M} classification, there is evidence of at least three components present at LOFAR frequencies with the first leading component being the strongest; see Fig.~\ref{figA107}. At higher frequencies, the components seem to merge with the middle component remaining the strongest.  While the fluctuation spectra give no guidance, the structure seems to be that of a conal triple c{\bf T} profile.  The pulsar is detected at 111 MHz \citep{prao} but no measurements are possible.
\vskip 0.1in

\noindent\textit{\textbf{J1022+1001}}: This 16-ms pulsar seems to have a core-cone triple {\bf T} configuration in our profiles.  The central components have widths compatible with an orthogonal geometry---but narrow substantially at 4.9 GHz \citep{kkwj98}---and the conal widths, though difficult to measure accurately, show a consistent inner conal behavior.  The PPA rate is difficult to interpret and measure as different tracks are seen at different frequencies (see also the other polarimetry in \citep{xilouris98,stairs99,Dai2015}; however, a rate around +2\degr/\degr\ is compatible with an inner cone.  No core is discernible at 149 MHz or below, but the LWA profiles \citep{KTSD23} trace the overall width and onset of scattering at 35 MHz. The $t_{scatt}$ value seems much larger than the pulsar experiences.  The profile asymmetry could be the result of aberration/retardation.
\vskip 0.1in

\noindent\textit{\textbf{J1046+0304}}: The pulsar shows weak emission over much of its rotation cycle especially at the lower frequency.  Our poor 1.4-GHz profile seems to confirm the double form of the main feature but not the weaker possible postcursor seen in the \citet{Burgay+06} image.  Several 327-MHz observations show two broad dissimilar features, and we are at a loss about how these correspond to the 1.4-GHz profiles---and there is no reliable PPA signature to go on.  The pulsar needs further study before any model can be attempted.  
\vskip 0.1in

\noindent\textit{\textbf{J1147+0829}}: The 1.4-GHz profiles are of poor quality but have about the same narrow widths as the lower frequency one, where the PPA rate can also be estimated.  All show some evidence of bifurcation.  We model the profiles using an inner conal single/double beam. 

\vskip 0.1in

\begin{figure}
\begin{center}
\includegraphics[width=65mm,angle=-90.]{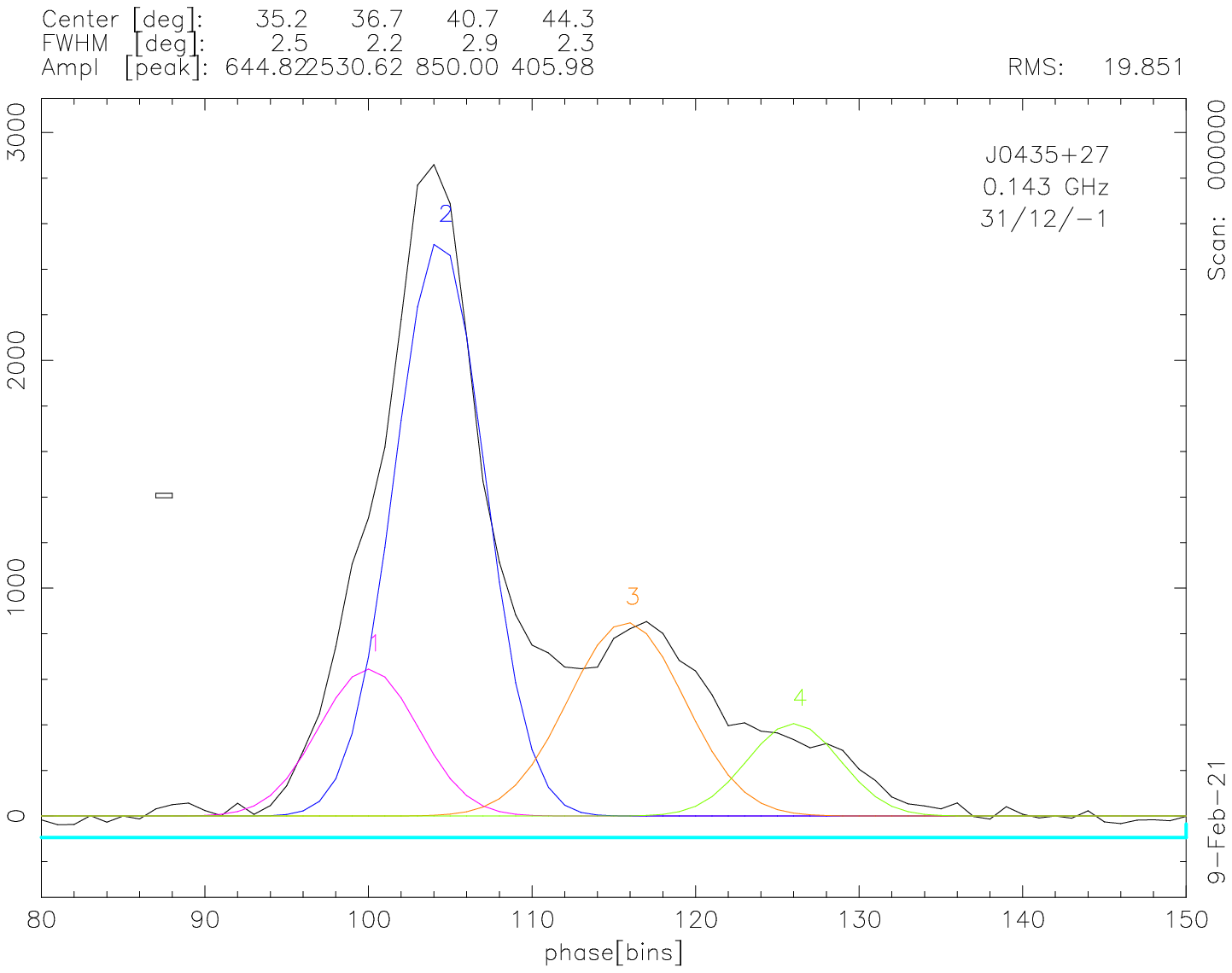}
\caption{J1238+2152: Fitted Gaussian-component model for PHS+'s 149-MHz observation.}
\label{figA108}
\end{center}
\end{figure}
\noindent\textit{\textbf{J1238+2152}}: The pulsar has three components that are clearly seen at LOFAR frequencies (see Fig.~\ref{figA108}), visible at 327 MHz, and not so at 1.4 GHz.  Single pulses show what appear to be weak drift bands with a $P_3$ of some 16 $P$.  We model it using a conal triple c{\bf T} beam geometry.   We see no evidence of scattering at the lowest frequencies.  The pulsar is detected at both 111 MHz \citep{prao} and 20/25 MHz \citep{ZVKU13} but no measurements are possible.
\vskip 0.1in

\noindent\textit{\textbf{J1246+2253}}: We note \citet{brinkman_freire_rankin_stovall}'s Brinkman  {\bf S$_t$} classification, and at LOFAR frequencies, we see a single component with perhaps some structure.  However, the 1.4-GHz profile has a triple structure as may the 327-MHz one as well. Fluctuation spectra show a 12-$P$ amplitude modulations, so we again find that a conal triple c{\bf T} beam model is most appropriate.  The pulsar is detected at 111 MHz \citep{prao} but no measurements are possible.
\vskip 0.1in

\begin{figure}
\begin{center}
\includegraphics[height=85mm,angle=-90.]{PQJ1313+0931_modfold.ps}
\caption{Modulation-folded display of pulse intensity over the 11.6-$P$ modulation cycle of pulsar J1313+0931.  Note that the leading and trailing components are also modulated.}
\label{figA109}
\end{center}
\end{figure}
\noindent\textit{\textbf{J1313+0931}}: The LOFAR profiles of this pulsar show leading and trailing components straddling a bright, asymmetric central component, and at the higher frequencies we see a similar structure.  The pulse sequences are modulated with an 11.6-$P$ period as shown in Fig.~\ref{figA109} and an apparently non-commensurate modulation at 3.8 $P$ is also seen.  This overall structure is seen in many conal quadruple c{\bf Q} profiles and we thus model it as such.  The pulsar is detected at 111 MHz \citep{prao} but no measurements are possible.  It is also detected in three LWA bands \citep{KTSD23} but the profiles are too narrow to reflect the full profile structure, so we cannot know just what survives.  
\vskip 0.1in

\begin{figure}
\begin{center}
\includegraphics[height=85mm,angle=-90.]{PQJ1404+1159.57566p_4.7P_modfold.ps}
\caption{Modulation-folded display of pulse intensity over the 4.7-$P$ modulation cycle of pulsar J1404+1159.  In other parts of the 3435-pulse sequence, the period in more like 4.9 $P$.}
\label{figA110}
\end{center}
\end{figure}
\noindent\textit{\textbf{J1404+1159}}: The profile is single at 1.4-GHz but develops weak ``outriders'' at 327 MHz.  No doubt that the pulsar has a conal triple configuration, given the strong, regular drifting subpulses shown in Fig.~\ref{figA110}.  The pulsar is strongly detected at 111 MHz \citep{prao} and the width seems compatible with those at higher frequency.
\vskip 0.1in

\noindent\textit{\textbf{J1503+2111}}: seems to show a fairly usual outer conal single {\bf S$_d$} evolution down to 129 MHz.  Scattering seems to be negligible.
\vskip 0.1in

\noindent\textit{\textbf{J1538+2345}}: There seems to be some question about whether this is an RRAT or a pulsar.  In any case, we find no published profiles at frequencies other than 1.4 GHz.  Therefore, we model it with a nominal conal double {\bf D} geometry.  The pulsar is detected at 111 MHz \citep{prao}, but not in a manner that the width can be determined.
\vskip 0.1in

\noindent\textit{\textbf{J1549+2113}}: Throughout the three bands, only one component is present in this pulsar's profile, so we model it as an inner conal single profile.  Scattering seems to have little effect.  The PPA rate, and thus $\alpha$ is poorly determined.  
\vskip 0.1in

\begin{figure}
\begin{center}
\includegraphics[width=65mm,angle=-90.]{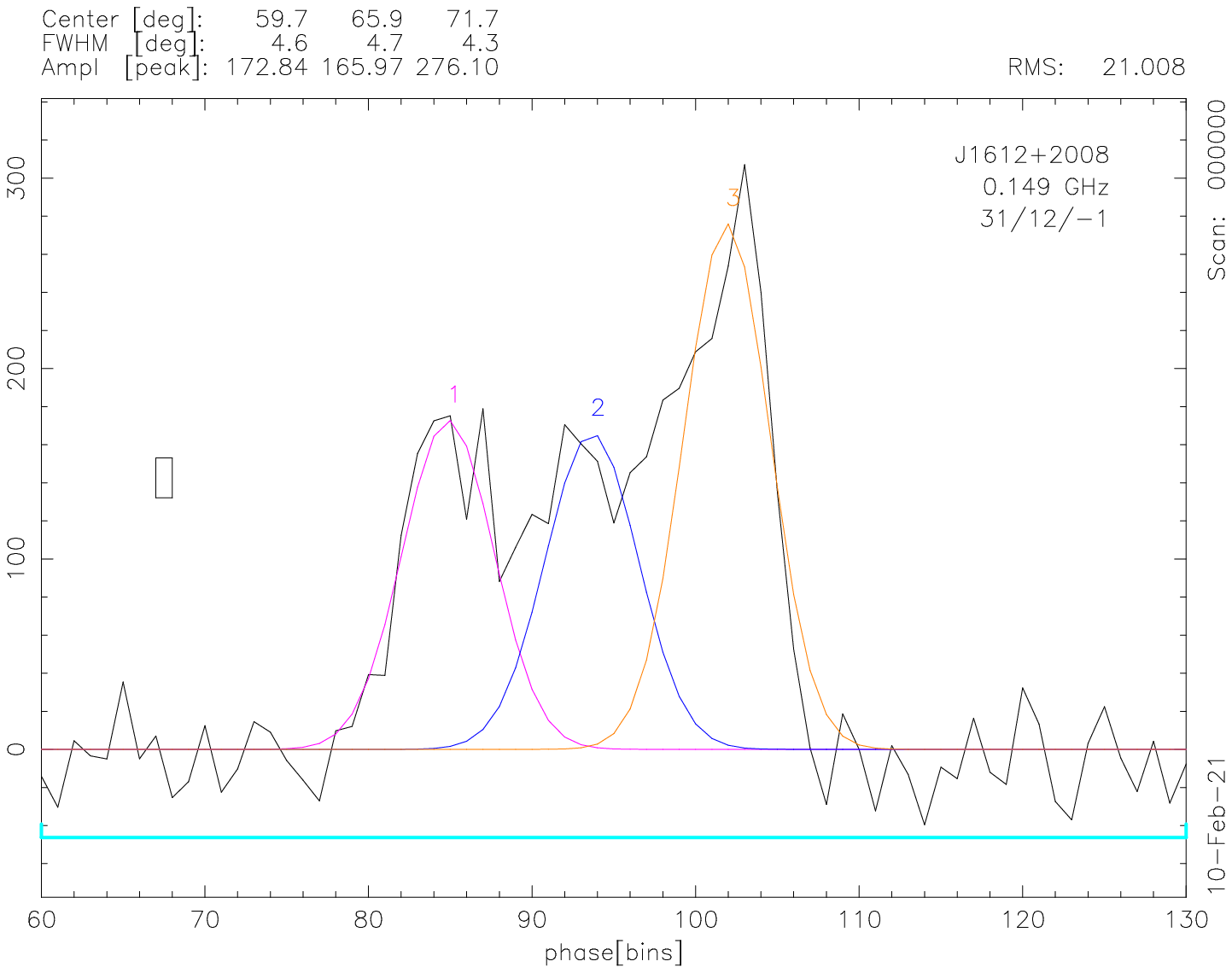}
\caption{J1612+2008: Fitted Gaussian-component model for BKK+'s 149-MHz observation.}
\label{figA111}
\end{center}
\end{figure}
\vskip 0.1in
\noindent\textit{\textbf{J1612+2008}}: The LOFAR profile of this pulsar has two components with the trailing stronger than the leading; see Fig.~\ref{figA111}.  As frequency increases, the strength of the leading component weakens.  We see a hint of a central component with a roughly 4\degr\ width at 1.4 GHz.   We thus model it as a triple {\bf T} profile.  The inner cone geometry requires an $\alpha$ of 90\degr, and the above core width is compatible with this geometry.  The model shows some inner cone increase with wavelength and little effect of scattering.
\vskip 0.1in

\noindent\textit{\textbf{J1627+1419}}: This pulsar's profile shows three or four barely resolved components both in the AO and LOFAR profiles.  The structure suggests a c{\bf T} or c{\bf Q} classification, and we model it as such.  The pulsar is detected at 111 MHz \citep{prao} but no measurements are possible.  Scattering may slightly broaden the LOFAR profile.
\vskip 0.1in

\noindent\textit{\textbf{J1645+1012}}: At LOFAR frequencies, two merged components are present in this pulsar, the leading slightly stronger than the trailing; as frequency increases, the leading component becomes ever less prominent.  We model this as an inner cone double {\bf D} profile, despite the poorly determined PPA rate.  The pulsar is detected at 111 MHz \citep{prao} but no measurements are possible.  Scattering gradually broadens the profiles in the LWA observations down \citep{KTSD23} to 50 MHz.
\vskip 0.1in

\begin{figure}
\begin{center}
\includegraphics[height=85mm,angle=-90.]{PQJ1649+2533l_prof_1-200.ps}
\caption{Pulsar J1649+2533: 200-pulse sequence at 1.4 GHz showing its 2.5-period modulation and burst/null structure.}
\label{figA112}
\end{center}
\end{figure}
\noindent\textit{\textbf{J1649+2533}}: PSR J1649+2533 has a single component with some structure in all three bands. Fluctuation spectra show a strong 2.6-$P$ phase modulation as well as a strong amplitude modulation at about 30 $P$---the first associated with the prominent drifting subpulses and the latter the structure of bursts and nulls as seen in Fig.~\ref{figA112}.  It appears compatible with an inner conal single {\bf S$_d$} evolution, and we model it as such.  The pulsar is detected at 111 MHz \citep{prao} but no measurements are possible.  Effects of scattering seem to be minimal.
\vskip 0.1in

\noindent\textit{\textbf{J1652+2651}}: Two main components are seen in this pulsar, with the trailing having double peaks.  We see evidence of a double structure at 327 MHz but not at the other frequencies. Moreover, fluctuation spectra show an 18-$P$ stationary periodicity that modulates the entire profile.  This seems to reflect a conal quadruple c{\bf Q} geometry, and we model it as such; however, the inner conal dimensions are poorly determined.  The pulsar is detected at 111 MHz \citep{prao} but no measurements are possible.
\vskip 0.1in

\noindent\textit{\textbf{J1713+0747}}: For this 4.6-ms pulsar, a standard core-cone model fails completely.  The central putative core component is far narrower than the polar cap diameter, and here the very shallow PPA rate confounds any chance of finding an adequate model.  See also \citet{kkwj98} and \citet{Dai2015}.
\vskip 0.1in

\begin{figure}
\begin{center}
\includegraphics[width=65mm,angle=-90.]{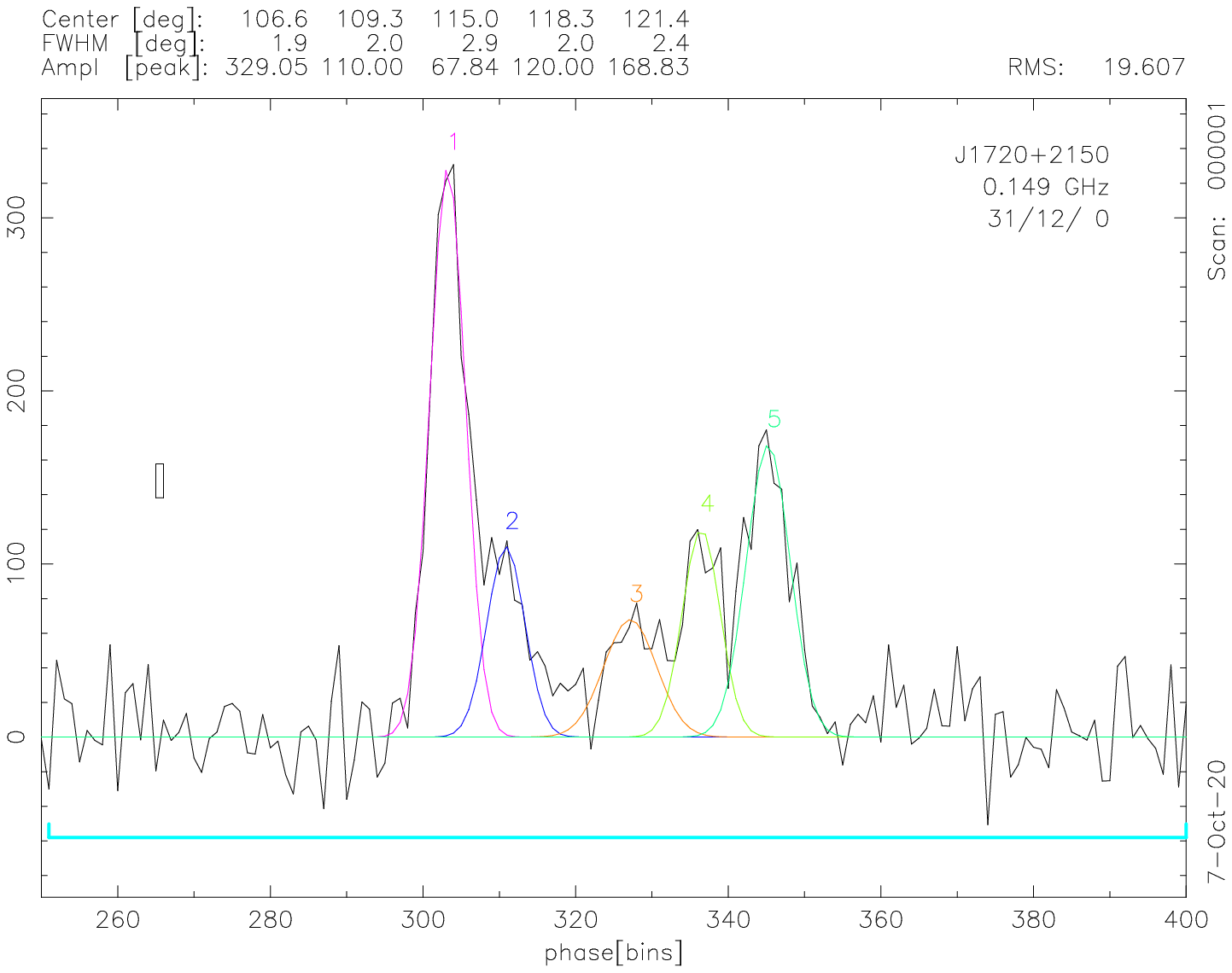}
\caption{Gaussian-component fitting of J1720+2150 at 149 MHz. The width of the second component is constrained to have the same width as that of component 4---so as to be consistent with what is usually observed.}
\label{figA113}
\end{center}
\end{figure}

\noindent\textit{\textbf{J1720+2150}}: This pulsar exhibits a striking resemblance to the five-component exemplar B1237+25.  Like this pulsar the central core component is weak and conflated with the inner conal features, and the PPAs indicate a highly central sightline traverse. A plot showing fitted Gaussian components at 149 MHz is given in Fig.~\ref{figA113}.  We model it as a double cone {\bf M} structure, but the inner conal dimensions are estimates.


\vskip 0.1in

\noindent\textit{\textbf{J1739+0612}}: the four ``components" of linear polarization at 1.4 GHz are suggestive of a conal quadruple {c\bf Q} configuration, and the quantitative geometry seems to bear this out.  Moreover, we see an 11-$P$ fluctuation feature in the single pulses again indicating that the emission we see is conal.  The \citet{ATNF} 692-MHz profile, though noisy, may have a similar structure, though only guesses at the widths are possible.  In any case, we model it.
\vskip 0.1in

\noindent\textit{\textbf{J1741+2758}}: Three components over our full frequency range in this pulsar, and its single pulses seem to show little regularity.  The middle component seems to be a core feature, and its width at just over 2\degr\ gives an orthogonal geometry and a core/outer conal structure.  The beams show little growth at low frequency as can be seen at 111 MHz \citep{prao}; however, the Kharkiv 20/25-MHz profile \citep{ZVKU13} is clearly scattered. The evolution is seen in the LWA observations \citep{KTSD23} down to 50 MHz where the triple structure can be discerned.
\vskip 0.1in

\noindent\textit{\textbf{J1746+2245}}: The profiles of this pulsar in the three bands show evidence of both inner and outer conal pairs with the former weaker in the LOFAR band per the usual evolution.  Fluctuation spectra show features at some 5.3 and 32 $P$.  The inner cone seems discernible at 149 MHz, but its dimensions cannot be measured reliably.  We thus model the profiles as having a conal quadruple c{\bf Q} structure, and we note that any core would probably be missed in the resulting geometry because $\beta$ is larger than the core beam halfwidth.  We see no clear effect of scattering.
\vskip 0.1in

\noindent\textit{\textbf{J1746+2540}}: Though the pulsar's profile evolution from single at 1.4 GHz to double at the lower frequencies is usual in the conal single {\bf S$_d$} geometry, here the two lower frequency profiles have the same dimensions, so it seems that a part of the 1.4-GHz profile is missing, perhaps due to moding.  We have thus modeled the profile as an inner conal double.  Scattering seems minimal here.
\vskip 0.1in

\noindent\textit{\textbf{J1752+2359}}: This pulsar shows a single component with weak outriders and a flat PPA traverse. We thus model it as a core single {\bf S$_t$} profile.  However, the asymmetric core appears incomplete at both 1.4 GHz and 327 MHz, so we use the 149-MHz width that then squares with the usual inner cone geometry.  Other 327-MHz profiles show the trailing outrider better than this one with a poor baseline.  The pulsar is detected at 111 MHz \citep{prao} but no measurements are possible.  Scattering seems to have a minimal effect.
\vskip 0.1in

\noindent\textit{\textbf{J1758+3030}}: A single component is seen at all frequencies.  We then model it as an inner conal single {\bf S$_d$} profile, although \citet{Weltevrede2007} found it has a flat (featureless) fluctuation spectrum.  The pulsar is detected at 111 MHz \citep{prao} but no measurements are possible.  The LWA profiles \citep{KTSD23} trace the profile width down to 35 MHz where scattering is clearly discernible. 
\vskip 0.1in

\noindent\textit{\textbf{J1807+0756}}: The two somewhat broad profiles seem to show the usual conal evolution from single at high frequency to double at meter wavelengths---and the later is a little broader so we tilt to an outer conal geomettry.  Both profiles are so depolarized that no PPA traverse can be gleaned, but a guess of about 4\degr/\degr\ results in an appropriate grazing sightline geometry. 
\vskip 0.1in

\noindent\textit{\textbf{J1811+0702}}: This pulsar has a closely-spaced, highly-polarized double profile at 1.4 GHz and a more separated one in Foster \etal's (\citeyear{foster}) 430-MHz profile.  The PPA rate is slight, implying a grazing sightline traverse, which here may miss the pulsar's core beam.  The single-peaked feature in MM10's profile is hard to interpret in terms of the double profile at higher frequencies.  Possibly only the trailing component is being seen at 102 MHz in that MM10's scattering values seems too small to produce a full conflation of the profile there.
\vskip 0.1in
\vskip 0.1in

\noindent\textit{\textbf{J1813+1822}}: PSR J1813+1822 seems to have a conal profile.  We model it with a conal single {\bf S$_d$} geometry, but the 1.4-GHz profile may suggest a triple form and thus a conal triple configuration.  The pulsar was not detected by LOFAR at 149 MHz.
\vskip 0.1in

\noindent\textit{\textbf{J1814+1130}}: The profile of this pulsar seems to have an inner conal single {\bf S$_d$} profile.  The LOFAR non-detection seems likely to have been the result of a usual level of scattering at 149 MHz.
\vskip 0.1in

\begin{figure}
\begin{center}
\includegraphics[width=65mm,angle=-90.]{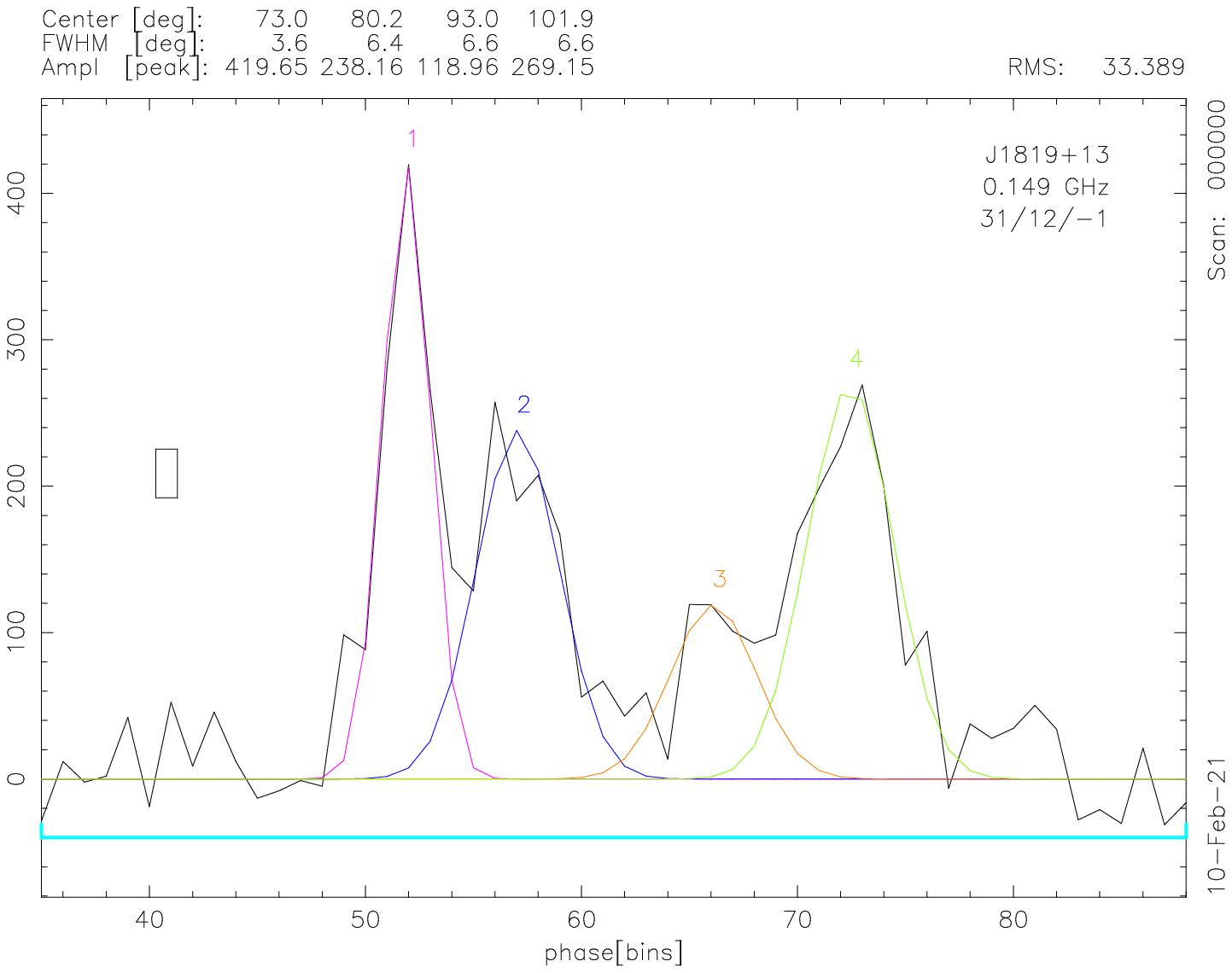}
\caption{J1819+1305: Fitted Gaussian-component model for BKK+'s 149-MHz observation.}
\label{}
\end{center}
\end{figure}
\noindent\textit{\textbf{J1819+1305}}: PSR J1819+1305 was studied by \citet{RankinWright} and is an excellent example of the conal quadruple c{\bf Q} configuration.  The pulsar exhibits two modes, one of which illuminates only the leading components, thus the third component is often missing in its profiles; however, it is quite clearly visible in the LOFAR observations; see Fig.~\ref{figA113}.  Scattering may have some effect at 149 MHz.
\vskip 0.1in

\noindent\textit{\textbf{J1821+1715}}: This pulsar's triplicity is most evident at 327 MHz, but it can be discerned both at 1.4 GHz and 149 MHz.  The central component is always conflated, but its width at 327 MHz must be about 5\degr.  The value is too wide for a core but comfortable for an inner cone.  We therefore model the pulsar as a conal triple c{\bf T} structure.  Scattering seems to have little effect at 149 MHz.
\vskip 0.1in

\noindent\textit{\textbf{J1822+0705}}: PSR J1822+0705 seems to have an inner conal double {\bf D} profile.  However, the magnetic geometry is poorly determined because the PPA rate is ambiguous.  We model the pulsar with a positive rate and use \citet{foster}'s 430-MHz profile.  MM10's 102-MHz detection seems compatible with the two other observations, and the scattering seems to be only a few degrees. 
\vskip 0.1in

\noindent\textit{\textbf{J1828+1359}}: An outer conal single profile is seen in this pulsar.  The LOFAR non-detection at 149 MHz could either be due to larger than average scattering or poor S/N.
\vskip 0.1in

\noindent\textit{\textbf{J1829+0000}}: despite this pulsar's wide profile, our timing seems adequate, and indeed the discovery profile \citet{lorimer2006} seems quite similar.  Its fluctuation spectra are featureless, so we tilt to modeling it as a core-single profile.  The poor \citet{mcewen} 350-MHz profile is narrower than the high frequency one, which also shows structure that can be interpreted as a core and outrider pair.  If the core width is a plausible 11-12\degr, and we interpret the PPA traverse as a central one, the overall width of the outrider pair is some 40\degr, and a core/inner conal {\bf S$_t$} results quantitatively.    
\vskip 0.1in

\noindent\textit{\textbf{J1837+1221}}: This pulsar is most likely an inner cone double {\bf D}.  There is little linear polarization but the $R$ value seen at 327 MHz is supported by the 21-cms profile by \cite{JK18}.  As shown in the model plot, scattering is very likely to have quenched the profile at 149 MHz, making the LOFAR detection impossible.
\vskip 0.1in

\noindent\textit{\textbf{J1837--0045}}: Our two 1.4-GHz profiles are tripartite with an unusually pointed central component and broad wings, and its single pulses show that it is dominated by a small number of very bright narrow subpulses; whereas the lower frequency profile shows a broad Gaussian form.  The PPA rate is unclear, a --6\degr/\degr\ value can be estimated from one of the higher frequency profiles, where the lower frequency suggests about half this.  Even using the steeper rate, the ``pointy'' putative central core features with widths of some 4\degr\ do not seem to measure $\alpha$ accurately.  We conjecture that these are incomplete and model the geometry with a 7\degr\ ``complete?'' (polarcap) width. The \citet{mcewen} 350-MHz halfwidth value is twice as large as our 1.4-GHz values, so we could also interpolate them to a 7\degr\ 1-GHz width.  Then, if the flared edges of the profile are conflated conal outriders, we can compute that they have about the right dimensions to represent an inner cone.  Our fluctuation spectra show ``red noise''  which also supports our interpretation of a core-single {\bf S$_t$} beam geometry.  
\vskip 0.1in

\noindent\textit{\textbf{J1838+1650}}: Two prominent components are seen in this pulsar at 149 and 327 MHz and a single one at 1.4 GHz.  This is a usual outer conal single {\bf S$_d$} evolution, and we model it as such.  Scattering seems to have little effect at 149 MHz.  The pulsar is detected at 111 MHz \citep{prao} but no measurements are possible.
\vskip 0.1in

\begin{figure}
\begin{center}
\includegraphics[height=85mm,angle=-90.]{PQJ1842+0257.57113_prof_1-256.ps}
\includegraphics[height=85mm,angle=-90.]{PQJ1842+0257.57113_modfold_120-170.ps}
\caption{Pulsar J1842+0257 emits a complex mixture of drifting-like emission 
and long nulls (upper panel), and the pulses 120-170 show a clear 3.-period 
drift modulation.}
\hspace{-0.55 cm}
\label{figA114}
\end{center}
\end{figure}
\noindent\textit{\textbf{J1842+0257}} shows a complex emission comprised of what appears to be drift modes and long nulls as shown in Fig.~\ref{figA114}---clearly attesting to its conal character.  Its 1.4-GHz profile is symmetrical with a well defined negative PPA sweep.  The \citet{mcewen} 350-MHz width may indicate an increase with wavelength, so we model the profile as having an outer conal single {\bf S$_d$} geometry. 
\vskip 0.1in

\begin{figure}
\begin{center}
\includegraphics[height=85mm,angle=-90.]{PQhrfJ1843+2024.ps}
\caption{Harmonic-resolved fluctuation spectrum of the 1.4 GHz observation of J1843+2024.  Note the strong narrow fluctuation feature that is the harmonic of a nearly even-odd response.}
\hspace{-0.55 cm}
\label{figA115}
\end{center}
\end{figure}
\noindent\textit{\textbf{J1843+2024}}: This very slow pulsar has a narrow conal double profile, where both conflated components are seen at 1.4 GHz but perhaps only one at 430 MHz \citet{Champion2005}.  Our harmonic resolved fluctuation spectrum shows both a near odd-even response and its stronger harmonic; see Figure~\ref{figA115}.  This response is very similar to what is seen in pulsar B0943+10 \citep{deshpande} and begs to be followed up.  The single spike in the LOFAR 149-MHz profile could then be a marginal detection.  Its width can only be roughly estimated as 11\degr\ or less.  
\vskip 0.1in

\noindent\textit{\textbf{J1848+0826}}: The broad 1.4-GHz profile has two conflated components and perhaps a hint of a weaker core component; whereas the 327-MHz profile seems to resolve them somewhat.  We model the profile as an outer conal double, though a core could have an 8\degr\ width.  A hint of the scattered profile at 149 MHz may be visible. 
\vskip 0.1in

\noindent\textit{\textbf{J1849+2423}}: Two components are discernible at all frequencies with hints of perhaps other conflated features in  this pulsar.  The model reflects an inner cone double {\bf D} configuration.  No evidence of 149-MHz scattering at the average level.
\vskip 0.1in

\noindent\textit{\textbf{J1850+0026}}: The asymmetric double profile shows a clear 9-$P$ drift modulation in the trailing component.  We model it as having as an outer conal double {\bf D} configuration approximating the PPA sweep rate as --4\degr/\degr.  The 350-MHz measurement \citep{mcewen} gives no more than a guess.
\vskip 0.1in

\noindent\textit{\textbf{J1851--0053}}: A strange profile with a shallow rise and steep decay as well as a flat PPA traverse.  \citet{mcewen} 350-MHz profile is too poor to be of help.  The pulsar is detected at 20/25 MHz \citep{KZUS22} but scattering prevents useful measurements.  No sensible model is possible.  Scattered emission is weakly detected at 20/25 MHz \citep{ZVKU13}.
\vskip 0.1in

\noindent\textit{\textbf{J1859+1526}}: PSR J1859+1526 shows three components at L-band that may be present at lower frequencies.  A core-cone triple {\bf T} model gives an outer cone structure, and if the central component at L-band is a core, its plausible width of 3.5\degr\ give a compatible $\alpha$ value of 46\degr.
\vskip 0.1in

\noindent\textit{\textbf{J1900+30}}: This pulsar has a single profile at L and P band as well as at 149 MHz. The disordered PPA traverse and significant Stokes $V$ suggests that this is a core component.  No spindown value has been measured for this pulsar, so we know nothing of the magnetic field strength, age or $\dot E$.  The pulsar was found at 1.4 GHz at a declination of 30\degr\ 56', and the AO feed has a beam size of about 3.5', so the improved declination error is about $\pm$2'.
\vskip 0.1in

\noindent\textit{\textbf{J1901+1306}}: Possibly a triple profile with a weak leading component.  This structure seems clearest in the 327-MHz profile, but the 1.4-GHz profile then seems incompatible.   Strangely the components are broader at 1.4 GHz, and this does not appear to be a timing issue.  If this model is correct, the core width shows a highly unusual behavior.

\vskip 0.1in

\noindent\textit{\textbf{J1903+2225}}: The pulsar has an asymmetric single profile at 1.4 GHz that evolves into a double form at 327 MHz, with the leading feature much weaker than the trailing one.  There is a hint of this leading feature at 1.4 GHz as well as in \citet{han2009} at 774 MHz, and this total width is roughly that at 327 MHz.  On this basis we model the profiles as an {\bf S$_d$} or {\bf D}, but more sensitive profiles are needed to confirm this interpretation.  Scattering may well account for the LOFAR non-detection at 149 MHz.  
\vskip 0.1in

\noindent\textit{\textbf{J1906+1854}}: The 1.4-GHz profile of this pulsar has a very different shape that the others, but its width is only about 10\% greater.  We model its beams as having a conal single {\bf S$_d$} configuration, which is confirmed by the strong regular drifting subpulses.  Scattering could contribute to the increased widths of the LOFAR profiles.
\vskip 0.1in

\noindent\textit{\textbf{J1908+0734}}: We have no Arecibo observations of this pulsar but profiles at 1.2, 1.5, 2.4 GHz \citep{Bhat04} and 430 MHz \citep{camilo_nice} have been published.  And at decameter wavelengths it has been detected at 20/25 MHz \citet{ZVKU13}.  The 1.5-GHz profile seems to have 4 features, but none of the observations are polarimetric, so no interpretation is possible.
\vskip 0.1in

\noindent\textit{\textbf{J1908+2351}}: One component is seen in all three bands, and the polarization is so slight that no PPA slope can be estimated.  We model its as an inner conal {\bf S$_d$} beam and take $\beta$ as 7\degr\ for a peripheral sightline traverse.  Scattering may be responsible for the broadening at low frequency.  
\vskip 0.1in

\begin{figure}
\begin{center}
\includegraphics[width=65mm,angle=-90.]{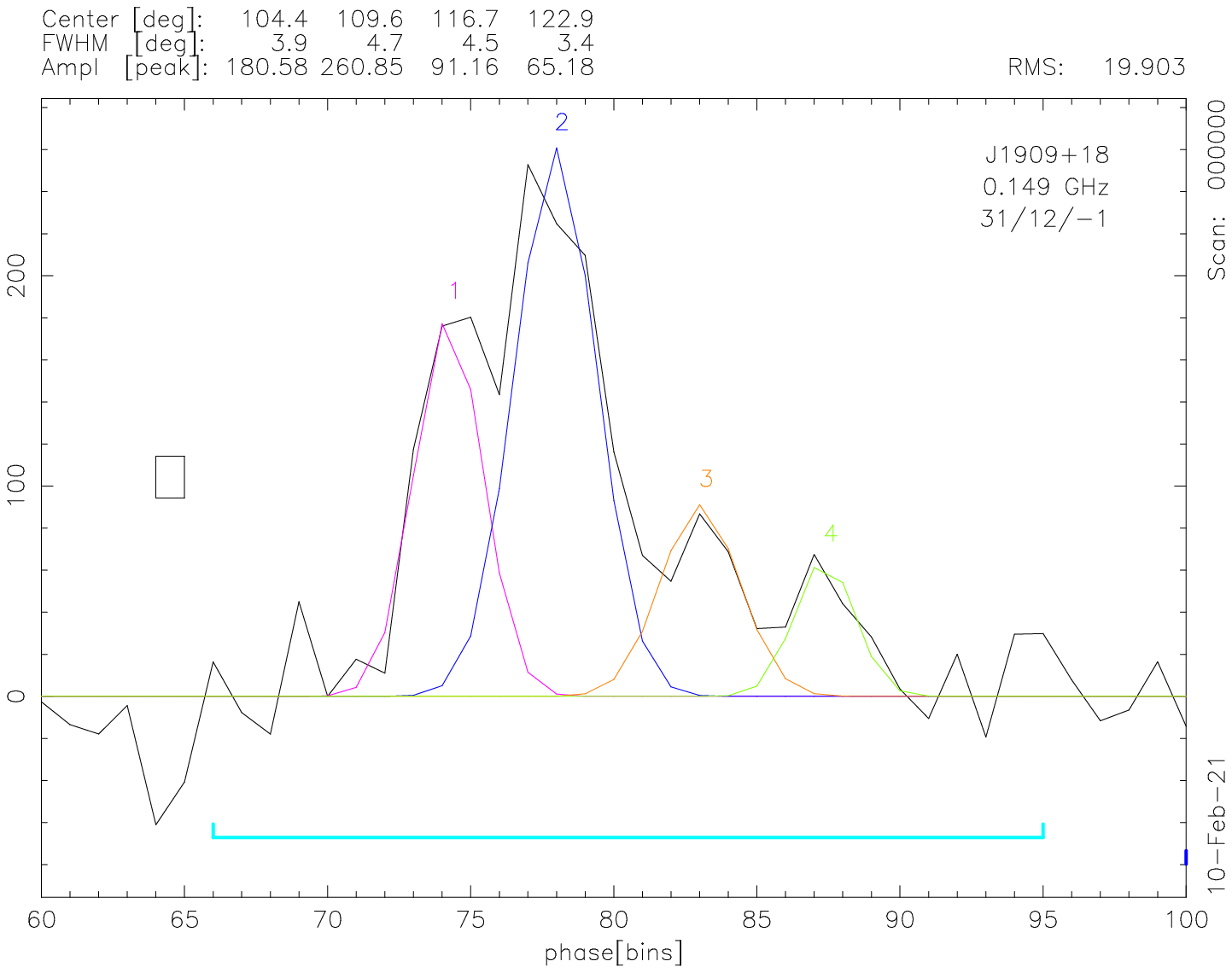}
\caption{J1909+1859: Fitted Gaussian-component model for BKK+'s 149-MHz observation.}
\label{figA116}
\end{center}
\end{figure}
\vskip 0.1in
\noindent\textit{\textbf{J1909+1859}}: Each of this pulsar's profiles suggests four conflated components, but all are difficult to measure accurately; see Fig.~\ref{figA116}.  We interpret the profiles as conal quadruple c{\bf Q} and estimate the inner and outer conal widths down to 149 MHz.  Scattering seems to have little effect on the LOFAR profile.  
\vskip 0.1in

\noindent\textit{\textbf{J1910+0714}}: Both of our 1.4-GHz profiles suggest three features in the $L/I$.  We provisionally model the beam structure using a conal triple {c\bf T} beam system, though as usual the inner cone dimensions are very approximate.
\vskip 0.1in

\noindent\textit{\textbf{J1911+1758}}: Despite the poor observation, this pulsar seems to have two unresolved features at lower frequencies [see also \citet{han2009}].  Using the patch of significant linear at L-band to estimate $R$, the outer conal single {\bf S$_d$} solution shows a small increase of $\rho$ with wavelength.  A high quality 430-MHz profile \citep{camilo_nice} shows a clear central component, so it is possible that a conal triple c{\bf T} model would be more correct, but our observations fail to show this structure clearly enough.  Scattering seems unimportant at 149 MHz.  
\vskip 0.1in

\noindent\textit{\textbf{J1912+2525}}: This pulsar has one component throughout the frequency spectrum [see also the \citet{han2009} 774-MHz profile and the 430-MHz from \citet{camilo_nice}], although those at 327 MHz and lower show a slight bifurcation.  We then model it as an inner conal single {\bf S$_d$} profile.  The conal radii increase somewhat with wavelength, and here an outer cone is impossible because the inner one requires an $\alpha$ of 90\degr.  A nearly odd-even drift fluctuation feature in our 327-MHz observation confirms the conal character of the emission.  Scattering seems to have little effect on the LOFAR profiles.  The pulsar is detected at 111 MHz \citep{prao} but no measurements are possible.
\vskip 0.1in

\begin{figure}
\begin{center}
\includegraphics[width=65mm,angle=-90.]{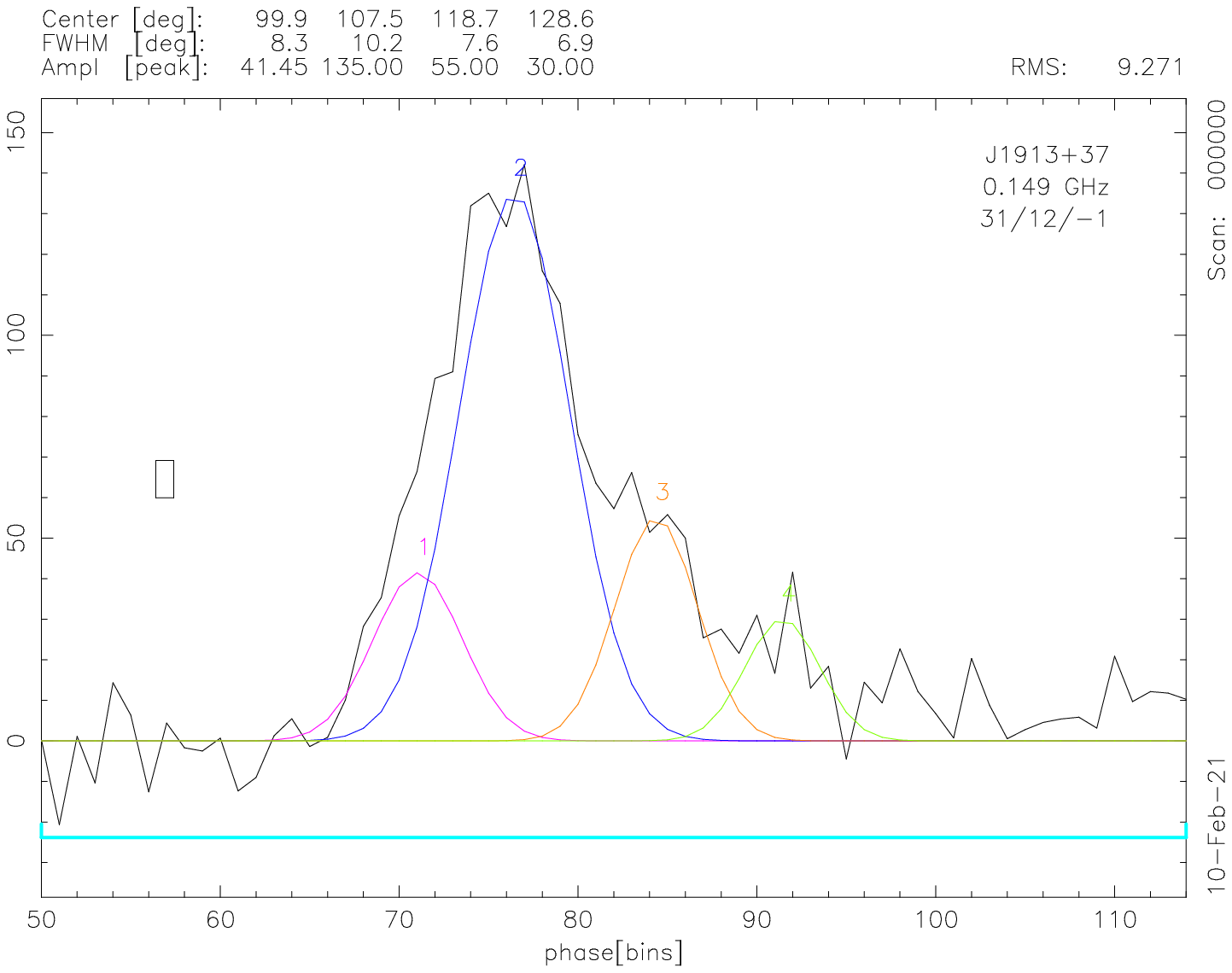}
\caption{J1913+3732: Fitted Gaussian-component model for BKK+'s 149-MHz observation.}
\label{figA117}
\end{center}
\end{figure}
\noindent\textit{\textbf{J1913+3732}}: This pulsar seems to be a canonical example of a {c\bf T} or {c\bf Q}, and the 149-MHz profile seems to have a structure similar to that at 327 MHz, though with poorer S/N and perhaps some scattering this is a difficult call. See Fig.~\ref{figA117}. We estimate the widths at 149 MHz by using the same measurement points.  The pulsar is detected at 111 MHz \citep{prao} but no measurements are possible.
\vskip 0.1in

\noindent\textit{\textbf{J1913+0446}}: The 1.4-GHz profile seems to show a bifurcated central core component and two pairs of conal features---all of which can be modeled quantitatively using a core/double-cone beam configuration. The 327-MHz profile then has three clearly resolved features and perhaps other weak or conflated ones, but the profile quality prevents clear tracing of their evolution.  
\vskip 0.1in

\noindent\textit{\textbf{J1914+0219}} is probably a core-cone triple where the core is conflated with the conal components.  The PPA rate is well defined so we model it as a core/inner conal triple.  The \citet{mcewen} 350-MHz profile is useless due to poor resolution.  The putative width of the core is about 6\degr\ which squares well with the above.  An outer conal model would require a core width of some 4\degr, and this seems too narrow.
\vskip 0.1in

\noindent\textit{\textbf{J1915+0227}} seems to reflect a conal beam as 7.8-$P$ amplitude modulation is seen in the single pulses.  We model the geometry using an inner cone because the \citet{mcewen} poorly resolved 350-MHz profile indicated little growth with wavelength.  
\vskip 0.1in

\noindent\textit{\textbf{J1915+0738}}: The two profiles have about the same widths but are asymmetric with a fast rise and slower falloff.  Its profiles almost look scattered, but the pulsar's small dispersion and \citet{camilo_nice} 430-MHz observation that shows a similar form and width make this unlikely.  One of the profiles also has a very flat PPA traverse.  Overall, the profiles appear to represent the leading portions of a conal double pair, but possible features trailing by 90\degr\ or more are too weak to interpret reliably.  It emits in clumps of a few to perhaps 20 pulses, at times alternating between emission in the leading and trailing parts of the profile.  Occasional drift bands may also be seen.   However, it is the flat PPA traverse that prevents modeling.  
\vskip 0.1in

\noindent\textit{\textbf{J1919+0134}}: The pulsar seems to have a standard outer conal double configuration.  The PPA rate is estimated from the 1.4-GHz profile as the polarimetry was faulty at the lower frequency.  
\vskip 0.1in

\begin{figure}
\begin{center}
\includegraphics[width=65mm]{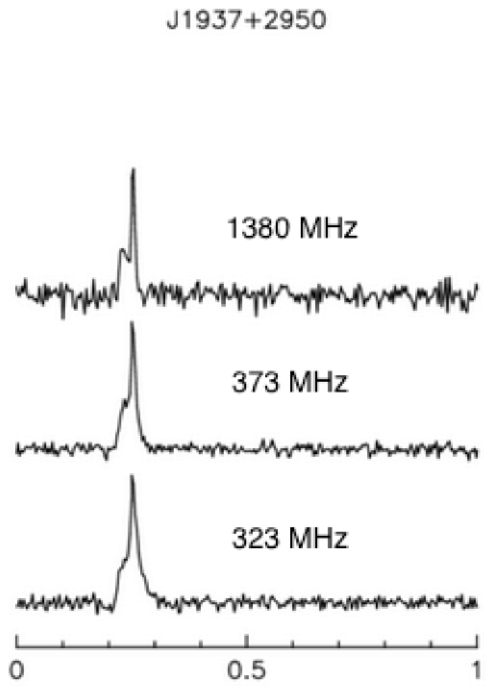}
\caption{J1937+2950: Profiles at three frequencies from \citet{Janssen09}.}
\label{figA117a}
\end{center}
\end{figure}
\noindent\textit{\textbf{J1937+2950}}: We never had opportunity to observe this pulsar in our Arecibo programs.  It was included in the LOFAR High Band Survey but was not detected.  The only available observations are those of \citet{Janssen09} and are not polarimetric (Fig.~\ref{figA117a}). Most probably the pulsar has a conal double geometry with perhaps a hint of scattering at the lowest frequency---but without polarimetry no model is possible.
\vskip 0.1in

\noindent\textit{\textbf{J1938+0650}}: Both profiles are Gaussian-shaped with the lower frequency one much broader.  The pulsar appears to reflect a standard outer conal single beam configuration. 
\vskip 0.1in

\noindent\textit{\textbf{J1941+1026}}: The pulsar seems to have a triple {\bf T} beam geometry.  The 1.4-GHz core width squares with an outer cone, but at 327 MHz the core may be wider, and its width could suggest an inner cone.  The triple profile structure is reiterated in \citet{han2009} and \citet{camilo_nice}.  An average scattering level at 149 MHz probably explains the LOFAR nondetection.  
\vskip 0.1in

\noindent\textit{\textbf{J1951+1123}}: Remarkably, both a core single and conal single model work for this very slow 5.1-s pulsar.  We show the former model in the plot and table, but don't necessarily favor it.  For the $R$ of --18\degr/\degr\ even $\alpha$ would be very similar in an {\bf S$_d$} model.  A fluctuation spectrum shows a weak feature, but apparently not a drift fluctuation, so it is unclear on which classification is correct.  Scattering may be responsible for the LOFAR nondetection.
\vskip 0.1in

\noindent\textit{\textbf{J1953+1149}}: This pulsar seems to have an inner conal single {\bf S$_d$} profile with a high magnetic inclination such that no outer conal model is possible.  Here then is a case where the inner cone shows a substantial increase with wavelength.  Scattering may be a factor in the LOFAR non-detection.
\vskip 0.1in

\noindent\textit{\textbf{J1956+0838}}: We have modeled this wide profile as an inner conal single, but without conviction.  The profile width grows substantially with wavelength and the inclination is already such that an outer conal model is excluded.  The PPA rate is well determined, and scattering seems to have minimal effect at 149 MHz.  
\vskip 0.1in

\noindent\textit{\textbf{J1957+2831}}: This energetic pulsar might well have a core component given the $V$ signature, but the apparent shallow PPA traverse suggests that the sightline might miss it.  We have tried to model the profile with a number of double-cone/core configurations, and none seem satisfactory.   The \citet{han2009} 774-MHz profile is similar to ours and seems to have 4 or 5 features, whereas the 350-MHz \citet{mcewen} profile may be single.  Another approach would be to interpret the five features as a core/double-cone beam configuration.  The core width could be close to the the 4.4\degr\ polar cap diameter as the 350-MHz latter profile suggests.  This produces the usual conal widths if $\beta$ is some 5/degr\ which in turn requires an 11\degr/degr\ PPA rate; however, this seems incompatible with the profiles that \textit{might} have a rate as low as 4\degr/\degr.  This is an interesting pulsar that would reward more investigation.  
\vskip 0.1in

\noindent\textit{\textbf{J1959+3620}}: The broad 1.4-GHz profile shows hints of a triple structure in the polarization; but the profile's quality does not facilitate pursuing this interpretation.  The \citet{barr} 21-cm profile seems to have a similar form and width, so we model it using a conal double {\bf D} geometry.  The pulsar was not detected by LOFAR at 149 MHz, and scattering may be an important reason.
\vskip 0.1in

\noindent\textit{\textbf{J2002+1637}}: seems to have a triple {\bf T} profile, comprised of the central component and weaker conal features on its edges.  This said, the core widths in our profiles are too narrow to support this interpretation; however, the higher quality 430-MHz \citep{camilo_nice} shows a core width of 10.5\degr, the core in our profiles may be partial as is suggested particularly at 327 MHz.    Scattering at the average level seems to have little effect on the 149-MHz observation.
\vskip 0.1in

\noindent\textit{\textbf{J2007+0809}}: The broad profiles of this pulsar seem to have an inner conal single {\bf S$_d$} geometry, and a narrow fluctuation feature in our 327-MHz observation may well confirm this classification.  Scattering at the normal level seems to have little effect.  The pulsar is detected at 111 MHz \citep{prao} but no measurements are possible.
\vskip 0.1in

\noindent\textit{\textbf{J2007+0910}}: Evidence of only one component is present throughout the frequency spectrum in this pulsar.  The pulsar is detected at 111 MHz \citep{prao} but no measurements are possible.
\vskip 0.1in

\noindent\textit{\textbf{J2008+2513}}: This pulsar seems to be another {c\bf T} or {c\bf Q}.  The outer conal dimensions are easily measured at all three frequencies, and the inner ones as well at the upper two bands.  The problem is that the PPA traverse gives little useful information.  For the double cone geometry model to be satisfactory an $R$ value of about 6\degr/\degr\ is required, and fortunately the \citet{han2009}'s 774-MHz profile confirms this.  Scattering seems minimal at 149 MHz.  
\vskip 0.1in

\noindent\textit{\textbf{J2010+2845}}: Clearly a core-cone triple {\bf T} beam configuration.  Fluctuation spectra show weak evidence of periodicity in the conal outer pair of components.  The PPA traverse is steep, and the core width can be estimated at about 4\degr---the linear peak has about this width.  This implies an inner cone geometry, and there is no solution for an outer conal one with the observed PPA rate. 
\vskip 0.1in

\noindent\textit{\textbf{J2015+2524}}: PSR J2015+2524 appears to have an outer conal double {\bf D} geometry.  The \citet{camilo_nice} 430-MHz profile confirms the width values.  Scattering seems unusually low for this pulsar, so this does not explain why LOFAR failed to detect it at 149 MHz. 
\vskip 0.1in

\noindent\textit{\textbf{J2016+1948}}: This 65-ms pulsar is plausibly conal dominated.  The weak fluctuation spectra show no features, the PPA traverse is unclear and the profile width quite narrow.  We model it aspirationally as a conal single profile, guessing that the PPA rate has to be shallower than about --3\degr/\degr.
\vskip 0.1in

\begin{figure}
\begin{center}
\includegraphics[width=65mm,angle=-90.]{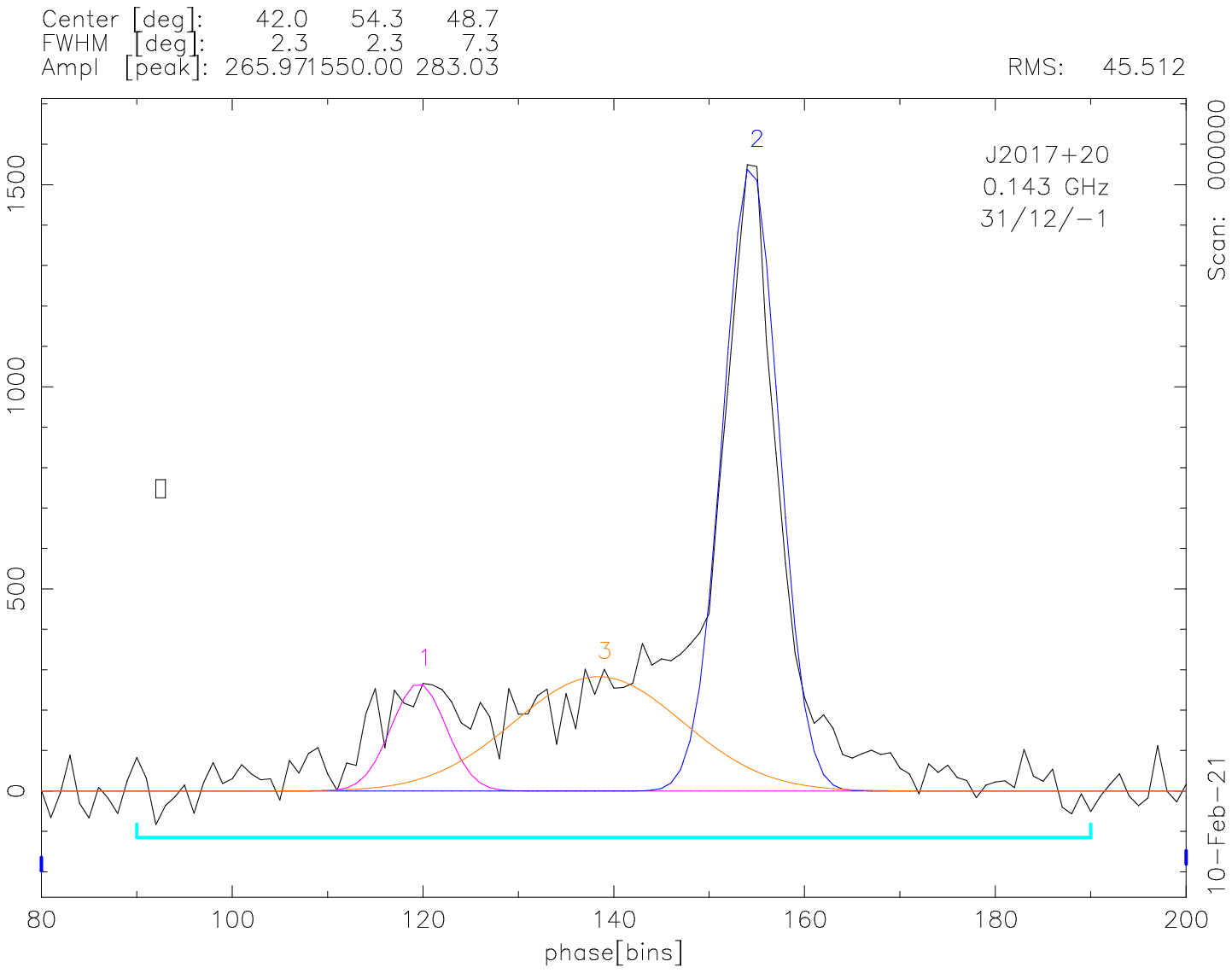}
\caption{J2017+2043: Fitted Gaussian-component model for BKK+'s 149-MHz observation.}
\label{figA118}
\end{center}
\end{figure}
\noindent\textit{\textbf{J2017+2043}}: This pulsar has two components over the bands, the trailing increasing relative to the leading one with wavelength.  The central region is filled, and higher quality LOFAR profiles might show a third component; see Fig.~\ref{figA118}.  We model the geometry as an inner conal double configuration, but an outer model might be preferable as there is some escalation in the beam size.  Scattering seems to be of little effect at LOFAR frequencies.  The pulsar is detected at 111 MHz \citep{prao} but no measurements are possible.
\vskip 0.1in
\vskip 0.1in

\noindent\textit{\textbf{J2033+0042}}: This 5-s pulsar seems to be mostly or entirely conal.  The two widths are impossible to compare accurately but seemingly similar.  The PPA rate is a guestimate given the early turndown.  We model it using an inner conal single beam system.  
\vskip 0.1in

\noindent\textit{\textbf{J2036+2835}}: A double profile is seen in our Arecibo observations of this pulsar with a progressively stronger trailing component.  At LOFAR frequencies, the trailing is much stronger than the leading, and there may be a central component as well.  We model its geometry as an inner {\bf D} conal double, and the inclination is large enough that it cannot be modeled as a outer cone.  Scattering seems to have little effect at 149 MHz.
\vskip 0.1in

\noindent\textit{\textbf{J2040+1657}}: This pulsar has a 4 or 5 component profile that is difficult to classify.  As 327 MHz there appears to be a bifurcated trailing component and a hint of bifurcation in the leading one with a further feature in between.  These five features can be interpreted as two cones and possibly a core.  The dimensions of this outer and inner cone can be estimated, and a satisfactory model results for an $\alpha$ of some 22\degr.  This implies a core width of about 7\degr, and the putative core seems to have this width.  At L-band, the inner and outer conal features merge into a double form, and the core all but disappears. A compatible structure can also be discerned at 149 MHz, but no inner cone width can be estimated.
\vskip 0.1in

\noindent\textit{\textbf{J2043+2740}}: This fast pulsar seems to have an inner cone triple {\bf T} profile with a delayed PPA inflection point. The core width can be estimated from the P-band profile, and if its width is about 8\degr\ it implies an $\alpha$ value that is compatible with an inner cone geometry for the outriders.
\vskip 0.1in

\noindent\textit{\textbf{J2045+0912}}: The profile of this pulsar has a bright central and trailing component and a weaker leading one at P- and L-band, and the LOFAR profile can be read this way as well.  No unique $R$ value could be obtained.  We model it as a conal triple (c{\bf T}) beam system.  Scattering seems minimal at 149 MHz. 
\vskip 0.1in

\noindent\textit{\textbf{J2048+2255}}: It seems to have a core-cone triple {\bf T} profile with the three components more or less discernible at each of the three frequencies.  Maybe the PPA rate is about +9\degr/\degr.  No scattering effect evident.
\vskip 0.1in

\begin{figure}
\begin{center}
\includegraphics[width=60mm,angle=-90.]{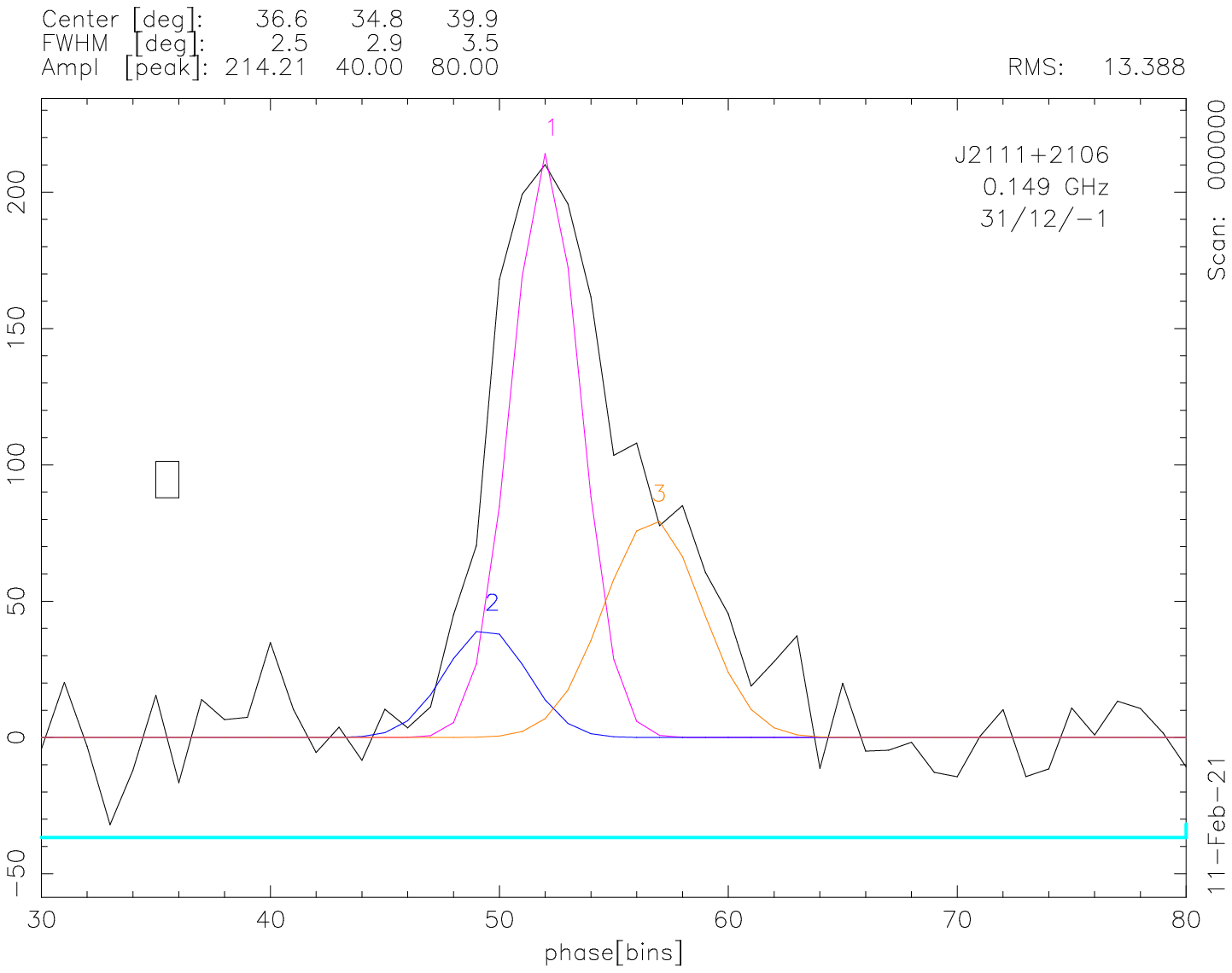}
\caption{J2111+2106: Fitted Gaussian-component model for BKK+'s 149-MHz observation.}
\label{figA119}
\end{center}
\end{figure}
\noindent\textit{\textbf{J2111+2106}}: This very slow pulsar seems to have three conflated {\bf T} components in the AO profiles; see Fig.~\ref{figA119}.  It is surprising to see a core feature in such a slow pulsar, but that the roughly 2\degr\ widths that can estimated from the L- and P-band profiles implies an $\alpha$ value of about 40\degr, and this in turn gives reasonable inner cone dimensions for the outriders.  The LOFAR profile then seems to have a conflated double structure with the leading component much brighter than the trailing one, and probably broadened by some scattering.
\vskip 0.1in

\begin{figure}
\begin{center}
\includegraphics[width=65mm,angle=-90.]{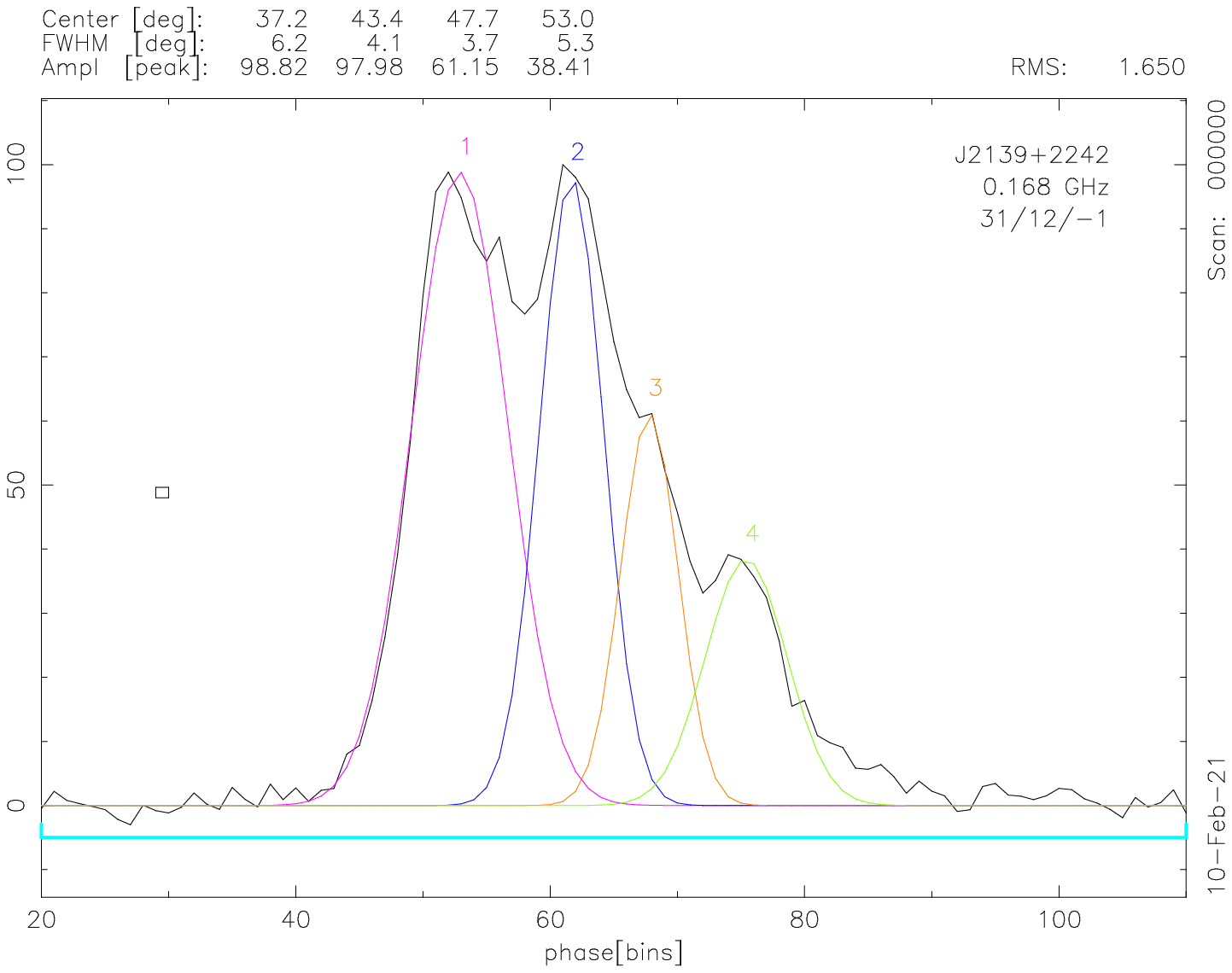}
\includegraphics[width=65mm,angle=-90.]{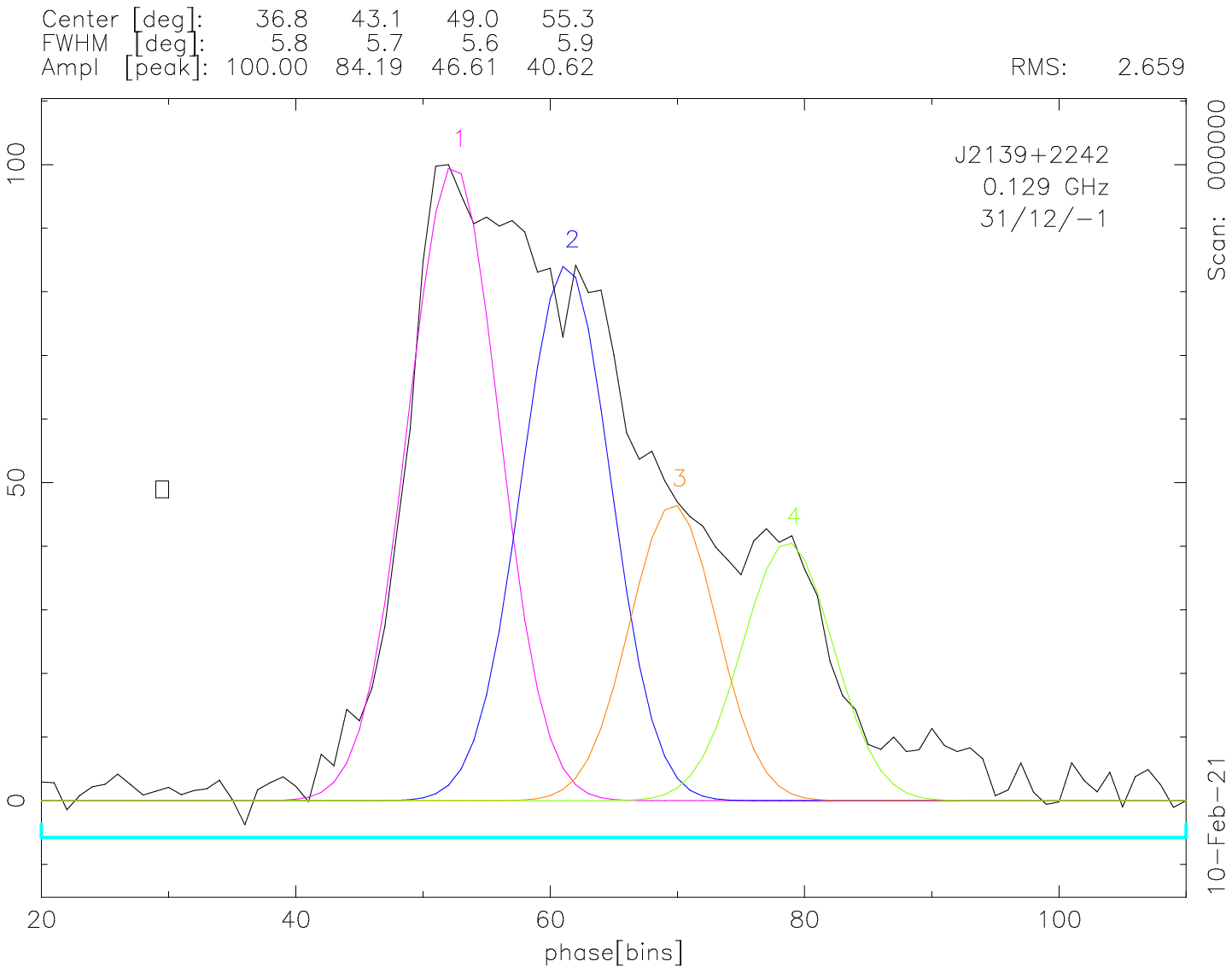}
\caption{J2139+2242: Fitted Gaussian-component model for BKK+'s 149-MHz observation.}
\label{figA120}
\end{center}
\end{figure}
\noindent\textit{\textbf{J2139+2242}}: This pulsar is a fine example of a {c\bf Q} profile. Four components are clearly delineated at LOFAR frequencies, with the trailing ones weaker than the leading as is usual for the class; see Fig.~\ref{figA120}.  The pulsar is also a bright drifter which is compatible with the classification.  The inner cone widths are estimates in each band.  Scattering may be a factor in the LOFAR profiles.  The pulsar is detected at 111 MHz \citep{prao} but no measurements are possible.
\vskip 0.1in

\noindent\textit{\textbf{J2151+2315}} seems to have an inner conal single {\bf S$_d$} geometry, and the scattering seems severe enough that the LOFAR profile was washed out.  
\vskip 0.1in

\noindent\textit{\textbf{J2155+2813}}: A single component is seen at all frequencies with some asymmetry as is often found in conal singles profiles.  We thus model it with inner cone {\bf S$_d$} geometry, and see that scattering broadens the LOFAR profiles little if any.  The pulsar is detected at 111 MHz \citep{prao} but no measurements are possible.
\vskip 0.1in

\noindent\textit{\textbf{J2156+2618}}: A single profile is seen at all three frequencies in this pulsar, though the 327-MHz profile may have some structure (due perhaps to drifing or moding).  We model it with an inner conal single {\bf S$_d$} geometry. The width increase of the LOFAR profile may be due to scattering.
\vskip 0.1in

\noindent\textit{\textbf{J2205+1444}}: This pulsar's profile retains a similar double form over the entire range from L-band to LOFAR frequencies.  As such it may have an inner cone despite its well resolved profiles; however, the model reflects an outer conal geometry.  The pulsar is detected at 111 MHz \citep{prao} but no measurements are possible.
\vskip 0.1in

\noindent\textit{\textbf{J2215+1538}}: At 149 MHz, the profile of this pulsar has a resolved double form with a stronger trailing component, and we see this profile structure at 430 MHz \citep{camilo_nice} with a closer spacing and at 1.4 GHz with the two components almost merged.  However, our 327-MHz observation has such a different shape that one wonders if it can be reconciled with the others.  Overall, this is a usual conal single evolution.  The steep PPA rate at P-band strongly suggests an OPM transition, and if so $R$ would be much flatter.   Were $R$ about --4\degr/\degr\ $\alpha$ would be about 30\degr, $\beta/\rho$ grazing at L-band, and perhaps enough less at lower frequencies to account for the increasing resolution of the components. Scattering seems minimal in the LOFAR profile.  The pulsar is detected at 111 MHz \citep{prao} but no measurements are possible.  The pulsar may have an interpulse.  
\vskip 0.1in

\noindent\textit{\textbf{J2222+2923}}: PSR J2222+2923's P-band profile suggests a {\bf T} structure with a weak trailing component and perhaps a weak core.  An inner cone model seems to obtain, with an $\alpha$ of 18\degr.  If the core had a width of about 15\degr, as seems quite plausible, it would give this $\alpha$ value.  Scattering seems to be negligible at 150 MHz.
\vskip 0.1in

\noindent\textit{\textbf{J2234+2114}}: This pulsar's strange profile is challenging; however, there is evidence of double conal structure in the P-band profile, and we have modeled the profile successfully on this assumption.  Only the inner cone, or a part of it, is discernible in the LOFAR profile.  The pulsar is detected at 111 MHz \citep{prao} but no measurements are possible.  Indeed, emission is detected with the LWA \citep{KTSD23} at 79, 64 and 50 MHz with a 6-8\degr\ width.  Scattering seems minimal in these observations.
\vskip 0.1in

\noindent\textit{\textbf{J2243+1516}}: A fairly usual {\bf S$_d$} profile is seen in this pulsar, and its evolution indicates an outer cone.  The double structure seems discernible in the LOFAR profile, but scattering contributes to its much larger width.
\vskip 0.1in

\noindent\textit{\textbf{J2248--0101}}: More work is needed to provide a suitable model for this pulsar.  The evolution does not support a core model, nor does a fluctuation spectrum indicate conal modulation.  And the profiles are too narrow assuming a --8\degr/\degr\ PPA rate to model as conal.
\vskip 0.1in

\noindent\textit{\textbf{J2253+1516}}: Over the spectrum, this pulsar's profile has two well resolved components, the leading stronger than the latter.  We model it with an inner cone {\bf D} geometry.  Scattering at the 10X average level does no seem to affect the LOFAR profile.  A double form is detected at 111 MHz \citep{prao} but no measurements are possible.
\vskip 0.1in

\noindent\textit{\textbf{J2307+2225}}: The rough single profile of this pulsar and its flat PPA traverse (also seen in \citet{han2009} at 774 MHz) make this pulsar's geometry difficult to model.  It seems to have two conal components with the second one weak over the entire band.  However, the LOFAR profile's poor S/N makes identification of the trailing component difficult. This double interpretation---together with assuming a central sightline traverse---provides a means of computing a model.  Scattering, even at the 10X average level, does not affect the LOFAR profile.  The pulsar is detected at 20/25 MHz \citep{ZVKU13}, but no measurements are possible.
\vskip 0.1in

\noindent\textit{\textbf{J2355+2246}}: The pulsar seems to have a steep spectrum and was detected in only a single 86-MHz band at 1420 MHz, showing a single form and a hint of a shallow PPA rate.  The 327-MHz profile is more complex with perhaps two conflated features---and thus similar to many conal single profiles.  While the PPA rate is poorly determined, we model the pulsar using an inner cone {\bf S$_d$} geometry.  

\onecolumn
\setlength{\tabcolsep}{3pt}
\begin{longtable}{lc|ccc|ccccc|ccccc|cccc}
\caption{Core/Double Cone Emission-Beam Model Information\\For each pulsar, its profile class is given together with the magnetic colatitude, PPA rate and sightline impact angle.  The core and inner/outer conal widths and beam radii are then given at 1.4 GHz, 327 MHz and the lowest measurable frequency around or below 100 MHz.}   \label{tabA3}  \\
    \toprule
    Pulsar &  Class & $\alpha$ & $R$ & $\beta$ & $W_c $ & $W_i$ & $\rho_i$ & $W_o$  & $\rho_o$ & $W_c $ & $W_i$ & $\rho_i$ & $W_o$  & $\rho_o$ & $W_i$ & $\rho_i$ & $W_o$  & $\rho_o$  \\
          &   & (\degr) & (\degr/\degr) & (\degr) & (\degr) & (\degr) & (\degr) & (\degr) & (\degr) & (\degr) & (\degr) & (\degr) & (\degr) & (\degr) & (\degr) & (\degr) & (\degr) & (\degr) \\
    \midrule
    & & \multicolumn{3}{c|}{(1-GHz  Geometry)} & \multicolumn{5}{c|}{(1.4-GHz Beam Sizes)} & \multicolumn{5}{c|}{(327-MHz Beam Sizes)} & \multicolumn{4}{c}{($\lesssim$100-MHz Beam Sizes)} \\
    \midrule
    \midrule
\endfirsthead    \toprule
    Pulsar &  Class & $\alpha$ & $R$ & $\beta$ & $W_c $ & $W_i$ & $\rho_i$ & $W_o$  & $\rho_o$ & $W_c $ & $W_i$ & $\rho_i$ & $W_o$  & $\rho_o$ & $W_i$ & $\rho_i$ & $W_o$  & $\rho_o$  \\
          &   & (\degr) & (\degr/\degr) & (\degr) & (\degr) & (\degr) & (\degr) & (\degr) & (\degr) & (\degr) & (\degr) & (\degr) & (\degr) & (\degr) & (\degr) & (\degr) & (\degr) & (\degr) \\
    \midrule
    & & \multicolumn{3}{c|}{(1-GHz  Geometry)} & \multicolumn{5}{c|}{(1.4-GHz Beam Sizes)} & \multicolumn{5}{c|}{(327-MHz Beam Sizes)} & \multicolumn{4}{c}{($\lesssim$100-MHz Beam Sizes)} \\
    \midrule
    \midrule
\endhead
    J0006+1834 & D? & 7 & -1 & -7.0 &  --- &  --- &  --- & $\sim$45 & 7.0 &  --- &  --- &  --- & $\sim$180 & 7.0 &  --- &  --- & $\sim$180 & 7.0 \\
J0030+0451 & T?? & 55 & 1 & 55.0 & 20.0 & 47.7 & 59.5 &  --- &  --- &  --- & 0.0 &  --- &  --- &  --- & 88.0 & 69.1 &  --- &  --- \\
J0051+0423 & cT & 23 & +3.2 & +7.1 &  --- &  --- &  --- & $\sim$29 & 9.7 &  --- & 4.7 & 7.2 & 33.2 & 10.3 & 6.8 & 7.3 & $\sim$47 & 12.7 \\
J0122+1416 & Sd? & 35 & -10 & +3.3 &  --- & $\sim$6 & 3.7 &  --- &  --- &  --- & 6.0 & 3.7 &  --- &  --- & $\sim$17 & 6.0 &  --- &  --- \\
J0137+1654 & cQ?? & 18 & +3.6 & +4.8 &  --- &  --- &  --- & $\sim$45 & 9.0 &  --- &  --- &  --- & 60.4 & 11.2 & 15.6 &  --- &  --- &  --- \\
\\ 
J0139+3336 & D? & 23 & -6 & +3.7 &  --- & 4.8 & 3.9 &  --- &  --- &  --- & $\sim$5 & 3.9 & $\sim$13 &  --- &  --- &  --- &  --- &  --- \\
J0152+0948 & D & 42 & $\infty$ & 0.0 &  --- & 10.4 &  --- & $\sim$10 & 3.5 &  --- & 13.4 &  --- & 13.4 & 4.5 &  --- &  --- & 17.3 & 5.8 \\
\end{longtable}

\begin{figure*}
\begin{center}
\begin{tabular}{@{}ll@{}ll@{}}
{\mbox{\includegraphics[width=52mm,height=53mm]{J0006+1834_Bmodel.ps}}}&
{\mbox{\includegraphics[width=52mm,height=53mm]{J0030+0451_Bmodel.ps}}}& \ \ \ 
{\mbox{\includegraphics[width=52mm,height=53mm]{J0051+0423_Bmodel.ps}}}\\
{\mbox{\includegraphics[width=52mm]{J0006+1834P.ps}}}& 
&
{\mbox{\includegraphics[width=52mm]{J0051+0423P.ps}}}\\ 
{\mbox{\includegraphics[width=52mm]{J0006+1834L.ps}}}&
{\mbox{\includegraphics[width=52mm]{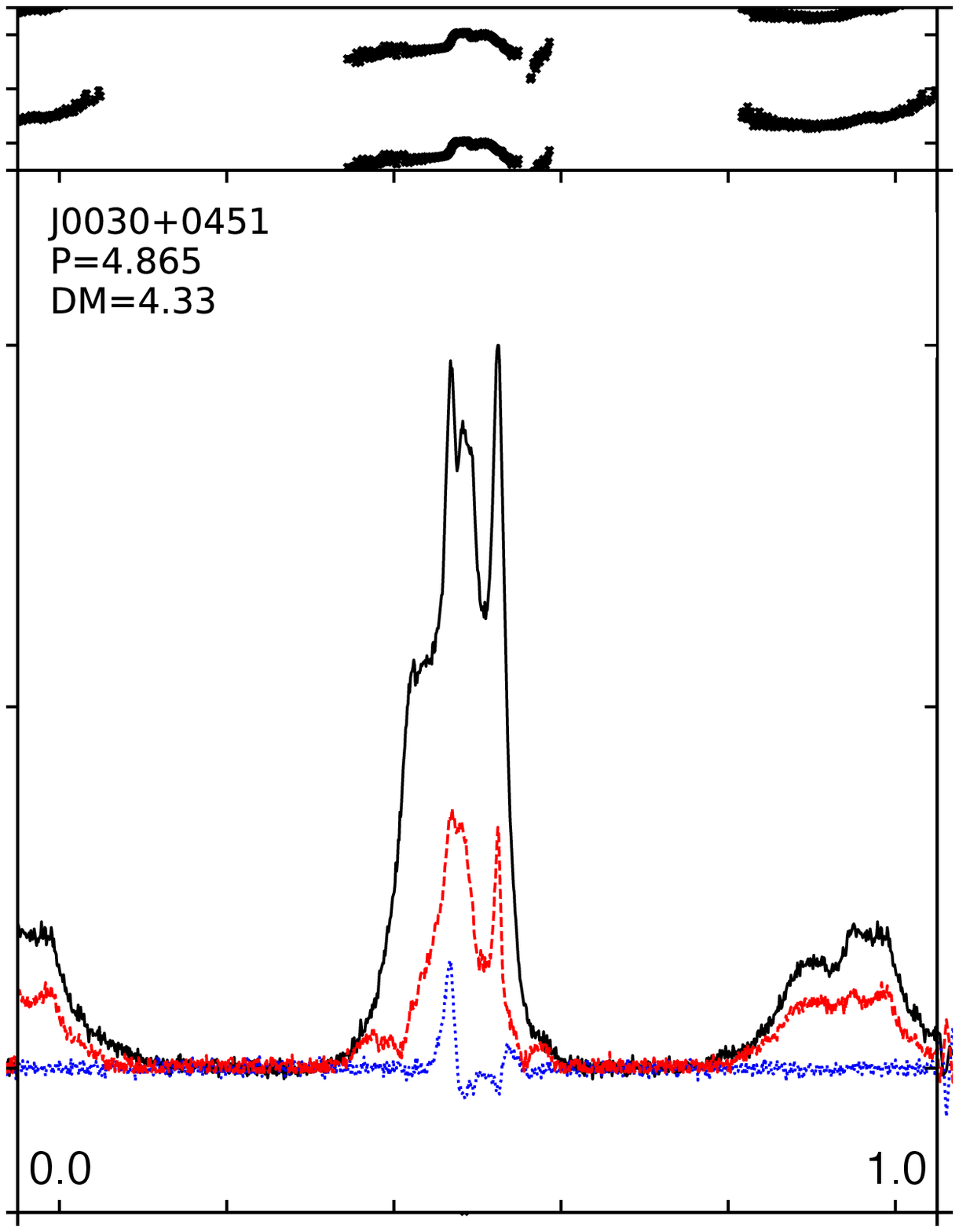}}}& \ \ \ 
{\mbox{\includegraphics[width=52mm]{J0051+0423L.ps}}}\\ 
\end{tabular}

\caption{Beam size models and average profiles for PSRs J0006+1834, J0030+0451 (1.3-GHz profile from \citet{SBM+22}) and J0051+0423. Plotted values represent the \textit{scaled} inner (blue) and outer (cyan) conal beam radii and the core angular diameter (red), respectively, and are logarithmic on both axes (see text). The nominal values of the three beam dimensions at 1 GHz are shown in each plot by a small triangle. The yellow hatching indicates an average scattering level (with an orange 10X line), and the orange hatching indicates measured scattering times (both where applicable). The top panel in each of the average profiles (second two rows) is the average profile, with the solid line (black) showing the total intensity, the dashed line (green) showing the linear polarization, and dotted (red) line showing the circular polarization. The bottom panels show the polarization angle against longitude.}
\label{figA1}
\end{center}
\end{figure*}

\begin{figure*}
\begin{center}
\begin{tabular}{@{}ll@{}ll@{}}
{\mbox{\includegraphics[width=52mm]{J0122+1416_Bmodel.ps}}}&
{\mbox{\includegraphics[width=52mm]{J0137+1654_Bmodel.ps}}}& \ \ \ 
{\mbox{\includegraphics[width=52mm]{J0139+3336_Bmodel.ps}}}\\
{\mbox{\includegraphics[width=52mm]{J0122+1416P.ps}}}& 
{\mbox{\includegraphics[width=52mm]{J0137+1654P.ps}}}& \ \ \ 
{\mbox{\includegraphics[width=52mm]{J0139+3336P.ps}}}\\ 
{\mbox{\includegraphics[width=52mm]{J0122+1416L.ps}}}&
{\mbox{\includegraphics[width=52mm]{J0137+1654L.ps}}}& \ \ \ 
{\mbox{\includegraphics[width=52mm]{J0139+3336L.ps}}}\\ 
\end{tabular}
\caption{Scaled beam dimensions and average profiles for PSRs J0122+1416, J0137+1654 and J0139+3336 as in Fig.~\ref{figA1}.}
\label{figA2}
\end{center}
\end{figure*}

\begin{figure*}
\begin{center}
\begin{tabular}{@{}ll@{}ll@{}}
{\mbox{\includegraphics[width=52mm]{J0152+0948_Bmodel.ps}}}&
{\mbox{\includegraphics[width=52mm]{J0329+1654_Bmodel.ps}}}& \ \ \ 
{\mbox{\includegraphics[width=52mm]{J0337+1715_Bmodel.ps}}}\\
{\mbox{\includegraphics[width=52mm]{J0152+0948P.ps}}}& 
{\mbox{\includegraphics[width=52mm]{J0329+1654P.ps}}}& \ \ \ 
{\mbox{\includegraphics[width=52mm]{pJ0337+1715u_shifted.ps}}}\\ 
{\mbox{\includegraphics[width=52mm]{J0152+0948L.ps}}}&
{\mbox{\includegraphics[width=52mm]{J0329+1654L.ps}}}& \ \ \ 
{\mbox{\includegraphics[width=52mm]{PQJ0337+1715.56760la.ps}}}\\
\end{tabular}
\caption{Scaled beam dimensions and average profiles for PSRs J0152+0948, J0329+1654 and J0337+1715 as in Fig.~\ref{figA1}.}
\label{figA3}
\end{center}
\end{figure*}

\begin{figure*}
\begin{center}
\begin{tabular}{@{}ll@{}ll@{}}
{\mbox{\includegraphics[width=52mm]{J0348+0432_Bmodel.ps}}}&
{\mbox{\includegraphics[width=52mm]{J0417+35_Bmodel.ps}}}& \ \ \ 
{\mbox{\includegraphics[width=52mm]{J0435+2749_Bmodel.ps}}}\\
{\mbox{\includegraphics[width=52mm]{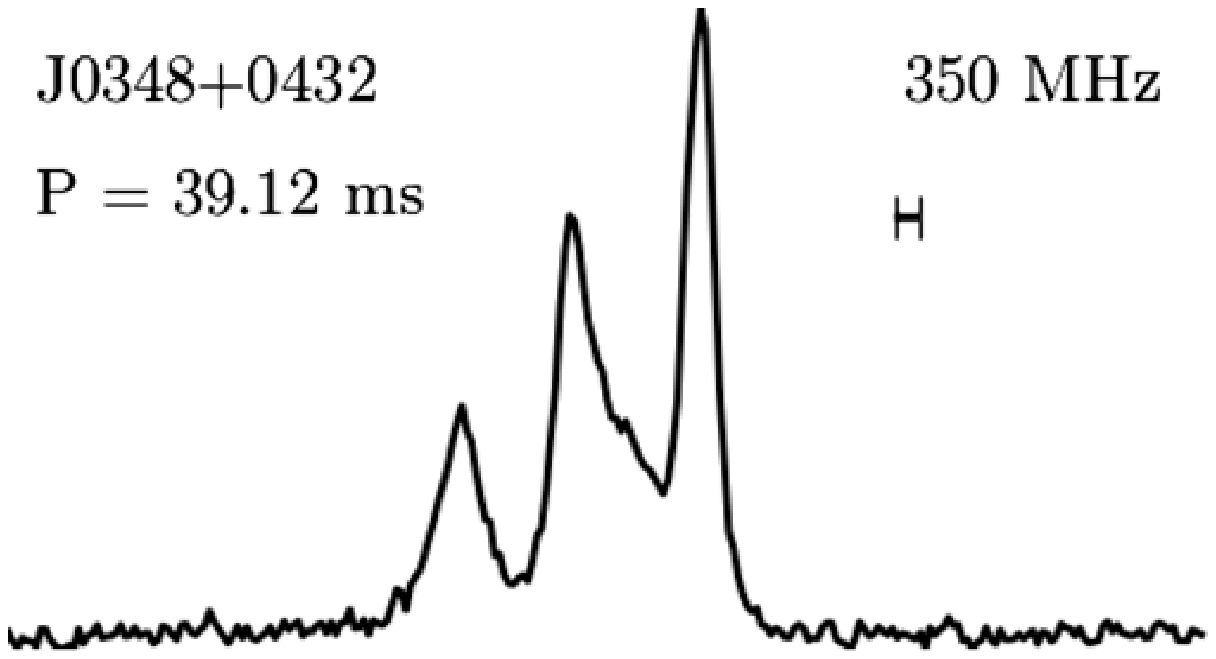}}}& 
{\mbox{\includegraphics[width=52mm]{J0417+35P.ps}}}& \ \ \ 
{\mbox{\includegraphics[width=52mm]{J0435+2749P.ps}}}\\ 
{\mbox{\includegraphics[width=52mm]{pcJ0348+0432.57131la.ps}}}&
{\mbox{\includegraphics[width=52mm]{J0417+35L.ps}}}& \ \ \ 
{\mbox{\includegraphics[width=52mm]{J0435+2749L.ps}}}\\ 
\end{tabular}
\caption{Scaled beam dimensions and average profiles for PSRs J0348+0432 [350-MHz profile from \citet{lynch2013}], J0417+35 and J0435+2749 as in Fig.~\ref{figA1}.}
\label{figA4}
\end{center}
\end{figure*}

\begin{figure*}
\begin{center}
\begin{tabular}{@{}ll@{}ll@{}}
{\mbox{\includegraphics[width=52mm]{J0457+2333_Bmodel.ps}}}&
{\mbox{\includegraphics[width=52mm]{J0533+0402_Bmodel.ps}}}& \ \ \ 
{\mbox{\includegraphics[width=52mm]{J0538+2817_Bmodel.ps}}}\\
{\mbox{\includegraphics[width=52mm]{J0457+2333P.ps}}}& 
{\mbox{\includegraphics[width=52mm]{J0533+0402P.ps}}}& \ \ \ 
{\mbox{\includegraphics[width=52mm]{J0538+2817P.ps}}}\\ 
{\mbox{\includegraphics[width=52mm]{J0457+2333L.ps}}}&
{\mbox{\includegraphics[width=52mm]{J0533+0402L.ps}}}& \ \ \ 
{\mbox{\includegraphics[width=52mm]{J0538+2817L.ps}}}\\ 
\end{tabular}
\caption{Scaled beam dimensions and average profiles for PSRs J0457+2333, J0533+0402 and J0538+2817 as in Fig.~\ref{figA1}.}
\label{figA5}
\end{center}
\end{figure*}

\begin{figure*}
\begin{center}
\begin{tabular}{@{}ll@{}ll@{}}
{\mbox{\includegraphics[width=52mm]{J0540+3207_Bmodel.ps}}}&
{\mbox{\includegraphics[width=52mm]{J0546+2441_Bmodel.ps}}}& \ \ \ 
{\mbox{\includegraphics[width=52mm]{J0609+2130_Bmodel.ps}}}\\
{\mbox{\includegraphics[width=52mm]{J0540+3207P.ps}}}& 
{\mbox{\includegraphics[width=52mm]{J0546+2441P.ps}}}& \ \ \ 
{\mbox{\includegraphics[width=52mm]{pJ0609+2130.55104pa+.ps}}}\\ 
{\mbox{\includegraphics[width=52mm]{J0540+3207L.ps}}}&
{\mbox{\includegraphics[width=52mm]{J0546+2441L.ps}}}& \ \ \ 
{\mbox{\includegraphics[width=52mm]{PQJ0609+2130.57130la.ps}}}\\ 
\end{tabular}
\caption{Scaled beam dimensions and average profiles for PSRs J0540+3207, J0546+2441 and J0609+2130 as in Fig.~\ref{figA1}.}
\label{figA6}
\end{center}
\end{figure*}

\begin{figure*}
\begin{center}
\begin{tabular}{@{}ll@{}ll@{}}
{\mbox{\includegraphics[width=52mm]{J0611+30_Bmodel.ps}}}&
{\mbox{\includegraphics[width=52mm]{J0613+3731_Bmodel.ps}}}& \ \ \ 
{\mbox{\includegraphics[width=52mm]{J0621+0336_Bmodel.ps}}}\\
{\mbox{\includegraphics[width=52mm]{J0611+30P.ps}}}&
{\mbox{\includegraphics[width=52mm]{pJ0613+3731.57123pa.ps}}}& \ \ \ 
{\mbox{\includegraphics[width=52mm]{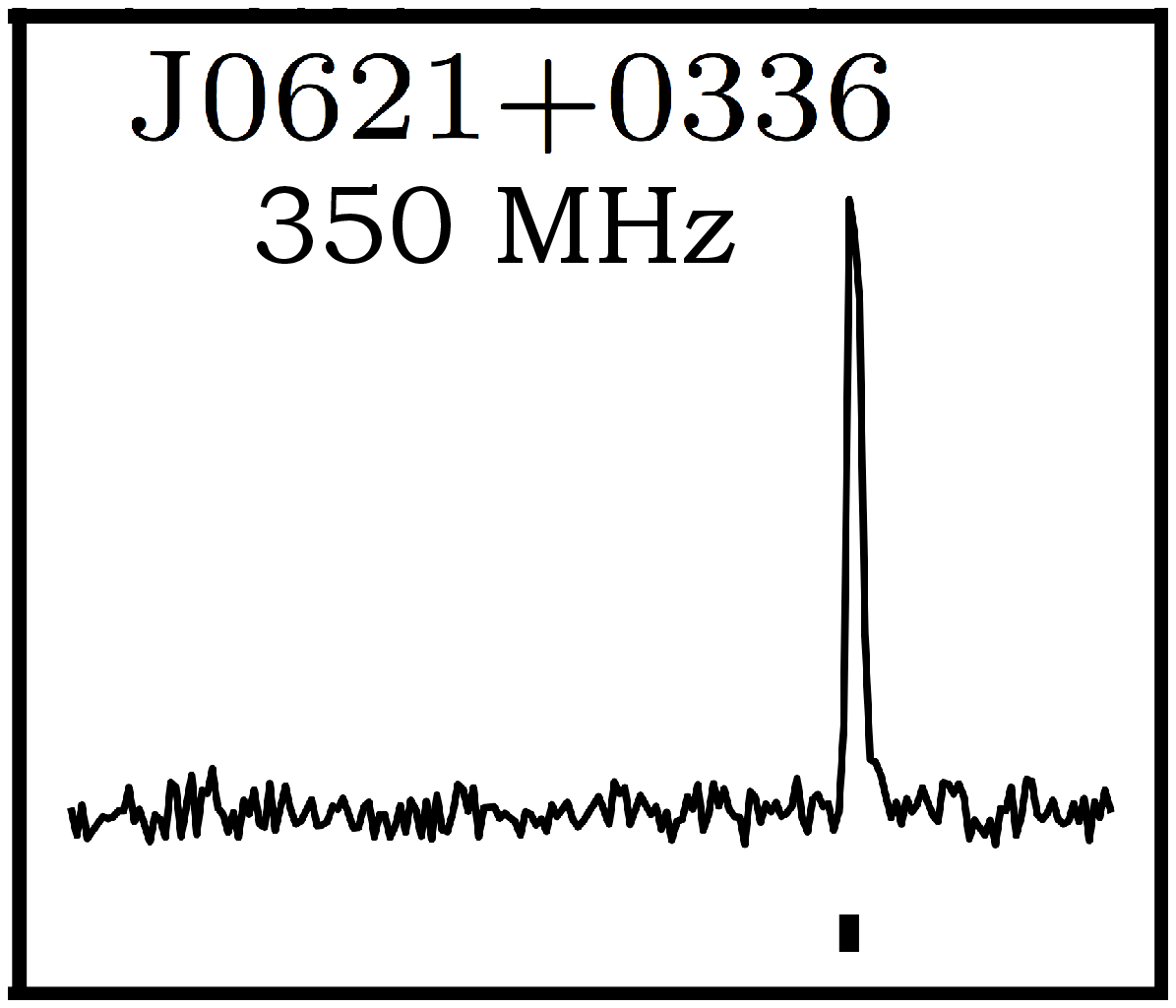}}}\\
{\mbox{\includegraphics[width=52mm]{J0611+30L.ps}}}& 
&
{\mbox{\includegraphics[width=52mm]{PQJ0621+0336.57004la.ps}}}\\
\end{tabular}
\caption{Scaled beam dimensions and average profiles for PSRs J0611+30 (1420-MHz CORRECTED), J0613+3731 and J0621+0336 [350-MHz profile from \citet{mcewen}] as in Fig.~\ref{figA1}.}
\label{figA6a}
\end{center}
\end{figure*}

\begin{figure*}
\begin{center}
\begin{tabular}{@{}ll@{}ll@{}}
{\mbox{\includegraphics[width=52mm]{J0621+1002_Bmodel.ps}}}&
&
{\mbox{\includegraphics[width=52mm]{J0625+1015_Bmodel.ps}}}\\
{\mbox{\includegraphics[width=52mm]{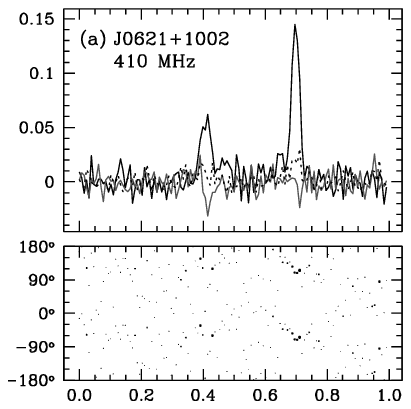}}}&
&
{\mbox{\includegraphics[width=52mm]{J0625+1015P.ps}}}\\
{\mbox{\includegraphics[width=52mm]{PQJ0621+1002.56572la.ps}}}& 
{\mbox{\includegraphics[width=52mm]{PQJ0623+0340.57005la.ps}}}&
{\mbox{\includegraphics[width=52mm]{J0625+1015L.ps}}}\\
\end{tabular}
\caption{Scaled beam dimensions and average profiles for PSRs J0621+1002 [410-MHz profile from \citet{stairs99}], J0623+0340 and J0625+1015 as in Fig.~\ref{figA1}.}
\label{figA7}
\end{center}
\end{figure*}

\begin{figure*}
\begin{center}
\begin{tabular}{@{}ll@{}ll@{}}
{\mbox{\includegraphics[width=52mm]{J0627+0649_Bmodel.ps}}}&
&
{\mbox{\includegraphics[width=52mm]{J0630-0046_Bmodel.ps}}}\\ 
{\mbox{\includegraphics[width=52mm]{J0627+0649P.ps}}}&
&
{\mbox{\includegraphics[width=52mm]{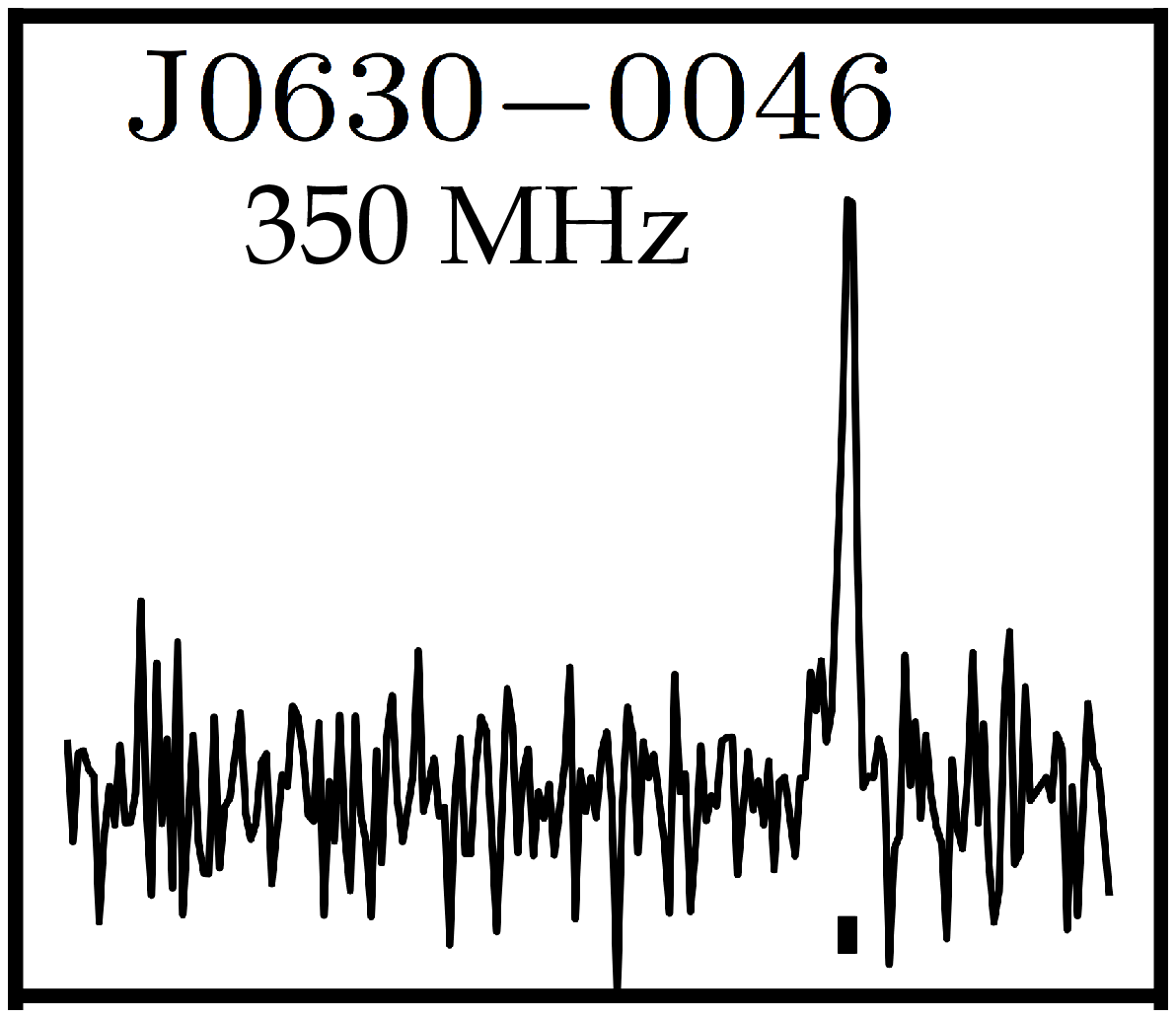}}}\\ 
{\mbox{\includegraphics[width=51mm]{J0627+0649L.ps}}}&
{\mbox{\includegraphics[width=52mm]{PQJ0628+0909.57131la.ps}}}& \ \ \ 
{\mbox{\includegraphics[width=51mm]{PQJ0630-0046.56932la.ps}}} \\ 
\end{tabular}
\caption{Scaled beam dimensions and average profiles for PSRs J0627+0649, J0628+0909 and J0630--0046 [350-MHz profile from \citet{mcewen}] as in Fig.~\ref{figA1}.}
\label{figA8}
\end{center}
\end{figure*}

\begin{figure*}
\begin{center}
\begin{tabular}{@{}ll@{}ll@{}}
&
&
\\
&
& \\
{\mbox{\includegraphics[width=51mm]{PQJ0646+0905.57004la.ps}}} &
{\mbox{\includegraphics[width=51mm]{PQJ0647+0913.57005la.ps}}}& \ \ \ 
{\mbox{\includegraphics[width=51mm]{PQJ0658+0022.57257la.ps}}} \\
\end{tabular}
\caption{Average profiles for PSRs J0646+0905, J0647+0913 and J0658+0022 as in Fig.~\ref{figA1}.}
\label{figA8a}
\end{center}
\end{figure*}

\begin{figure*}
\begin{center}
\begin{tabular}{@{}ll@{}ll@{}}
{\mbox{\includegraphics[width=52mm]{J0711+0931_Bmodel.ps}}}&
&
{\mbox{\includegraphics[width=52mm]{J0815+0939_Bmodel.ps}}}\\
{\mbox{\includegraphics[width=52mm]{J0711+0931P.ps}}}& 
&
{\mbox{\includegraphics[width=52mm]{J0815+0939P.ps}}}\\ 
{\mbox{\includegraphics[width=52mm]{J0711+0931L.ps}}}&
{\mbox{\includegraphics[width=52mm]{PQJ0806+08.57200la.ps}}}& \ \ \ 
{\mbox{\includegraphics[width=52mm]{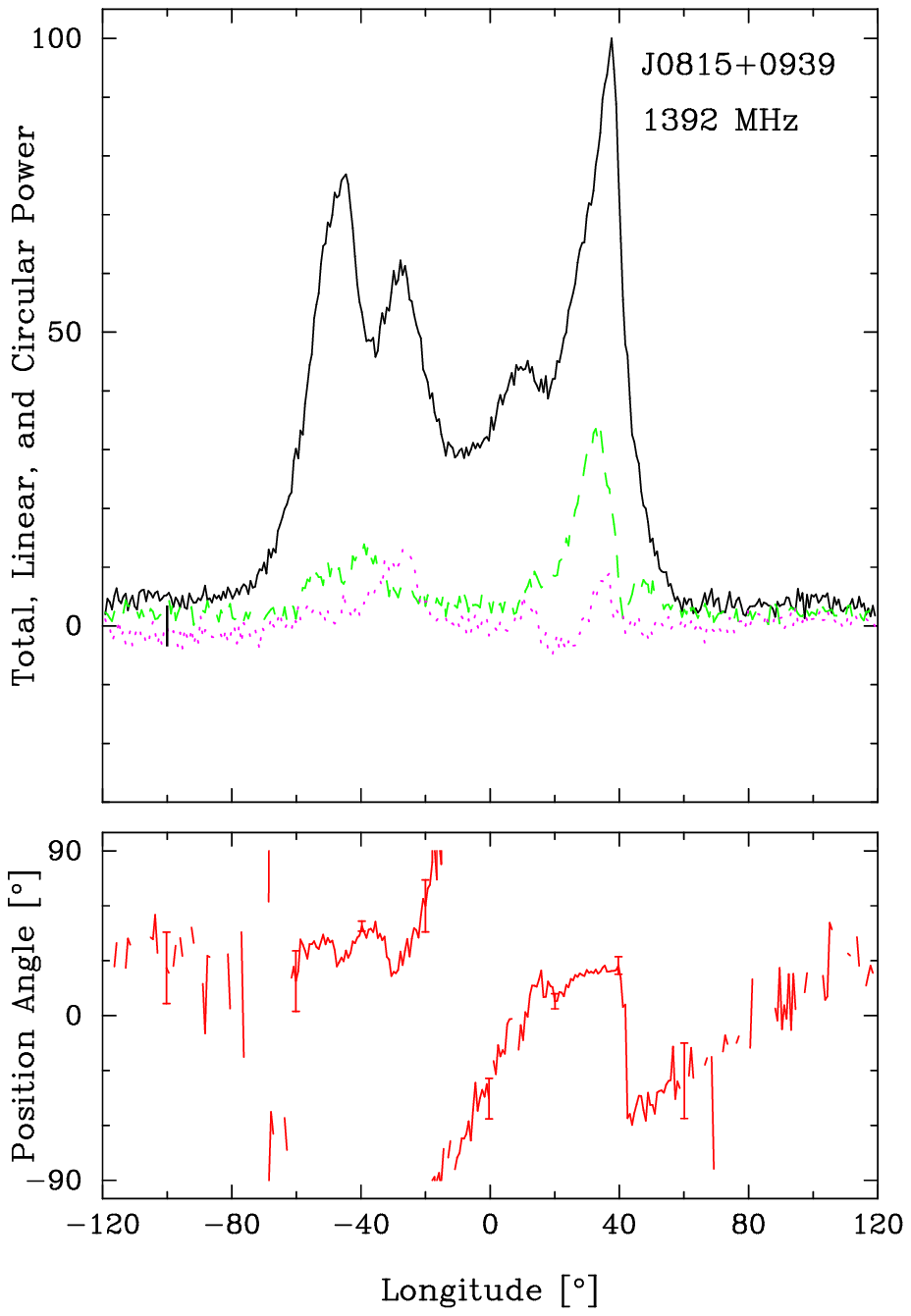}}}\\ 
\end{tabular}
\caption{Scaled beam dimensions and average profiles for PSRs J0711+0931, J0806+08 and J0815+0939 as in Fig.~\ref{figA1}.}
\label{figA9}
\end{center}
\end{figure*}

\begin{figure*}
\begin{center}
\begin{tabular}{@{}ll@{}ll@{}}
{\mbox{\includegraphics[width=52mm]{J0843+0719_Bmodel.ps}}}&
{\mbox{\includegraphics[width=52mm]{J0848+1640_Bmodel.ps}}}& \ \ \ 
{\mbox{\includegraphics[width=52mm]{J0928+0614_Bmodel.ps}}}\\
{\mbox{\includegraphics[width=52mm]{pJ0843+0719.56964pa.ps}}}& 
{\mbox{\includegraphics[width=52mm]{J0848+1640P.ps}}}& \ \ \ 
{\mbox{\includegraphics[width=52mm]{J0928+0614P.ps}}}\\ 
{\mbox{\includegraphics[width=52mm]{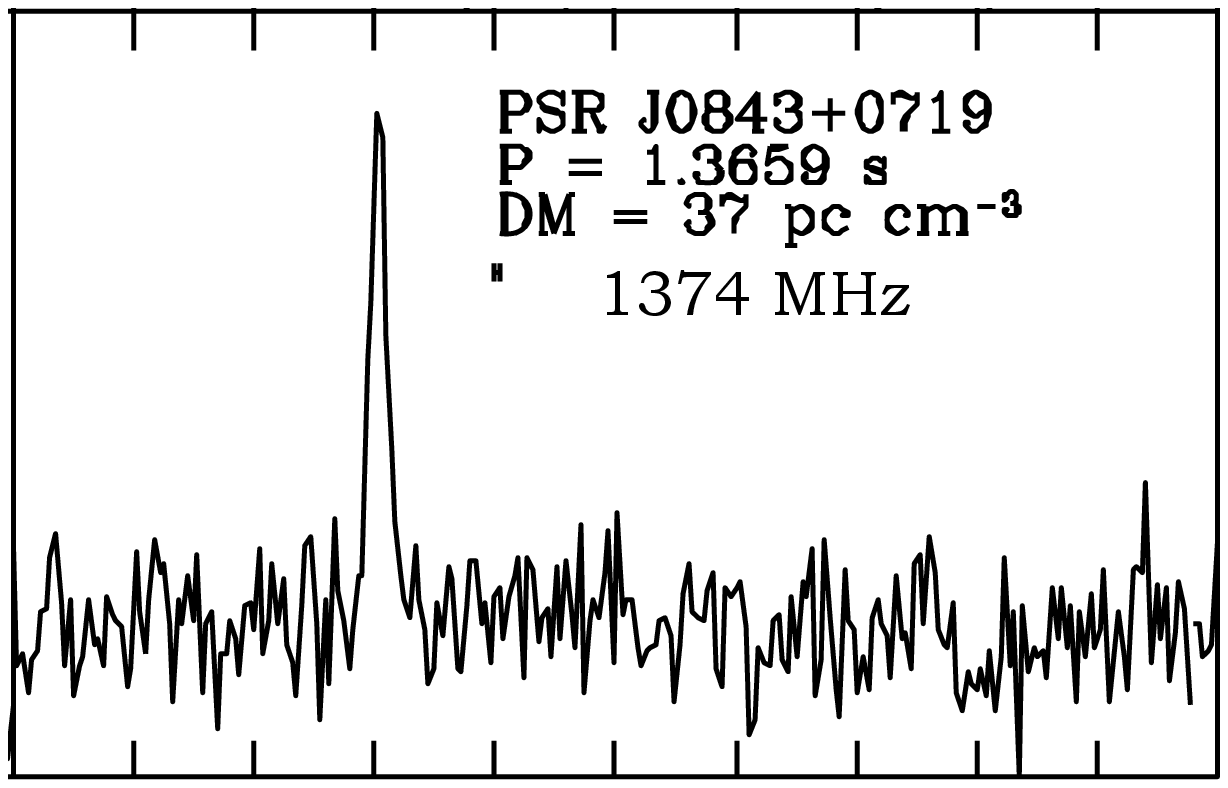}}}&
{\mbox{\includegraphics[width=52mm]{J0848+1640L.ps}}}& \ \ \ 
{\mbox{\includegraphics[width=52mm]{J0928+0614L.ps}}}\\ 
\end{tabular}
\caption{Scaled beam dimensions and average profiles for PSRs J0843+0719 [full period 1.4-GHz total power profile from \citet{Burgay+06}], J0848+1640 and J0928+0614 as in Fig.~\ref{figA1}.}
\label{figA9a}
\end{center}
\end{figure*}

\begin{figure*}
\begin{center}
\begin{tabular}{@{}ll@{}ll@{}}
{\mbox{\includegraphics[width=52mm]{J0943+2253_Bmodel.ps}}}&
{\mbox{\includegraphics[width=52mm]{J0947+2740_Bmodel.ps}}}& \ \ \ 
{\mbox{\includegraphics[width=52mm]{J1022+1001_Bmodel.ps}}}\\
{\mbox{\includegraphics[width=52mm]{J0943+2253P.ps}}}& 
{\mbox{\includegraphics[width=52mm]{J0947+2740P.ps}}}& \ \ \ 
{\mbox{\includegraphics[width=52mm]{J1022+1001P.ps}}}\\ 
{\mbox{\includegraphics[width=52mm]{J0943+2253L.ps}}}&
{\mbox{\includegraphics[width=52mm]{J0947+2740L.ps}}}& \ \ \ 
{\mbox{\includegraphics[width=52mm]{J1022+1001L.ps}}}\\ 
\end{tabular}
\caption{Scaled beam dimensions and average profiles for PSRs J0943+2253, J0947+2740 and J1022+1001 as in Fig.~\ref{figA1}.}
\label{figA10}
\end{center}
\end{figure*}

\begin{figure*}
\begin{center}
\begin{tabular}{@{}ll@{}ll@{}}
&
{\mbox{\includegraphics[width=52mm]{J1147+0829_Bmodel.ps}}}& \ \ \ 
{\mbox{\includegraphics[width=52mm]{J1238+2152_Bmodel.ps}}}\\
{\mbox{\includegraphics[width=52mm]{J1046+0304P.ps}}}& 
{\mbox{\includegraphics[width=52mm]{J1147+0829P.ps}}}& \ \ \ 
{\mbox{\includegraphics[width=52mm]{J1238+2152P.ps}}}\\ 
{\mbox{\includegraphics[width=52mm]{J1046+0304L.ps}}}&
{\mbox{\includegraphics[width=52mm]{J1147+0829L.ps}}}& \ \ \ 
{\mbox{\includegraphics[width=52mm]{J1238+2152L.ps}}}\\ 
\end{tabular}
\caption{Scaled beam dimensions and average profiles for PSRs J1046+0304, J1147+0829 and as J1238+2152 in Fig.~\ref{figA1}.}
\label{figA11}
\end{center}
\end{figure*}

\begin{figure*}
\begin{center}
\begin{tabular}{@{}ll@{}ll@{}}
{\mbox{\includegraphics[width=52mm]{J1246+2253_Bmodel.ps}}}&
{\mbox{\includegraphics[width=52mm]{J1313+0931_Bmodel.ps}}}& \ \ \ 
{\mbox{\includegraphics[width=52mm]{J1404+1159_Bmodel.ps}}}\\
{\mbox{\includegraphics[width=52mm]{J1246+2253P.ps}}}& 
{\mbox{\includegraphics[width=52mm]{J1313+0931P.ps}}}& \ \ \ 
{\mbox{\includegraphics[width=52mm]{pJ1404+1159.57566pa.ps}}}\\ 
{\mbox{\includegraphics[width=52mm]{J1246+2253L.ps}}}&
{\mbox{\includegraphics[width=52mm]{J1313+0931L.ps}}}& \ \ \ 
{\mbox{\includegraphics[width=52mm]{pJ1404+1159.57564l2a.ps}}}\\ 
\end{tabular}
\caption{Scaled beam dimensions and average profiles for PSRs J1246+2253, J1313+0931 and J1404+1159 as in Fig.~\ref{figA1}.}
\label{figA12}
\end{center}
\end{figure*}

\begin{figure*}
\begin{center}
\begin{tabular}{@{}ll@{}ll@{}}
{\mbox{\includegraphics[width=52mm]{J1503+2111_Bmodel.ps}}}& 
&
{\mbox{\includegraphics[width=52mm]{J1549+2113_Bmodel.ps}}}\\
{\mbox{\includegraphics[width=52mm]{J1503+2111P.ps}}}& 
&
{\mbox{\includegraphics[width=52mm]{J1549+2113P.ps}}}\\ 
{\mbox{\includegraphics[width=52mm]{J1503+2111L.ps}}}&
{\mbox{\includegraphics[width=52mm]{PQJ1538+2345.57312l3a.ps}}}& \ \ \ 
{\mbox{\includegraphics[width=52mm]{J1549+2113L.ps}}}\\
\end{tabular}
\caption{Scaled beam dimensions and average profiles for PSRs J1503+2111, J1538+2345 and J1549+2113 as in Fig.~\ref{figA1}.}
\label{figA13}
\end{center}
\end{figure*}

\begin{figure*}
\begin{center}
\begin{tabular}{@{}ll@{}ll@{}}
{\mbox{\includegraphics[width=52mm]{J1612+2008_Bmodel.ps}}}& 
{\mbox{\includegraphics[width=52mm]{J1627+1419_Bmodel.ps}}}& \ \ \
{\mbox{\includegraphics[width=52mm]{J1645+1012_Bmodel.ps}}}\\
{\mbox{\includegraphics[width=52mm]{J1612+2008P.ps}}}& 
{\mbox{\includegraphics[width=52mm]{J1627+1419P.ps}}}& \ \ \
{\mbox{\includegraphics[width=52mm]{J1645+1012P.ps}}}\\ 
{\mbox{\includegraphics[width=52mm]{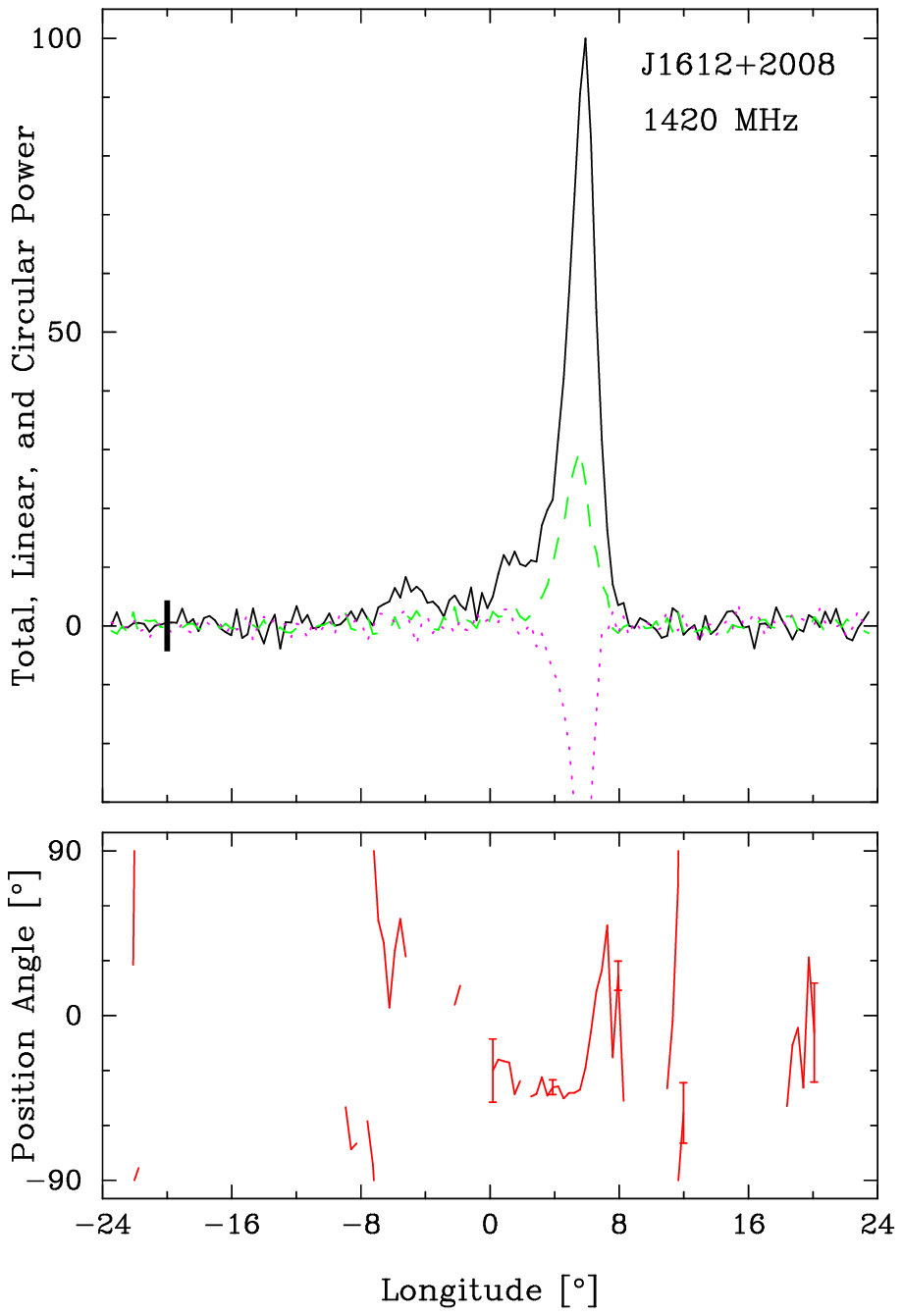}}}&
{\mbox{\includegraphics[width=52mm]{J1627+1419L.ps}}}& \ \ \ 
{\mbox{\includegraphics[width=52mm]{J1645+1012L.ps}}}\\
\end{tabular}
\caption{Scaled beam dimensions and average profiles for PSRs J1612+2008, J1627+1419 and J1645+1012 as in Fig.~\ref{figA1}.}
\label{figA14}
\end{center}
\end{figure*}

\begin{figure*}
\begin{center}
\begin{tabular}{@{}ll@{}ll@{}}
{\mbox{\includegraphics[width=52mm]{J1649+2533_Bmodel.ps}}}& 
{\mbox{\includegraphics[width=52mm]{J1652+2651_Bmodel.ps}}}& \ \ \
\\
{\mbox{\includegraphics[width=52mm]{J1649+2533P.ps}}}& 
{\mbox{\includegraphics[width=52mm]{J1652+2651P.ps}}}& \ \ \
\\
{\mbox{\includegraphics[width=52mm]{J1649+2533L.ps}}}&
{\mbox{\includegraphics[width=52mm]{J1652+2651L.ps}}}& \ \ \ 
{\mbox{\includegraphics[width=52mm]{PQJ1713+0747.56768la.ps}}}\\
\end{tabular}
\caption{Scaled beam dimensions and average profiles for J1649+2533, PSRs J1652+2651 and J1713+0747 as in Fig.~\ref{figA1}.}
\label{figA14a}
\end{center}
\end{figure*}

\begin{figure*}
\begin{center}
\begin{tabular}{@{}ll@{}ll@{}}
{\mbox{\includegraphics[width=52mm]{J1720+2150_Bmodel.ps}}}& 
{\mbox{\includegraphics[width=52mm]{J1739+0612_Bmodel.ps}}}& \ \ \
{\mbox{\includegraphics[width=52mm]{J1741+2758_Bmodel.ps}}}\\
{\mbox{\includegraphics[width=52mm]{J1720+2150P.ps}}}&
{\mbox{\includegraphics[width=52mm]{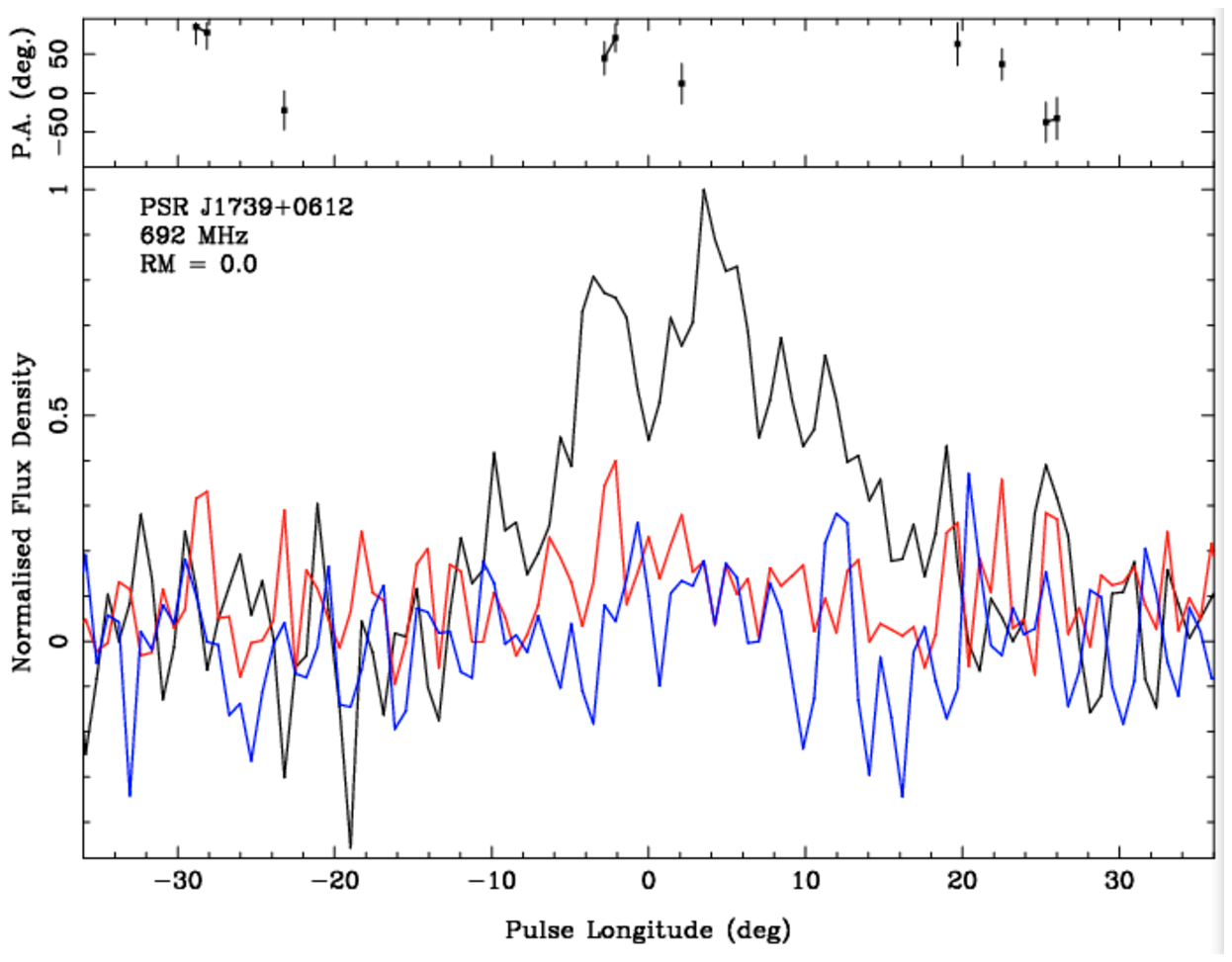}}}& \ \ \ 
{\mbox{\includegraphics[width=52mm]{J1741+2758P.ps}}}\\ 
{\mbox{\includegraphics[width=51mm]{J1720+2150L.ps}}}&
{\mbox{\includegraphics[width=52mm]{PQJ1739+0612.56605la.ps}}}& \ \ \
{\mbox{\includegraphics[width=52mm]{J1741+2758L.ps}}}\\
\end{tabular}
\caption{Scaled beam dimensions and average profiles for PSRs J1720+2150, J1739+0612 [692-MHz from \citet{ATNF}] and J1741+2758 as in Fig.~\ref{figA1}.}
\label{figA15}
\end{center}
\end{figure*}

\begin{figure*}
\begin{center}
\begin{tabular}{@{}ll@{}ll@{}}
{\mbox{\includegraphics[width=52mm]{J1746+2245_Bmodel.ps}}}&
{\mbox{\includegraphics[width=52mm]{J1746+2540_Bmodel.ps}}}& \ \ \ 
{\mbox{\includegraphics[width=52mm]{J1752+2359_Bmodel.ps}}}\\
{\mbox{\includegraphics[width=52mm]{J1746+2245P.ps}}}& 
{\mbox{\includegraphics[width=52mm]{J1746+2540P.ps}}}& \ \ \ 
{\mbox{\includegraphics[width=52mm]{J1752+2359P.ps}}}\\ 
{\mbox{\includegraphics[width=52mm]{J1746+2245L.ps}}}&
{\mbox{\includegraphics[width=52mm]{J1746+2540L.ps}}}& \ \ \ 
{\mbox{\includegraphics[width=52mm]{J1752+2359L.ps}}}\\ 
\end{tabular}
\caption{Scaled beam dimensions and average profiles for PSRs J1746+2245, J1746+2540 and J1752+2359 as in Fig.~\ref{figA1}.}
\label{figA16}
\end{center}
\end{figure*}

\begin{figure*}
\begin{center}
\begin{tabular}{@{}ll@{}ll@{}}
{\mbox{\includegraphics[width=52mm]{J1758+3030_Bmodel.ps}}}&
{\mbox{\includegraphics[width=52mm]{J1807+0756_Bmodel.ps}}}& \ \ \ 
{\mbox{\includegraphics[width=52mm]{J1811+0702_Bmodel.ps}}}\\
{\mbox{\includegraphics[width=52mm]{J1758+3030P.ps}}}& 
{\mbox{\includegraphics[width=52mm]{J1807+0756P.ps}}}& \ \ \ 
{\mbox{\includegraphics[width=52mm]{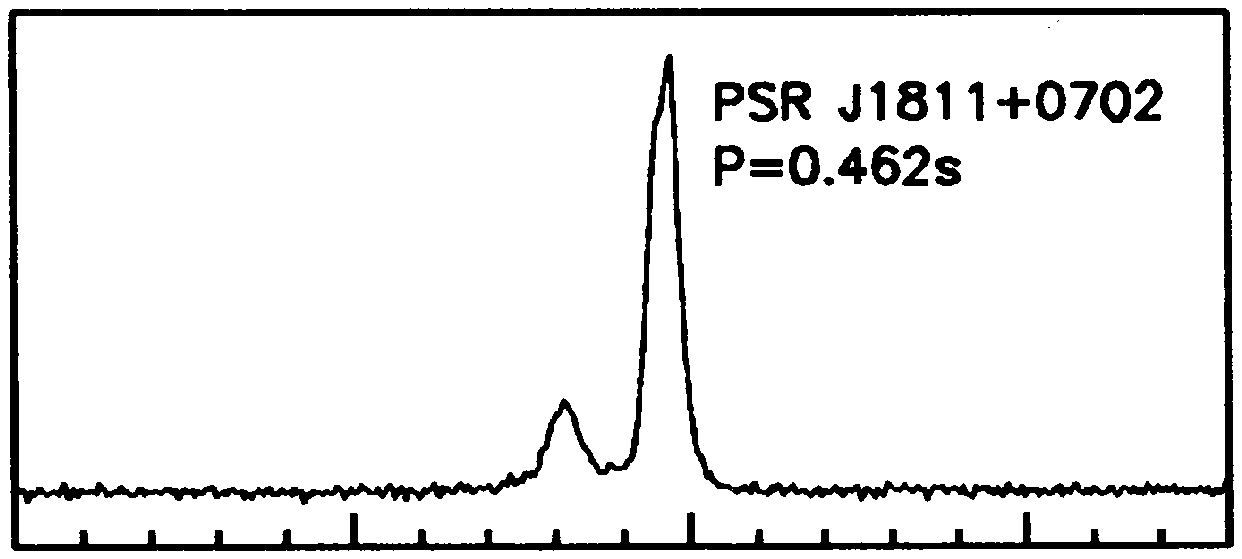}}}\\ 
{\mbox{\includegraphics[width=52mm]{J1758+3030L.ps}}}&
{\mbox{\includegraphics[width=52mm]{J1807+0756L.ps}}}& \ \ \ 
{\mbox{\includegraphics[width=52mm]{J1811+0702L.ps}}}\\ 
\end{tabular}
\caption{Scaled beam dimensions and average profiles for PSRs J1758+3030, J1807+0756 and J1811+0702 [full period, total power 430-MHz profile from \citet{foster}] as in Fig.~\ref{figA1}.}
\label{figA17}
\end{center}
\end{figure*}

\begin{figure*}
\begin{center}
\begin{tabular}{@{}ll@{}ll@{}}
{\mbox{\includegraphics[width=52mm]{J1813+1822_Bmodel.ps}}}&
{\mbox{\includegraphics[width=52mm]{J1814+1130_Bmodel.ps}}}& \ \ \ 
{\mbox{\includegraphics[width=52mm]{J1819+1305_Bmodel.ps}}}\\
{\mbox{\includegraphics[width=52mm]{J1813+1822P.ps}}}& 
{\mbox{\includegraphics[width=52mm]{J1814+1130P.ps}}}& \ \ \ 
{\mbox{\includegraphics[width=52mm]{J1819+1305P.ps}}}\\ 
{\mbox{\includegraphics[width=52mm]{J1813+1822L.ps}}}&
{\mbox{\includegraphics[width=52mm]{J1814+1130L.ps}}}& \ \ \ 
{\mbox{\includegraphics[width=52mm]{J1819+1305L.ps}}}\\ 
\end{tabular}
\caption{Scaled beam dimensions and average profiles for PSRs J1813+1822, J1814+1130 and J1819+1305 (327-MHz CORRECTED) as in Fig.~\ref{figA1}.}
\label{figA18}
\end{center}
\end{figure*}

\begin{figure*}
\begin{center}
\begin{tabular}{@{}ll@{}ll@{}}
{\mbox{\includegraphics[width=52mm]{J1821+1715_Bmodel.ps}}}&
{\mbox{\includegraphics[width=52mm]{J1822+0705_Bmodel.ps}}}& \ \ \ 
{\mbox{\includegraphics[width=52mm]{J1828+1359_Bmodel.ps}}}\\
{\mbox{\includegraphics[width=52mm]{J1821+1715P.ps}}}& 
{\mbox{\includegraphics[width=52mm]{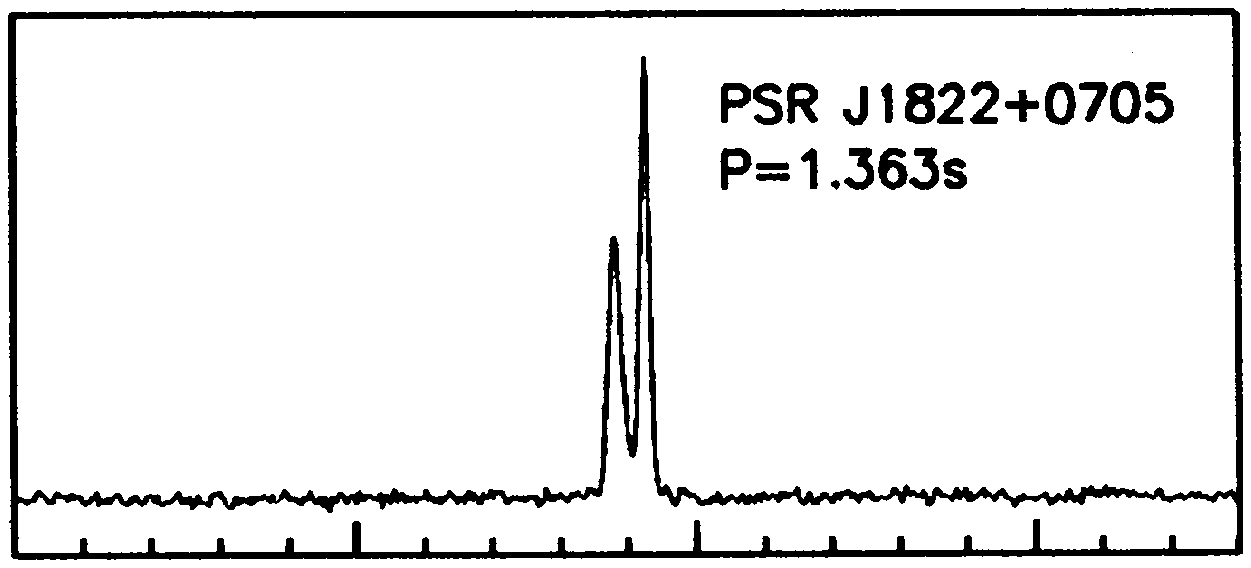}}}& \ \ \ 
{\mbox{\includegraphics[width=52mm]{J1828+1359P.ps}}}\\ 
{\mbox{\includegraphics[width=52mm]{J1821+1715L.ps}}}&
{\mbox{\includegraphics[width=52mm]{J1822+0705L.ps}}}& \ \ \ 
{\mbox{\includegraphics[width=52mm]{J1828+1359L.ps}}}\\ 
\end{tabular}
\caption{Scaled beam dimensions and average profiles for PSRs J1821+1715, J1822+0705 [full period, total power 430-MHz profile from \citet{foster}] and J1828+1359 as in Fig.~\ref{figA1}.}
\label{figA19}
\end{center}
\end{figure*}

\begin{figure*}
\begin{center}
\begin{tabular}{@{}ll@{}ll@{}}
{\mbox{\includegraphics[width=52mm]{J1829+0000_Bmodel.ps}}}&
{\mbox{\includegraphics[width=52mm]{J1837+1221_Bmodel.ps}}}& \ \ \ 
{\mbox{\includegraphics[width=52mm]{J1837-0045_Bmodel.ps}}}\\
{\mbox{\includegraphics[width=52mm]{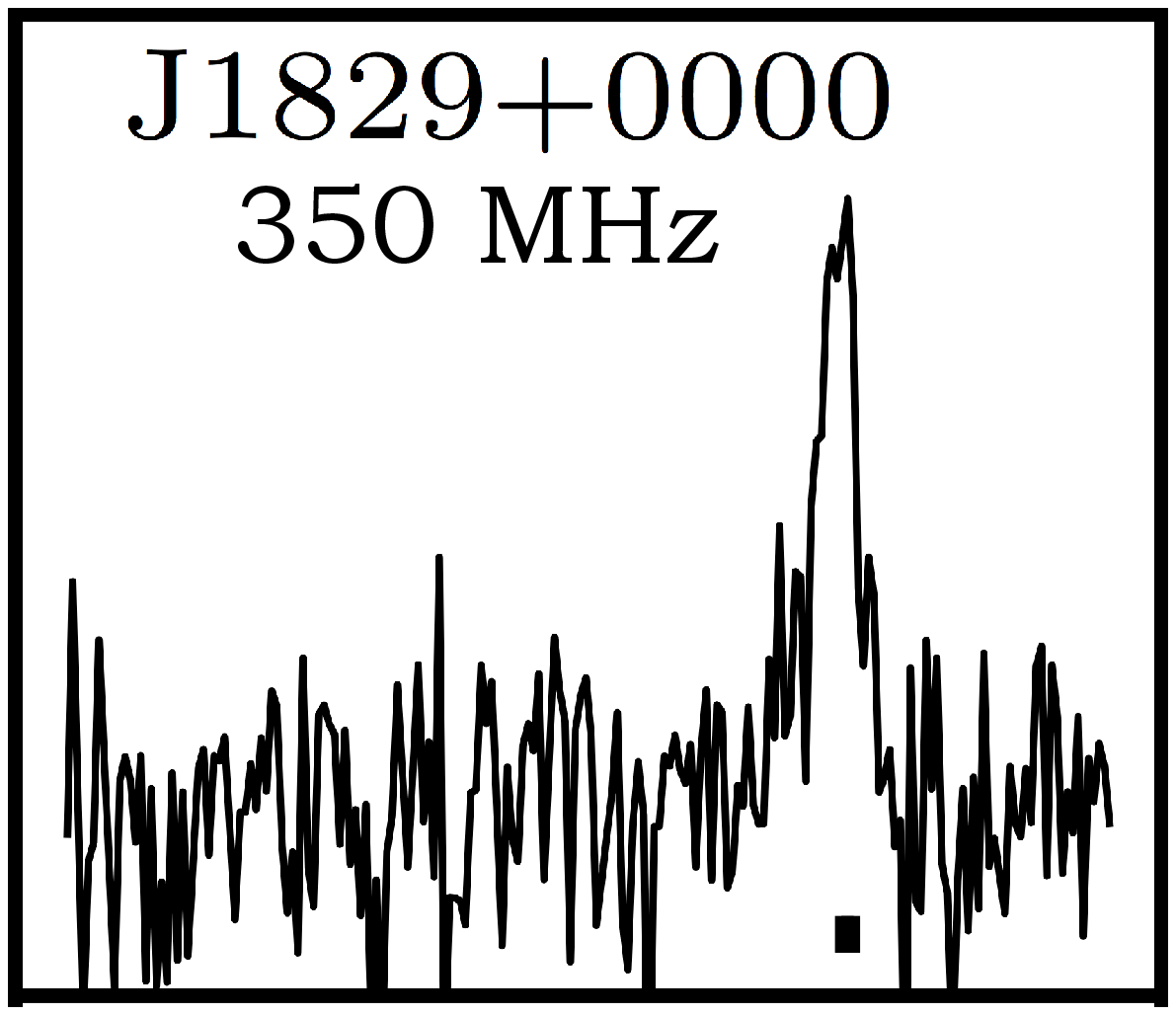}}}& 
{\mbox{\includegraphics[width=52mm]{J1837+1221P.ps}}}& \ \ \ 
{\mbox{\includegraphics[width=52mm]{J1837-0045P.ps}}}\\ 
{\mbox{\includegraphics[width=52mm]{PQJ1829+0000.56768la.ps}}}&
{\mbox{\includegraphics[width=52mm]{J1837+1221L.ps}}}& \ \ \ 
{\mbox{\includegraphics[width=52mm]{J1837-0045L.ps}}}\\ 
\end{tabular}
\caption{Scaled beam dimensions and average profiles for PSRs J1829+0000 [350-MHz profile from \citet{mcewen}], J1837+1221 and J1837--0045 as in Fig.~\ref{figA1}.}
\label{figA20}
\end{center}
\end{figure*}

\begin{figure*}
\begin{center}
\begin{tabular}{@{}ll@{}ll@{}}
{\mbox{\includegraphics[width=52mm]{J1838+1650_Bmodel.ps}}}&
{\mbox{\includegraphics[width=52mm]{J1842+0257_Bmodel.ps}}}& \ \ \ 
{\mbox{\includegraphics[width=52mm]{J1843+2024_Bmodel.ps}}}\\
{\mbox{\includegraphics[width=52mm]{J1838+1650P.ps}}}& 
{\mbox{\includegraphics[width=52mm]{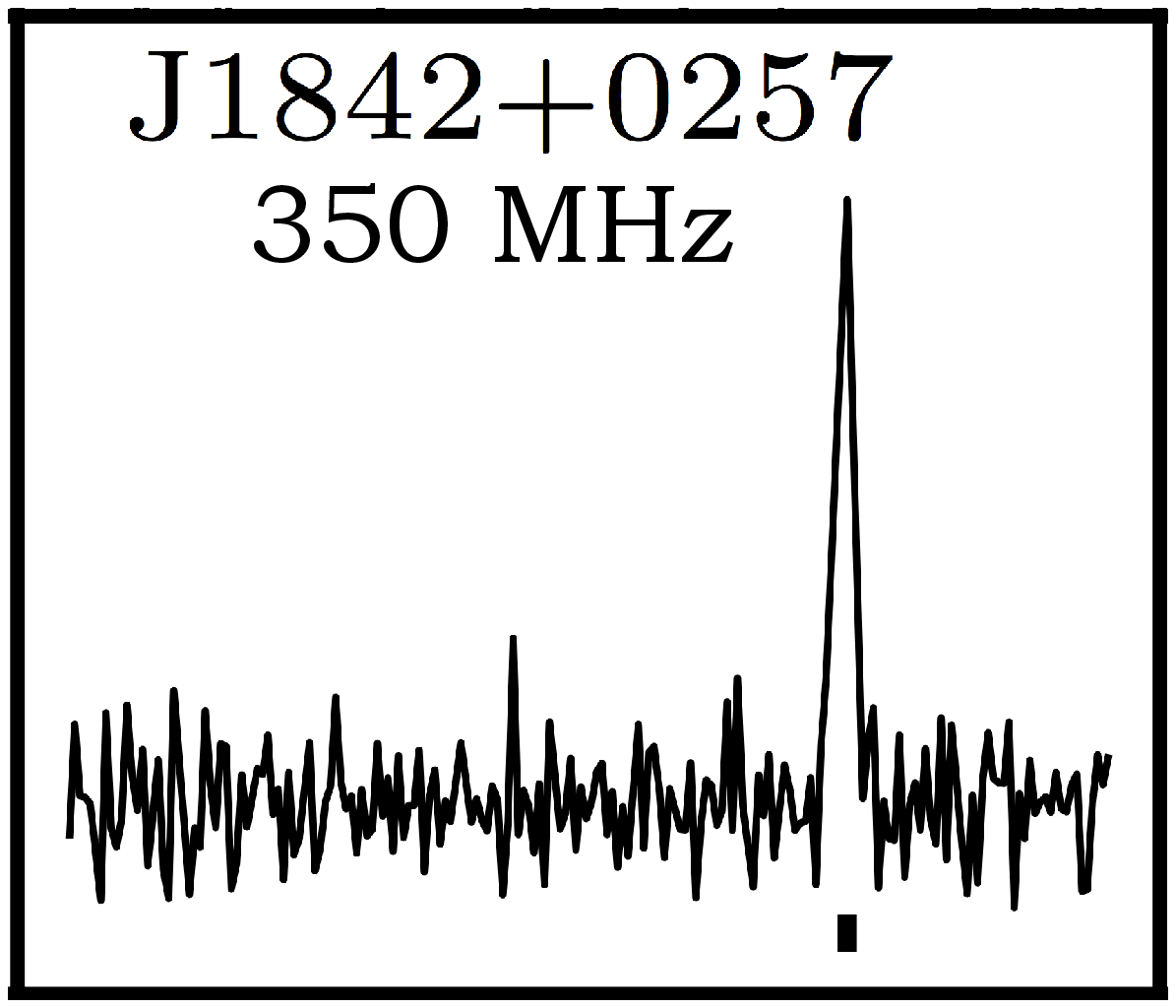}}}& \ \ \ 
{\mbox{\includegraphics[width=52mm]{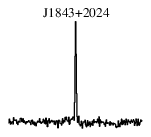}}}\\ 
{\mbox{\includegraphics[width=52mm]{J1838+1650L.ps}}}&
{\mbox{\includegraphics[width=52mm]{PQJ1842+0257.57113la.ps}}}& \ \ \ 
{\mbox{\includegraphics[width=52mm]{J1843+2024L.ps}}}\\ 
\end{tabular}
\caption{Scaled beam dimensions and average profiles for PSRs J1838+1650, J1842+0257 [350-MHz profile from \citet{mcewen}] and J1843+2024 [full period, total power profile from \citet{Champion2005}] as in Fig.~\ref{figA1}.}
\label{figA21}
\end{center}
\end{figure*}

\begin{figure*}
\begin{center}
\begin{tabular}{@{}ll@{}ll@{}}
{\mbox{\includegraphics[width=52mm]{J1848+0826_Bmodel.ps}}}&
{\mbox{\includegraphics[width=52mm]{J1849+2423_Bmodel.ps}}}& \ \ \ 
{\mbox{\includegraphics[width=52mm]{J1850+0026_Bmodel.ps}}}\\
{\mbox{\includegraphics[width=52mm]{J1848+0826P.ps}}}& 
{\mbox{\includegraphics[width=52mm]{J1849+2423P.ps}}}& \ \ \ 
{\mbox{\includegraphics[width=52mm]{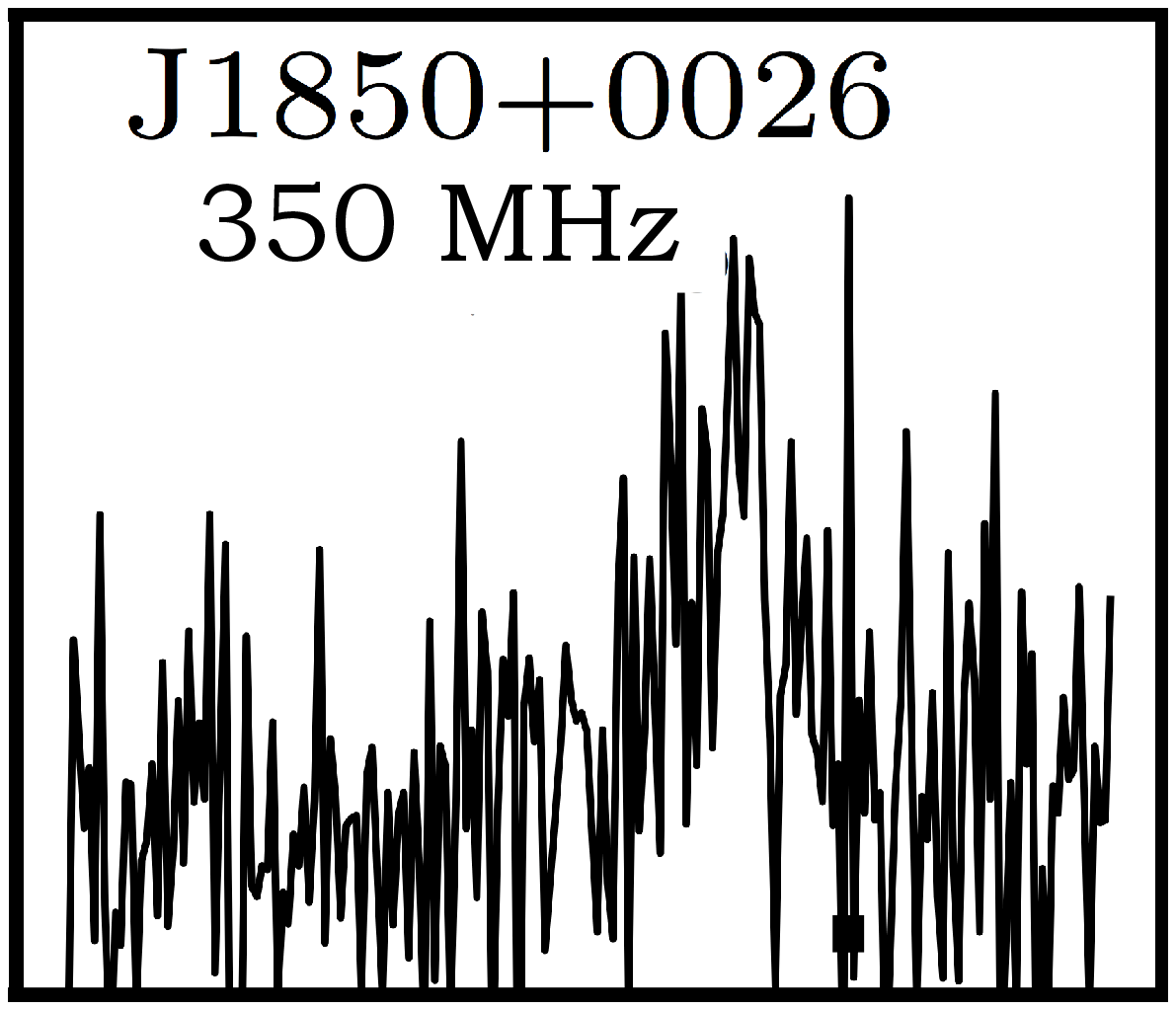}}}\\ 
{\mbox{\includegraphics[width=52mm]{J1848+0826L.ps}}}&
{\mbox{\includegraphics[width=52mm]{J1849+2423L.ps}}}& \ \ \ 
{\mbox{\includegraphics[width=52mm]{PQJ1850+0026.56415la.ps}}}\\ 
\end{tabular}
\caption{Scaled beam dimensions and average profiles for PSRs J1848+0826, J1849+2423 and J1850+0026 [350-MHz profile from \citet{mcewen}] as in Fig.~\ref{figA1}.}
\label{figA22}
\end{center}
\end{figure*}

\begin{figure*}
\begin{center}
\begin{tabular}{@{}ll@{}ll@{}}
& 
{\mbox{\includegraphics[width=52mm]{J1859+1526_Bmodel.ps}}}& \ \ \ 
{\mbox{\includegraphics[width=52mm]{J1900+30_Bmodel.ps}}}\\
{\mbox{\includegraphics[width=52mm]{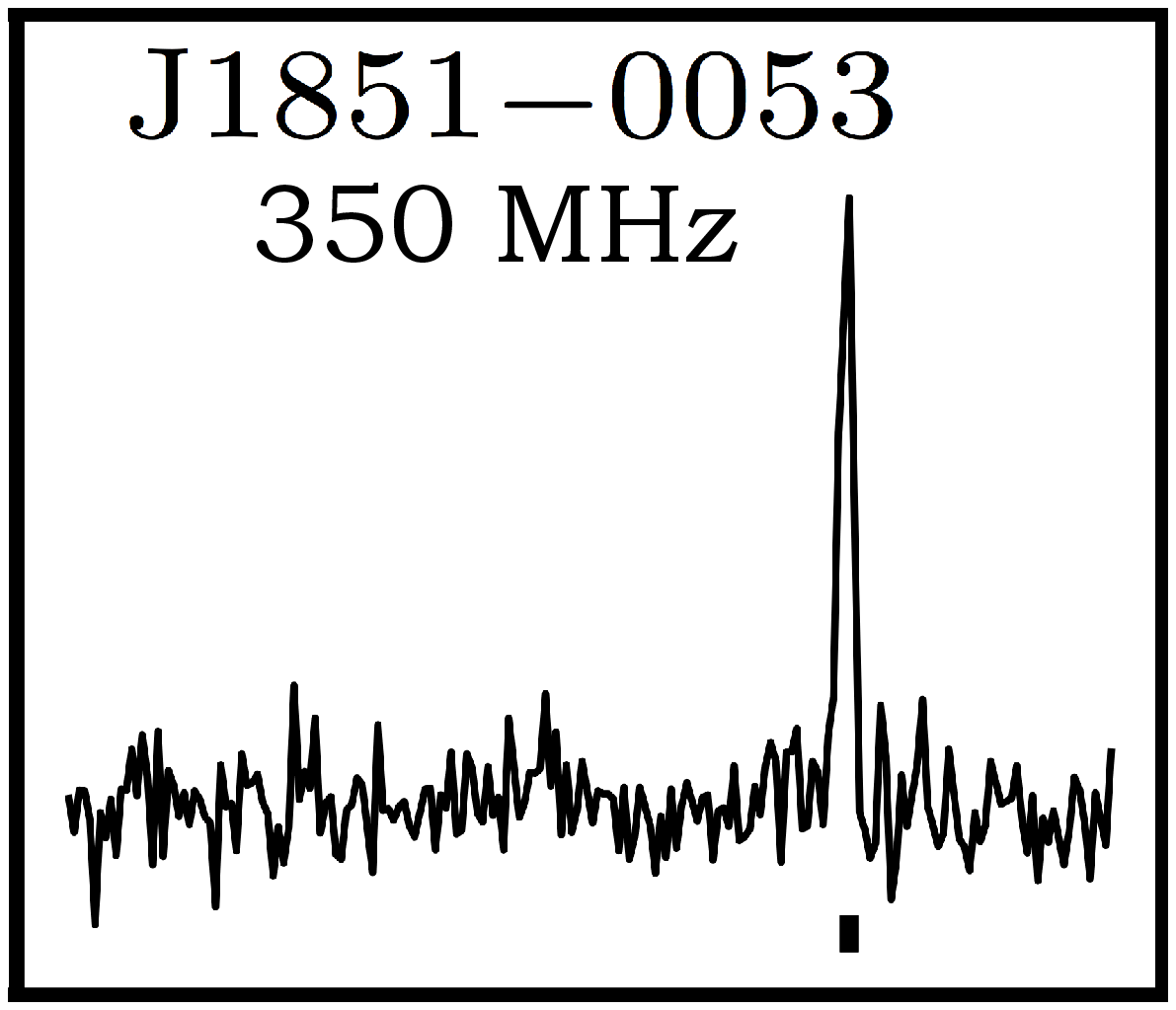}}}& 
{\mbox{\includegraphics[width=52mm]{J1859+1526P.ps}}}& \ \ \ 
{\mbox{\includegraphics[width=52mm]{J1900+30P.ps}}}\\ 
{\mbox{\includegraphics[width=52mm]{PQJ1851-0053.54540la.ps}}}&
{\mbox{\includegraphics[width=52mm]{J1859+1526L.ps}}}& \ \ \ 
{\mbox{\includegraphics[width=52mm]{J1900+30L.ps}}}\\ 
\end{tabular}
\caption{Scaled beam dimensions and average profiles for PSRs J1851$-$0053 [350-MHz profile from \citet{mcewen}], J1859+1526 and J1900+30 as in Fig.~\ref{figA1}.}
\label{figA23}
\end{center}
\end{figure*}

\begin{figure*}
\begin{center}
\begin{tabular}{@{}ll@{}ll@{}}
{\mbox{\includegraphics[width=52mm]{J1901+1306_Bmodel.ps}}}&
& 
{\mbox{\includegraphics[width=52mm]{J1906+1854_Bmodel.ps}}}\\
{\mbox{\includegraphics[width=52mm]{J1901+1306P.ps}}}& 
{\mbox{\includegraphics[width=52mm]{J1903+2225P.ps}}}& \ \ \ 
{\mbox{\includegraphics[width=52mm]{J1906+1854P.ps}}}\\ 
{\mbox{\includegraphics[width=52mm]{J1901+1306L.ps}}}&
{\mbox{\includegraphics[width=52mm]{J1903+2225L.ps}}}& \ \ \ 
{\mbox{\includegraphics[width=52mm]{J1906+1854L.ps}}}\\ 
\end{tabular}
\caption{Scaled beam dimensions and average profiles for PSRs J1901+1306, J1903+2225 and J1906+1854 as in Fig.~\ref{figA1}.}
\label{figA24}
\end{center}
\end{figure*}

\begin{figure*}
\begin{center}
\begin{tabular}{@{}ll@{}ll@{}}
{\mbox{\includegraphics[width=52mm]{J1908+2351_Bmodel.ps}}}&
{\mbox{\includegraphics[width=52mm]{J1909+1859_Bmodel.ps}}}& \ \ \ 
{\mbox{\includegraphics[width=52mm]{J1910+0714_Bmodel.ps}}}\\
{\mbox{\includegraphics[width=52mm]{J1908+2351P.ps}}}& 
{\mbox{\includegraphics[width=52mm]{J1909+1859P.ps}}}& \ \ \ 
{\mbox{\includegraphics[width=52mm]{J1910+0714P.ps}}}\\ 
{\mbox{\includegraphics[width=52mm]{J1908+2351L.ps}}}&
{\mbox{\includegraphics[width=52mm]{J1909+1859L.ps}}}& \ \ \ 
{\mbox{\includegraphics[width=52mm]{J1910+0714L.ps}}}\\ 
\end{tabular}
\caption{Scaled beam dimensions and average profiles for PSRs J1908+2351, J1909+1859 and J1910+0714 as in Fig.~\ref{figA1}.}
\label{figA25}
\end{center}
\end{figure*}

\begin{figure*}
\begin{center}
\begin{tabular}{@{}ll@{}ll@{}}
{\mbox{\includegraphics[width=52mm]{J1911+1758_Bmodel.ps}}}&
{\mbox{\includegraphics[width=52mm]{J1912+2525_Bmodel.ps}}}& \ \ \ 
{\mbox{\includegraphics[width=52mm]{J1913+3732_Bmodel.ps}}}\\
{\mbox{\includegraphics[width=52mm]{J1911+1758P.ps}}}& 
{\mbox{\includegraphics[width=52mm]{J1912+2525P.ps}}}& \ \ \ 
{\mbox{\includegraphics[width=52mm]{J1913+3732P.ps}}}\\ 
{\mbox{\includegraphics[width=52mm]{J1911+1758L.ps}}}&
{\mbox{\includegraphics[width=52mm]{J1912+2525L.ps}}}& \ \ \ 
{\mbox{\includegraphics[width=52mm]{J1913+3732L.ps}}}\\ 
\end{tabular}
\caption{Scaled beam dimensions and average profiles for PSRs J1911+1758, J1912+2525 and J1913+3732 as in Fig.~\ref{figA1}.}
\label{figA26}
\end{center}
\end{figure*}

\begin{figure*}
\begin{center}
\begin{tabular}{@{}ll@{}ll@{}}
{\mbox{\includegraphics[width=52mm]{J1913+0446_Bmodel.ps}}}&
{\mbox{\includegraphics[width=52mm]{J1914+0219_Bmodel.ps}}}& \ \ \ 
{\mbox{\includegraphics[width=52mm]{J1915+0227_Bmodel.ps}}}\\
{\mbox{\includegraphics[width=52mm]{J1913+0446P.ps}}}& 
{\mbox{\includegraphics[width=52mm]{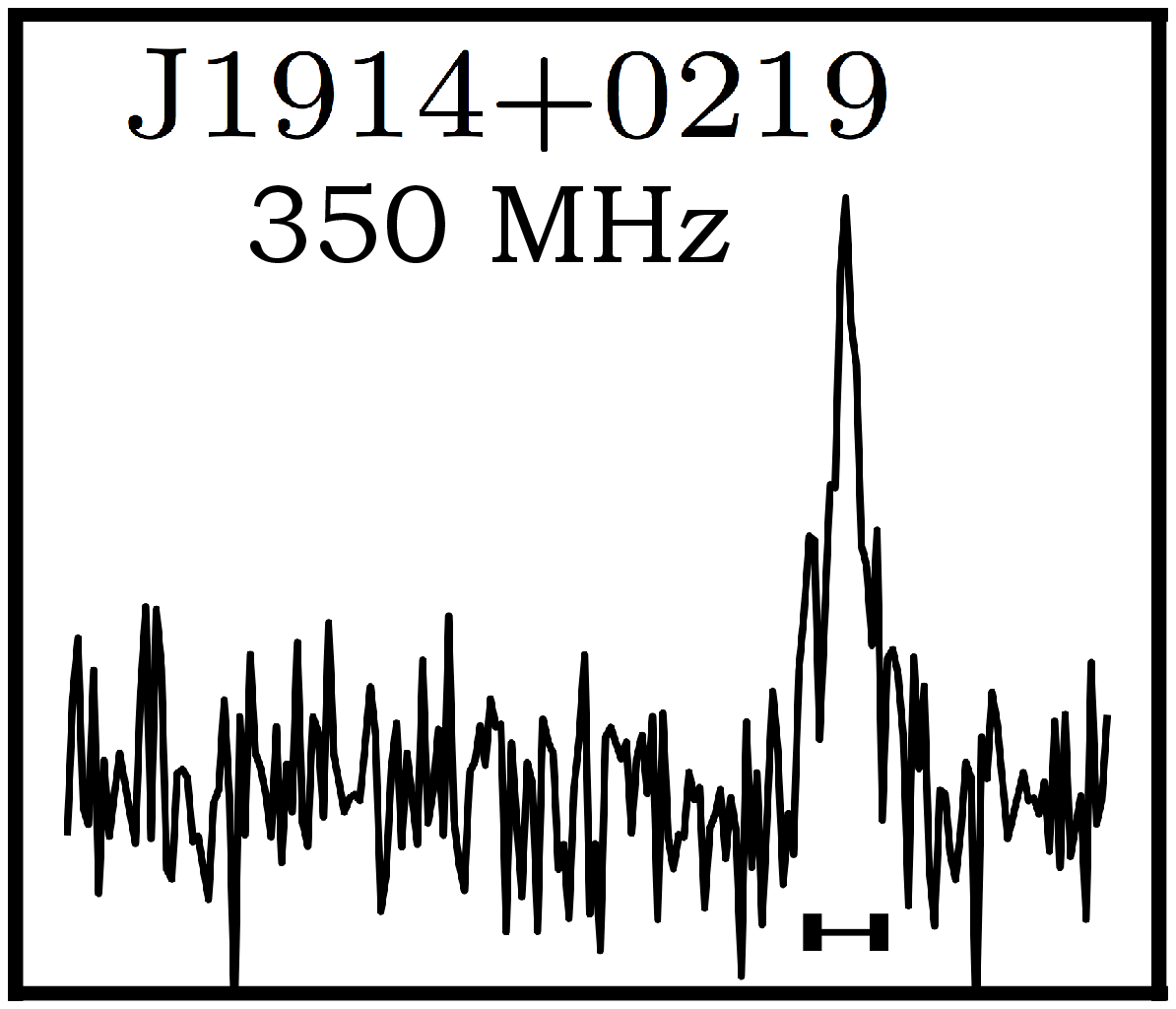}}}& \ \ \ 
{\mbox{\includegraphics[width=52mm]{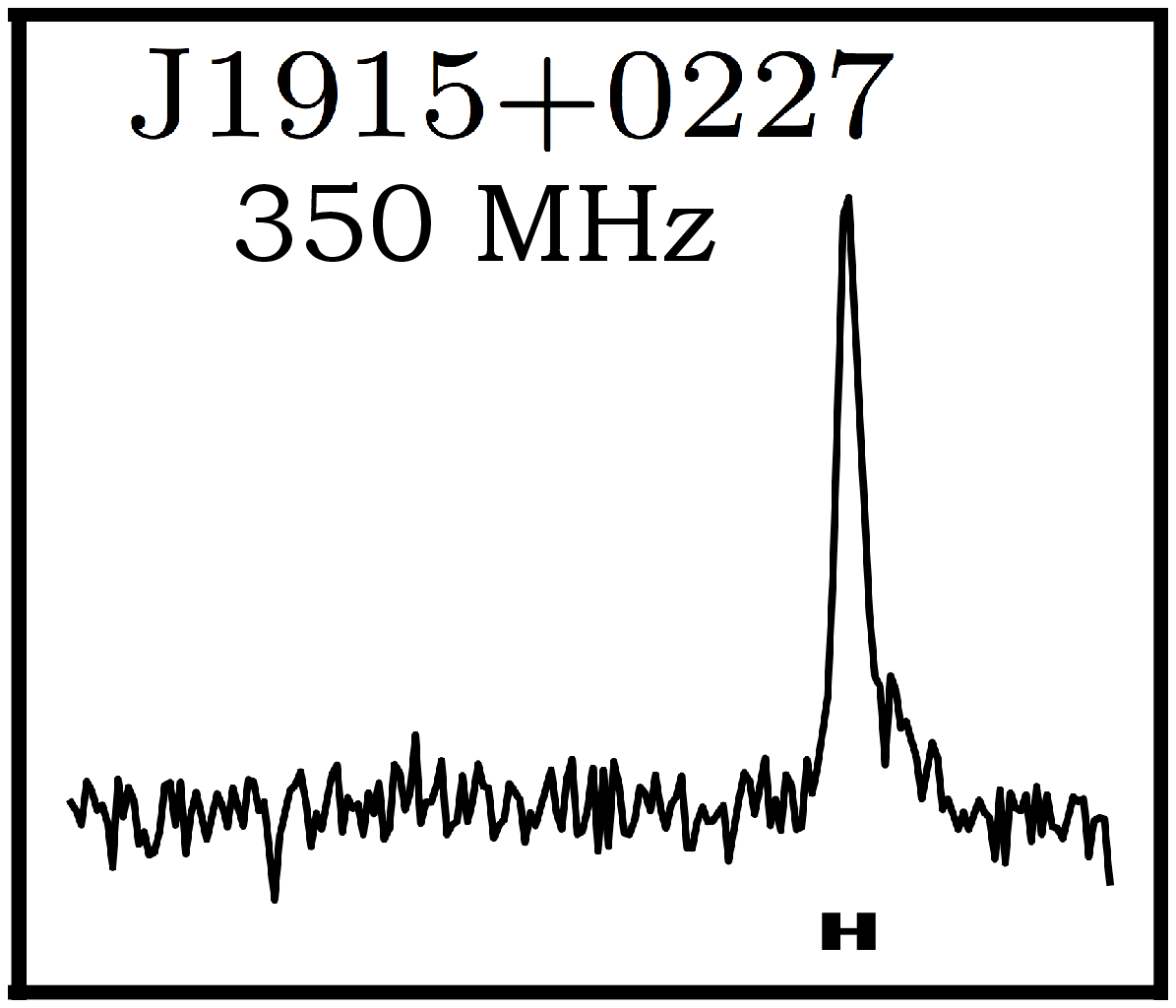}}}\\ 
{\mbox{\includegraphics[width=52mm]{J1913+0446L.ps}}}&
{\mbox{\includegraphics[width=52mm]{PQJ1914+0219.57131la.ps}}}& \ \ \ 
{\mbox{\includegraphics[width=52mm]{PQJ1915+0227.57131la.ps}}}\\ 
\end{tabular}
\caption{Scaled beam dimensions and average profiles for PSRs J1913+0446, J1914+0219 and J1915+0227 [350-MHz profiles from \citet{mcewen}] as in Fig.~\ref{figA1}.}
\label{figA27}
\end{center}
\end{figure*}

\begin{figure*}
\begin{center}
\begin{tabular}{@{}ll@{}ll@{}}
&
{\mbox{\includegraphics[width=52mm]{J1919+0134_Bmodel.ps}}}& \ \ \ 
{\mbox{\includegraphics[width=52mm]{J1938+0650_Bmodel.ps}}}\\
{\mbox{\includegraphics[width=52mm]{J1915+0738P.ps}}}& 
{\mbox{\includegraphics[width=52mm]{J1919+0134P.ps}}}& \ \ \ 
{\mbox{\includegraphics[width=52mm]{J1938+0650P.ps}}}\\ 
{\mbox{\includegraphics[width=52mm]{J1915+0738L.ps}}}&
{\mbox{\includegraphics[width=52mm]{J1919+0134L.ps}}}& \ \ \ 
{\mbox{\includegraphics[width=52mm]{J1938+0650L.ps}}}\\ 
\end{tabular}
\caption{Scaled beam dimensions and average profiles for PSRs J1915+0738, J1919+0134 (both 327-MHz CORRECTED) and J1938+0650 as in Fig.~\ref{figA1}.}
\label{figA28}
\end{center}
\end{figure*}

\begin{figure*}
\begin{center}
\begin{tabular}{@{}ll@{}ll@{}}
{\mbox{\includegraphics[width=52mm]{J1941+1026_Bmodel.ps}}}&
{\mbox{\includegraphics[width=52mm]{J1951+1123_Bmodel.ps}}}& \ \ \ 
{\mbox{\includegraphics[width=52mm]{J1953+1149_Bmodel.ps}}}\\
{\mbox{\includegraphics[width=52mm]{J1941+1026P.ps}}}& 
{\mbox{\includegraphics[width=52mm]{J1951+1123P.ps}}}& \ \ \ 
{\mbox{\includegraphics[width=52mm]{J1953+1149P.ps}}}\\ 
{\mbox{\includegraphics[width=52mm]{J1941+1026L.ps}}}&
{\mbox{\includegraphics[width=52mm]{J1951+1123L.ps}}}& \ \ \ 
{\mbox{\includegraphics[width=52mm]{J1953+1149L.ps}}}\\ 
\end{tabular}
\caption{Scaled beam dimensions and average profiles for PSRs J1941+1026, J1951+1123 and J1953+1149 (327-MHz CORRECTED) as in Fig.~\ref{figA1}.}
\label{figA29}
\end{center}
\end{figure*}

\begin{figure*}
\begin{center}
\begin{tabular}{@{}ll@{}ll@{}}
{\mbox{\includegraphics[width=52mm]{J1956+0838_Bmodel.ps}}}&
{\mbox{\includegraphics[width=52mm]{J1957+2831_Bmodel.ps}}}& \ \ \ 
{\mbox{\includegraphics[width=52mm]{J1959+3620_Bmodel.ps}}}\\
{\mbox{\includegraphics[width=52mm]{J1956+0838P.ps}}}& 
{\mbox{\includegraphics[width=52mm]{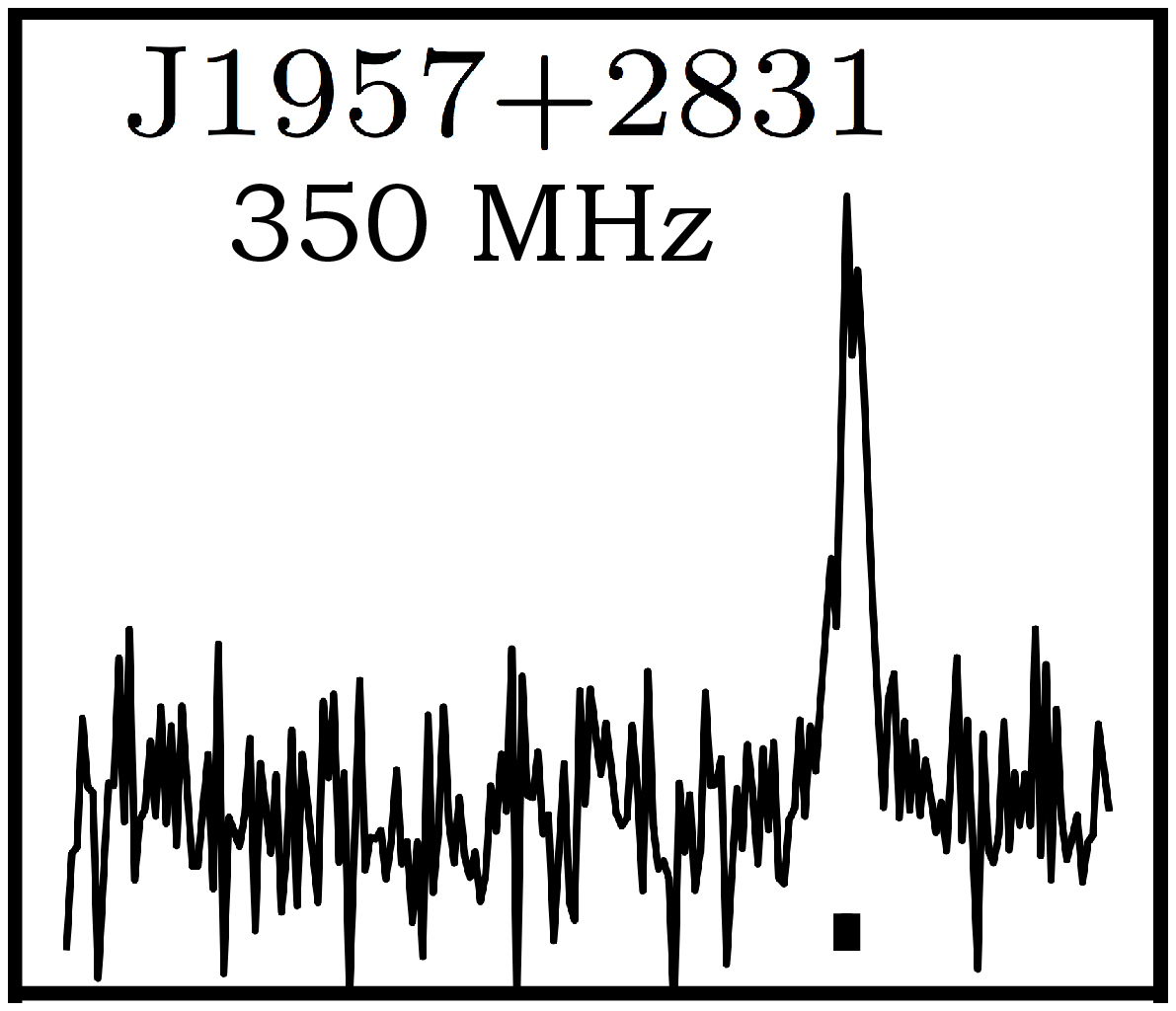}}}& \ \ \ 
{\mbox{\includegraphics[width=52mm]{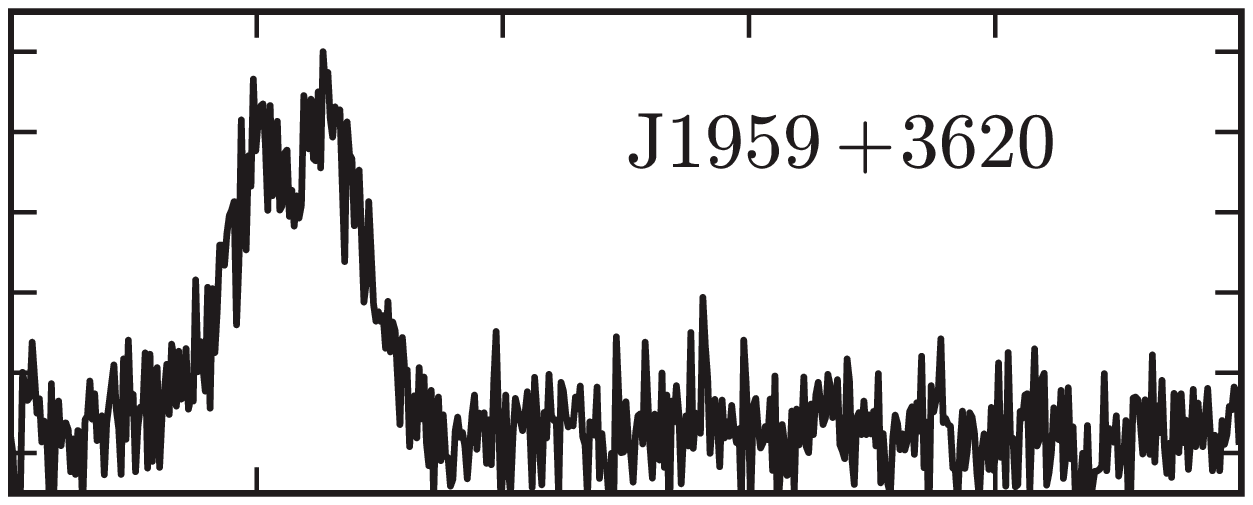}}}\\ 
{\mbox{\includegraphics[width=52mm]{J1956+0838L.ps}}}&
{\mbox{\includegraphics[width=52mm]{PQJ1957+2831.57981l3a.ps}}}& \ \ \ 
{\mbox{\includegraphics[width=52mm]{J1959+3620L.ps}}}\\ 
\end{tabular}
\caption{Scaled beam dimensions and average profiles for PSRs J1956+0838, J1957+2831 [350-MHz profile from \citet{mcewen}] and J1959+3620 [1.4-GHz profile from \citet{barr}] as in Fig.~\ref{figA1}.}
\label{figA30}
\end{center}
\end{figure*}

\begin{figure*}
\begin{center}
\begin{tabular}{@{}ll@{}ll@{}}
{\mbox{\includegraphics[width=52mm]{J2002+1637_Bmodel.ps}}}&
{\mbox{\includegraphics[width=52mm]{J2007+0809_Bmodel.ps}}}& \ \ \ 
{\mbox{\includegraphics[width=52mm]{J2007+0910_Bmodel.ps}}}\\
{\mbox{\includegraphics[width=52mm]{J2002+1637P.ps}}}& 
{\mbox{\includegraphics[width=52mm]{J2007+0809P.ps}}}& \ \ \ 
{\mbox{\includegraphics[width=52mm]{J2007+0910P.ps}}}\\ 
{\mbox{\includegraphics[width=52mm]{J2002+1637L.ps}}}&
{\mbox{\includegraphics[width=52mm]{J2007+0809L.ps}}}& \ \ \ 
{\mbox{\includegraphics[width=52mm]{pJ2007+0910.57639la0.ps}}}\\ 
\end{tabular}
\caption{Scaled beam dimensions and average profiles for PSRs J2002+1637, J2007+0809 and J2007+0910 (327-MHz CORRECTED) as in Fig.~\ref{figA1}.}
\label{figA31}
\end{center}
\end{figure*}

\begin{figure*}
\begin{center}
\begin{tabular}{@{}ll@{}ll@{}}
{\mbox{\includegraphics[width=52mm]{J2008+2513_Bmodel.ps}}}&
{\mbox{\includegraphics[width=52mm]{J2010+2845_Bmodel.ps}}}& \ \ \ 
{\mbox{\includegraphics[width=52mm]{J2015+2524_Bmodel.ps}}}\\
{\mbox{\includegraphics[width=52mm]{J2008+2513P.ps}}}& 
{\mbox{\includegraphics[width=52mm]{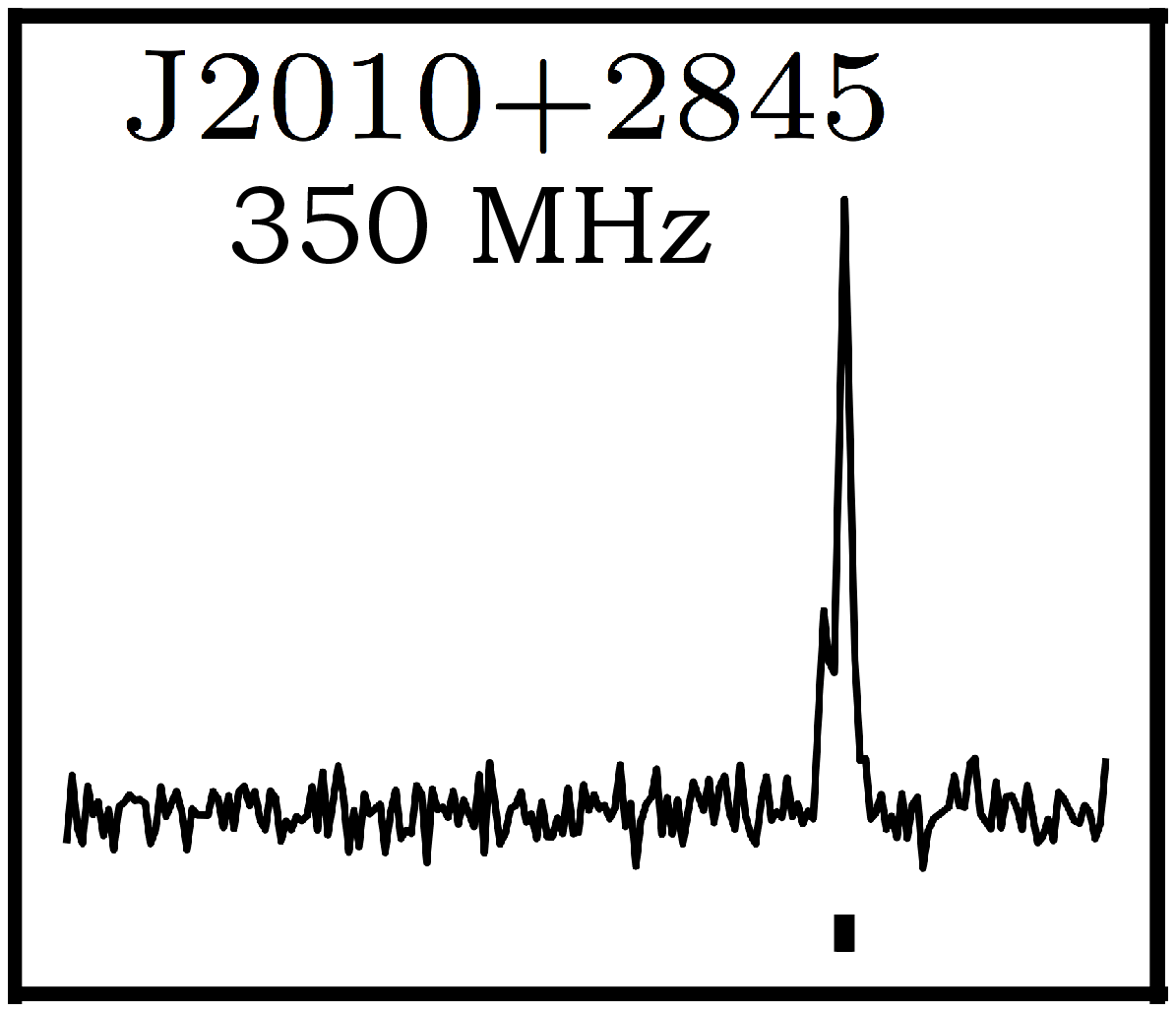}}}& \ \ \ 
{\mbox{\includegraphics[width=52mm]{J2015+2524P.ps}}}\\ 
{\mbox{\includegraphics[width=52mm]{J2008+2513L.ps}}}&
{\mbox{\includegraphics[width=52mm]{PQJ2010+2845.57115la.ps}}}& \ \ \ 
{\mbox{\includegraphics[width=52mm]{J2015+2524L.ps}}}\\ 
\end{tabular}
\caption{Scaled beam dimensions and average profiles for PSRs J2008+2513 (327-MHz CORRECTED), J2010+2845 [350-MHz profile from \citet{mcewen}] and J2015+2524 as in Fig.~\ref{figA1}.}
\label{figA32}
\end{center}
\end{figure*}

\begin{figure*}
\begin{center}
\begin{tabular}{@{}ll@{}ll@{}}
& 
{\mbox{\includegraphics[width=52mm]{J2017+2043_Bmodel.ps}}}& \ \ \ 
{\mbox{\includegraphics[width=52mm]{J2033+0042_Bmodel.ps}}}\\
{\mbox{\includegraphics[width=52mm]{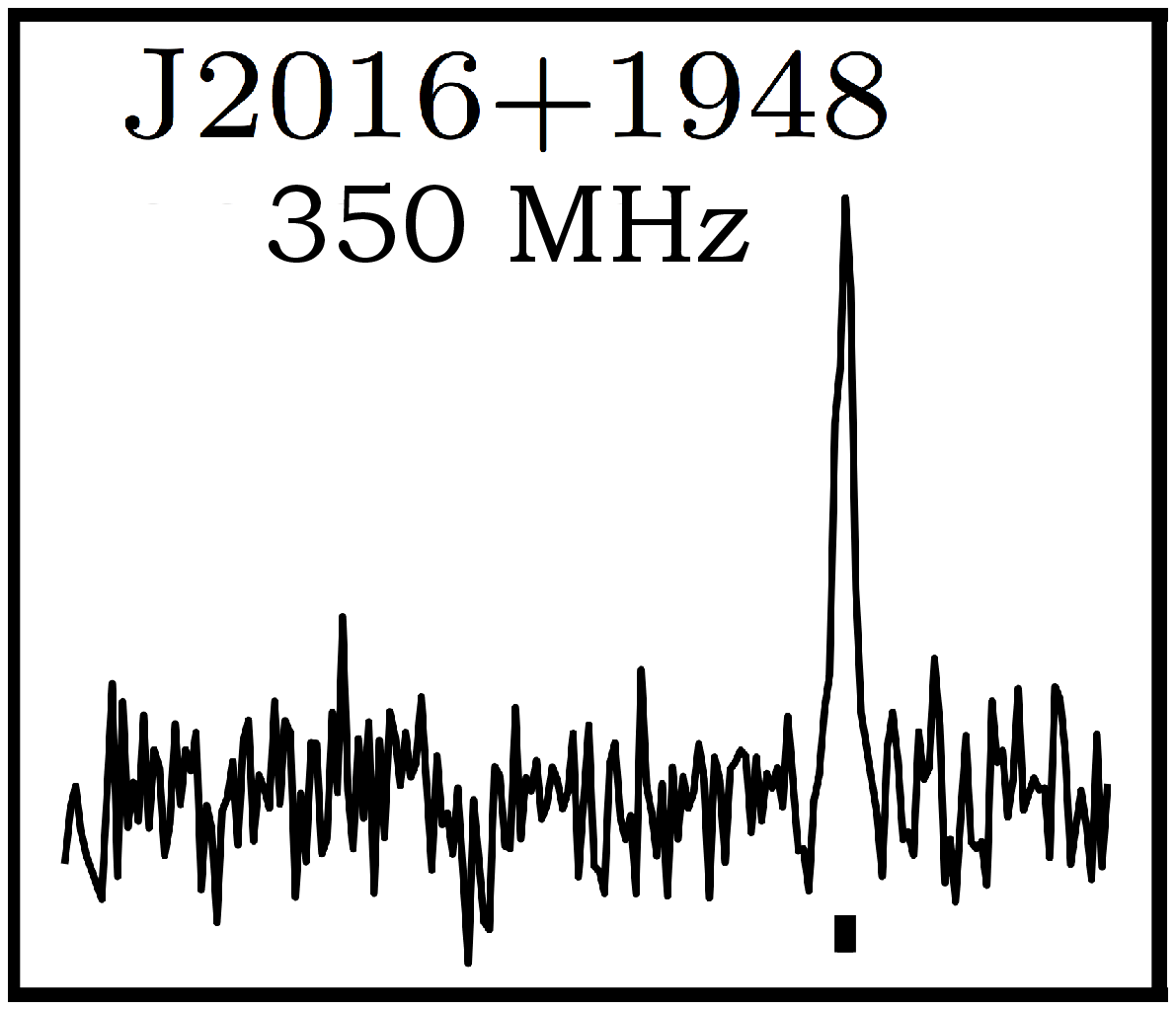}}}& 
{\mbox{\includegraphics[width=52mm]{J2017+2043P.ps}}}& \ \ \ 
{\mbox{\includegraphics[width=52mm]{J2033+0042P.ps}}}\\ 
{\mbox{\includegraphics[width=52mm]{PQJ2016+1948.57115la.ps}}}&
{\mbox{\includegraphics[width=52mm]{J2017+2043L.ps}}}& \ \ \ 
{\mbox{\includegraphics[width=52mm]{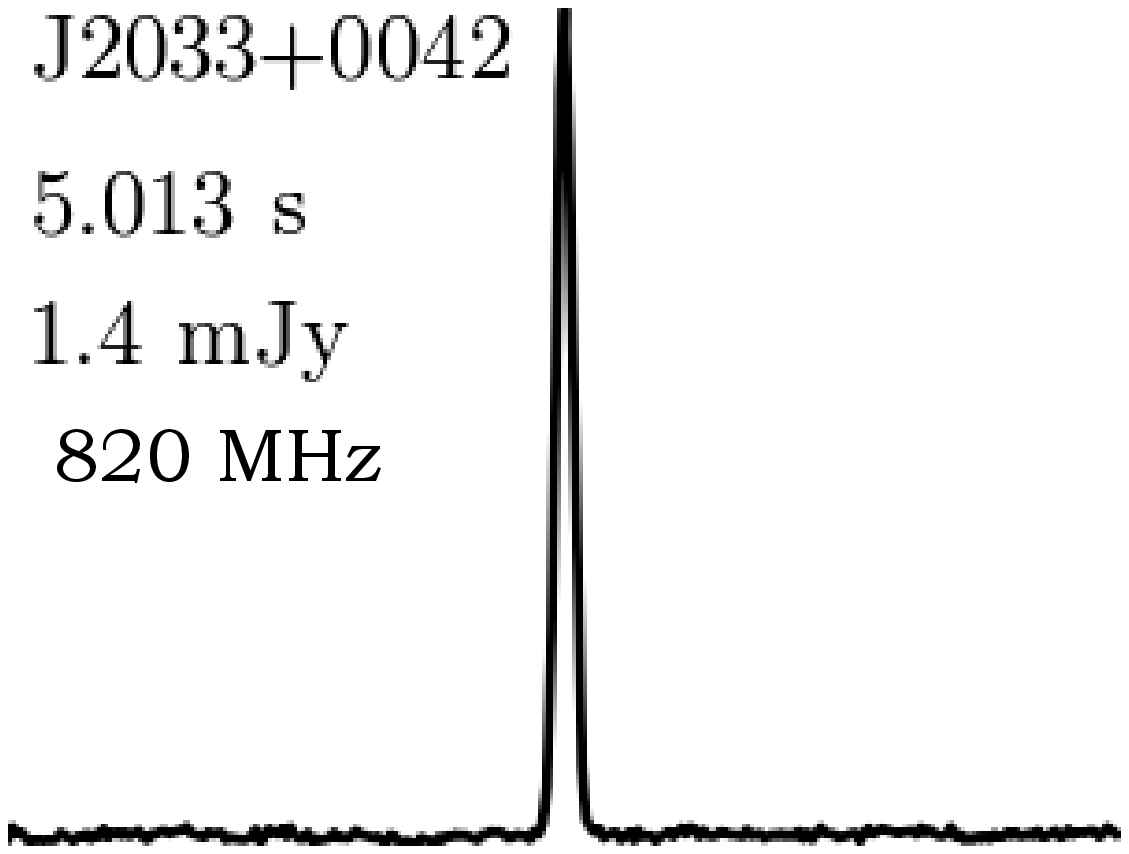}}}\\ 
\end{tabular}
\caption{Scaled beam dimensions and average profiles for PSRs J2016+1948 [350-MHz profile from \citet{mcewen}], J2017+2043 and J2033+0042 [820-MHz profile from \citet{lynch2013}] as in Fig.~\ref{figA1}.}
\label{figA33}
\end{center}
\end{figure*}

\begin{figure*}
\begin{center}
\begin{tabular}{@{}ll@{}ll@{}}
{\mbox{\includegraphics[width=52mm]{J2036+2835_Bmodel.ps}}}&
{\mbox{\includegraphics[width=52mm]{J2040+1657_Bmodel.ps}}}& \ \ \ 
{\mbox{\includegraphics[width=52mm]{J2043+2740_Bmodel.ps}}}\\
{\mbox{\includegraphics[width=52mm]{J2036+2835P.ps}}}& 
{\mbox{\includegraphics[width=52mm]{J2040+1657P.ps}}}& \ \ \ 
{\mbox{\includegraphics[width=52mm]{J2043+2740P.ps}}}\\ 
{\mbox{\includegraphics[width=52mm]{J2036+2835L.ps}}}&
{\mbox{\includegraphics[width=52mm]{J2040+1657L.ps}}}& \ \ \ 
{\mbox{\includegraphics[width=52mm]{J2043+2740L.ps}}}\\ 
\end{tabular}
\caption{Scaled beam dimensions and average profiles for PSRs J2036+2835, J2040+1657 and J2043+2740 as in Fig.~\ref{figA1}.}
\label{figA34}
\end{center}
\end{figure*}

\begin{figure*}
\begin{center}
\begin{tabular}{@{}ll@{}ll@{}}
{\mbox{\includegraphics[width=52mm]{J2045+0912_Bmodel.ps}}}&
{\mbox{\includegraphics[width=52mm]{J2048+2255_Bmodel.ps}}}& \ \ \ 
{\mbox{\includegraphics[width=52mm]{J2111+2106_Bmodel.ps}}}\\
{\mbox{\includegraphics[width=52mm]{J2045+0912P.ps}}}& 
{\mbox{\includegraphics[width=52mm]{J2048+2255P.ps}}}& \ \ \ 
{\mbox{\includegraphics[width=52mm]{J2111+2106P.ps}}}\\ 
{\mbox{\includegraphics[width=52mm]{J2045+0912L.ps}}}&
{\mbox{\includegraphics[width=52mm]{J2048+2255L.ps}}}& \ \ \ 
{\mbox{\includegraphics[width=52mm]{J2111+2106L.ps}}}\\ 
\end{tabular}
\caption{Scaled beam dimensions and average profiles for PSRs J2045+0912 (327-MHz CORRECTED), J2048+2255 and J2111+2106 as in Fig.~\ref{figA1}.}
\label{figA35}
\end{center}
\end{figure*}

\begin{figure*}
\begin{center}
\begin{tabular}{@{}ll@{}ll@{}}
{\mbox{\includegraphics[width=52mm]{J2139+2242_Bmodel.ps}}}&
{\mbox{\includegraphics[width=52mm]{J2151+2315_Bmodel.ps}}}& \ \ \ 
{\mbox{\includegraphics[width=52mm]{J2155+2813_Bmodel.ps}}}\\
{\mbox{\includegraphics[width=52mm]{J2139+2242P.ps}}}& 
{\mbox{\includegraphics[width=52mm]{J2151+2315P.ps}}}& \ \ \ 
{\mbox{\includegraphics[width=52mm]{J2155+2813P.ps}}}\\ 
{\mbox{\includegraphics[width=52mm]{J2139+2242L.ps}}}&
{\mbox{\includegraphics[width=52mm]{J2151+2315L.ps}}}& \ \ \ 
{\mbox{\includegraphics[width=52mm]{J2155+2813L.ps}}}\\ 
\end{tabular}
\caption{Scaled beam dimensions and average profiles for PSRs J2139+2242, J2151+2315 and J2155+2813 as in Fig.~\ref{figA1}.}
\label{figA36}
\end{center}
\end{figure*}

\begin{figure*}
\begin{center}
\begin{tabular}{@{}ll@{}ll@{}}
{\mbox{\includegraphics[width=52mm]{J2156+2618_Bmodel.ps}}}&
{\mbox{\includegraphics[width=52mm]{J2205+1444_Bmodel.ps}}}& \ \ \ 
{\mbox{\includegraphics[width=52mm]{J2215+1538_Bmodel.ps}}}\\
{\mbox{\includegraphics[width=52mm]{PQpJ2156+2618.53377pa0.ps}}}& 
{\mbox{\includegraphics[width=52mm]{J2205+1444P.ps}}}& \ \ \ 
{\mbox{\includegraphics[width=52mm]{J2215+1538P.ps}}}\\ 
{\mbox{\includegraphics[width=52mm]{J2156+2618L.ps}}}&
{\mbox{\includegraphics[width=52mm]{J2205+1444L.ps}}}& \ \ \ 
{\mbox{\includegraphics[width=52mm]{J2215+1538L.ps}}}\\ 
\end{tabular}
\caption{Scaled beam dimensions and average profiles for PSRs J2156+2618 (327-MHz CORRECTED), J2205+1444 and J2215+1538 as in Fig.~\ref{figA1}.}
\label{figA37}
\end{center}
\end{figure*}

\begin{figure*}
\begin{center}
\begin{tabular}{@{}ll@{}ll@{}}
{\mbox{\includegraphics[width=52mm]{J2222+2923_Bmodel.ps}}}&
{\mbox{\includegraphics[width=52mm]{J2234+2114_Bmodel.ps}}}& \ \ \ 
{\mbox{\includegraphics[width=52mm]{J2243+1518_Bmodel.ps}}}\\
{\mbox{\includegraphics[width=52mm]{J2222+2923P.ps}}}& 
{\mbox{\includegraphics[width=52mm]{J2234+2114P.ps}}}& \ \ \ 
{\mbox{\includegraphics[width=52mm]{J2243+1518P.ps}}}\\ 
{\mbox{\includegraphics[width=52mm]{J2222+2923L.ps}}}&
{\mbox{\includegraphics[width=52mm]{J2234+2114L.ps}}}& \ \ \ 
{\mbox{\includegraphics[width=52mm]{J2243+1518L.ps}}}\\ 
\end{tabular}
\caption{Scaled beam dimensions and average profiles for J2222+2923, PSRs J2234+2114 and J2243+1518 (1400-MHz CORRECTED) as in Fig.~\ref{figA1}.}
\label{figA38}
\end{center}
\end{figure*}

\begin{figure*}
\begin{center}
\begin{tabular}{@{}ll@{}ll@{}}
&
{\mbox{\includegraphics[width=52mm]{J2253+1516_Bmodel.ps}}} \ \ \ 
{\mbox{\includegraphics[width=52mm]{J2307+2225_Bmodel.ps}}}\\
{\mbox{\includegraphics[width=52mm]{J2248-0101P.ps}}}&
{\mbox{\includegraphics[width=52mm]{J2253+1516P.ps}}} \ \ \  
{\mbox{\includegraphics[width=52mm]{J2307+2225P.ps}}}\\ 
{\mbox{\includegraphics[width=52mm]{J2248-0101L.ps}}}&
{\mbox{\includegraphics[width=52mm]{J2253+1516L.ps}}} \ \ \ 
{\mbox{\includegraphics[width=52mm]{J2307+2225L.ps}}}\\ 
\end{tabular}
\caption{Scaled beam dimensions and average profiles for PSRs J2248--0101 (324-MHz CORRECTED), J2253+1516 and J2307+2225 as in Fig.~\ref{figA1}.}
\label{figA39}
\end{center}
\end{figure*}

\begin{figure}
\begin{center}
\begin{tabular}{@{}ll@{}ll@{}}
{\mbox{\includegraphics[width=52mm]{J2355+2246_Bmodel.ps}}}\\
{\mbox{\includegraphics[width=52mm]{J2355+2246P.ps}}}\\ 
{\mbox{\includegraphics[width=52mm]{J2355+2246L.ps}}}\\ 
\end{tabular}
\caption{Scaled beam dimensions and average profiles for PSRs J2355+2246 as in Fig.~\ref{figA1}.}
\label{figA40}
\end{center}
\end{figure}

\label{lastpage}

\end{document}